\documentclass[12pt,onecolumn]{article}

\usepackage{color}
\usepackage{amsmath}
\usepackage{amssymb}
\usepackage{amsfonts}
\usepackage{geometry}
\usepackage{setspace}
\usepackage{amsmath,bm}
\usepackage{graphicx}
\usepackage[section]{placeins}% avoid figures cross the section
\usepackage{subfigure}
\usepackage[sort&compress]{natbib}
\usepackage{booktabs}  
\usepackage{makecell}
\usepackage{caption}
\usepackage[colorlinks, citecolor=blue]{hyperref}% link to the superlink
\usepackage{lineno}% line number
\author{Shuwei Zhou$^{1,2}$, Xiaoying Zhuang$^{1,2,*}$, Timon Rabczuk$^3$}
\title {Phase field method for quasi-static hydro-fracture in porous media under stress boundary condition considering the effect of initial stress field}
\begin{document}
%\linenumbers % line numbers
% Figure title
\captionsetup[figure]{labelfont={bf},name={Fig.},labelsep=space}
% referecne style
\bibliographystyle{elsarticle-num-names}
\setcitestyle{numbers,square,aysep={},yysep={,},citesep={,}}
% delte the date
\date{}
\maketitle

\spacing {2}
\noindent
1 Department of Geotechnical Engineering, College of Civil Engineering, Tongji University, Shanghai 200092, P.R. China\\
2 Institute of Continuum Mechanics, Leibniz University Hannover, Hannover 30167, Germany\\
3 Institute of Structural Mechanics, Bauhaus University Weimar, Weimar 99423, Germany\\
$*$ Corresponding author: Xiaoying Zhuang (zhuang@ikm.uni-hannover.de)

\begin{abstract}
\noindent Phase field model (PFM) is an efficient fracture modeling method and has high potential for hydraulic fracturing (HF). However, the current PFMs in HF do not consider well the effect of in-situ stress field and the numerical examples of porous media with stress boundary conditions were rarely presented. The main reason is that if the remote stress is applied on the boundaries of the calculation domain, there will be relatively large deformation induced on these stress boundaries, which is not consistent with the engineering observations. To eliminate this limitation, this paper proposes a new phase field method to describe quasi-static hydraulic fracture propagation in porous media subjected to stress boundary conditions, and the new method is more in line with engineering practice. A new energy functional, which considers the effect of initial in-situ stress field, is established and then it is used to achieve the governing equations for the displacement and phase fields through the variational approach. Biot poroelasticity theory is used to couple the fluid pressure field and the displacement field while the phase field is used for determining the fluid properties from the intact domain to the fully broken domain. In addition, we present several 2D and 3D examples to show the effects of in-situ stress on hydraulic fracture propagation. The numerical examples indicate that under stress boundary condition our approach obtains correct displacement distribution and it is capable of capturing complex hydraulic fracture growth patterns.
\end{abstract}

 
%\twocolumn
\section {Introduction}\label{Introduction}

Hydraulic fracture propagation in porous media is one of the most attractive and significant research topics in mechanical, geological, energy, and environmental engineering. In particular, hydraulic fracturing (HF) \citep{settari1980simulation} has been widely used to exploit oil, tight gas, and shale gas from reservoirs that were unexploitable in past decades. The main reason for this is that the injection of highly pressurized fluid into a reservoir forms a fracture network for resource transportation. Another application of HF is the measurement of in-situ stress \citep{bredehoeft1976hydraulic}. In addition, HF can be also applied in an enhanced geothermal system to accelerate heat extraction \citep{haring2008characterisation}. Nevertheless, despite its advantages, HF still brings some controversies because the fracturing fluid may leak and further contaminate the underground space and surface due to unfavorable fracture growth paths \citep{osborn2011methane, vidic2013impact}. Therefore, an accurate numerical tool is critically important in the prediction of complex hydraulic fracture propagation in porous media.

However, correctly modeling hydraulic fracture in porous media is challenging and full of complexity due to solid-fluid interaction, fracture network, and different boundary conditions. This has prompted the development of various numerical methods for modeling fracturing processes, and these approaches can be classified into two types in the continuum framework: the discrete method and the smeared method. The discrete methods introduce displacement discontinuity for fractures and among the most popular are the extended finite element method (XFEM) \citep{moes2002extended, chen2012extended}, generalized finite element method (GFEM) \citep{fries2010extended}, boundary element method \citep{fu2013boundary, fu2018singular, fu2018boundary}, phantom-node method \citep{chau2012phantom, rabczuk2008new}, and element-erosion method \citep{belytschko1987three, johnson1987eroding}. On the other hand, in the smeared methods the displacement is continuous across a fracture and the gradient damage model \citep{peerlings1996some}, the screened Poisson method \citep{areias2016damage}, and the phase field model (PFM) \citep{bourdin2008variational, miehe2010thermodynamically, miehe2010phase, zhou2018phase3, zhou2019phase2} are the best-known.

Among all the fracture modeling methods, the PFMs are attracting more and more attention in recent years. In this method, an additional scalar field $\phi\in[0,1]$ is used to reflect the extent of fracture where $\phi=0$ represents an intact material and $\phi=1$ indicates a fully broken material (some literatures \citep{ambati2015review} use a phase field $s=1-\phi$ and $s=0$, 1 represents the fully broken state and intact state, respectively). In addition, the transition zone with $\phi\in(0,1)$ is controlled by an intrinsic length scale parameter $l_0$. After being first proposed by \citet{bourdin2008variational}, the PFM was further promoted by \citet{bourdin2008variational, miehe2010thermodynamically, miehe2010phase, borden2012phase, hofacker2012continuum, hofacker2013phase}. Compared with XIGA \citep{khatir2019fast}, XFEM \citep{martinez2017numerical, khatir2019computational}, cohesive zone model\citep{PEREIRA2020105899, pereira2018prediction}, and continuum damage model \citep{bhatti2018fretting}, PFM has ease of implementation. Furthermore, the recent development has shown that PFMs have a high efficiency in predicting complex fracture propagation patterns, even in 3D situations due to these reasons: i) simulations can be performed on a fixed mesh without any remeshing or adaptive technique; ii) complex fracture patterns such as branching and junction are automatically captured; iii) external fracture criteria or fracture surface tracking algorithms are not required; iv) PFMs can easily simulate fracture propagation in heterogeneous media, and v) no penetration criteria are required for hydraulic fracture when a layer interface is encountered.

The aforementioned advantages have also contributed to the development of PFMs in hydraulic fracturing. For example, in recent years many researchers have tried to couple the PFMs to HF and made some progress \citep{bourdin2012variational, wheeler2014augmented, mikelic2015quasi, mikelic2015phase, heister2015primal, lee2016pressure, wick2016fluid, yoshioka2016variational, miehe2015minimization, miehe2016phase, ehlers2017phase, santillan2017phase, zhou2018phase2}. However, the presented examples in these contributions all only established homogeneous Dirichlet boundary conditions for the displacement field. The current PFMs in HF therefore cannot consider well the effect of in-situ stress on fracture propagation. The main reason is that if the remote stress is applied on the boundaries of the calculation domain, there will be relatively large deformation induced on these stress boundaries, which is not consistent with the engineering observations. In fact, in geological environment, the displacement on the boundaries where the remote stress acts should be zero before HF. It should be noted that although a recently developed PFM \citep{shiozawaeffect} attempted to analyze the effect of stress boundary condition on fracture pattern, the main drawback remained unsolved because a large deformation was observed on the stress boundaries.

To eliminate the limitation from the initial stress field, this paper proposes a new phase field method to describe quasi-static hydraulic fracture propagation in porous media subjected to stress boundary conditions. A new energy functional, which fully considers the effect of initial in-situ stress field, is established and then it is used to achieve the governing equations of strong form for the displacement and phase fields through the variational approach. Biot poroelasticity theory is used to couple the fluid pressure field and the displacement field while the phase field is used for determining the fluid properties from the intact domain to the fully broken domain. In addition, several 2D and 3D examples are presented to show the effects of in-situ stress on hydraulic fracture propagation. The numerical examples indicate that under stress boundary conditions the new PFM obtains correct displacement distribution and it is capable of capturing well complex hydraulic fracture growth patterns.

This paper is organized as follows. Section \ref{mathematical model} describes the mathematical model for the new PFM and Section \ref{Numerical algorithm} shows the global numerical algorithm using a staggered scheme. In Sections \ref{2D examples} and \ref{3D example} the 2D and 3D examples are presented to demonstrate the capability of the new model. The present work and outlook for future development are concluded in Section \ref{Conclusions}.

\section {Mathematical model}\label{mathematical model}
\subsection {New energy functional}\label{subsection New energy functional}

\begin{figure}[htbp]
	\centering
	\subfigure[Sharp fracture]{\includegraphics[height = 5cm]{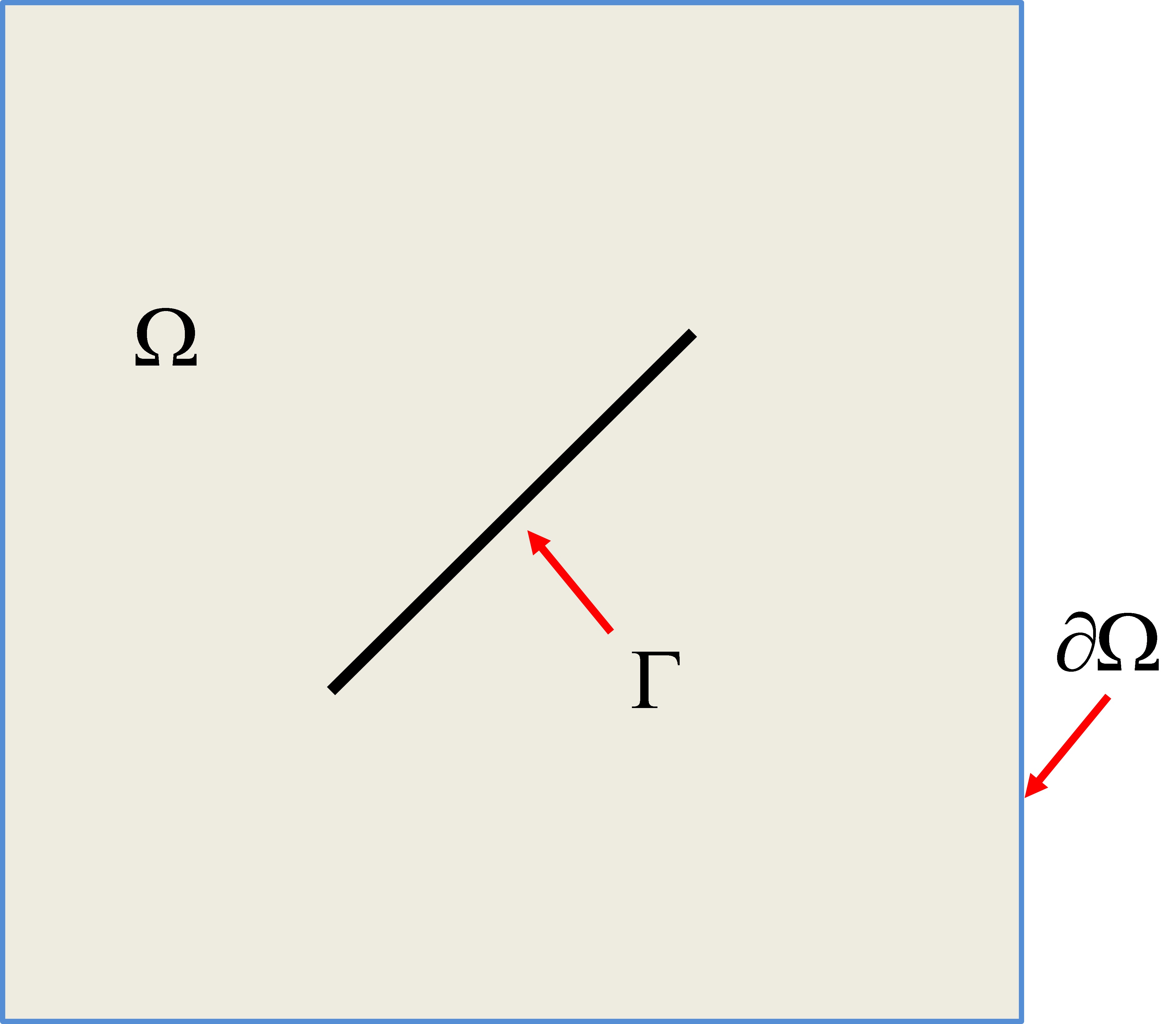}}
	\subfigure[Diffusive fracture]{\includegraphics[height = 5cm]{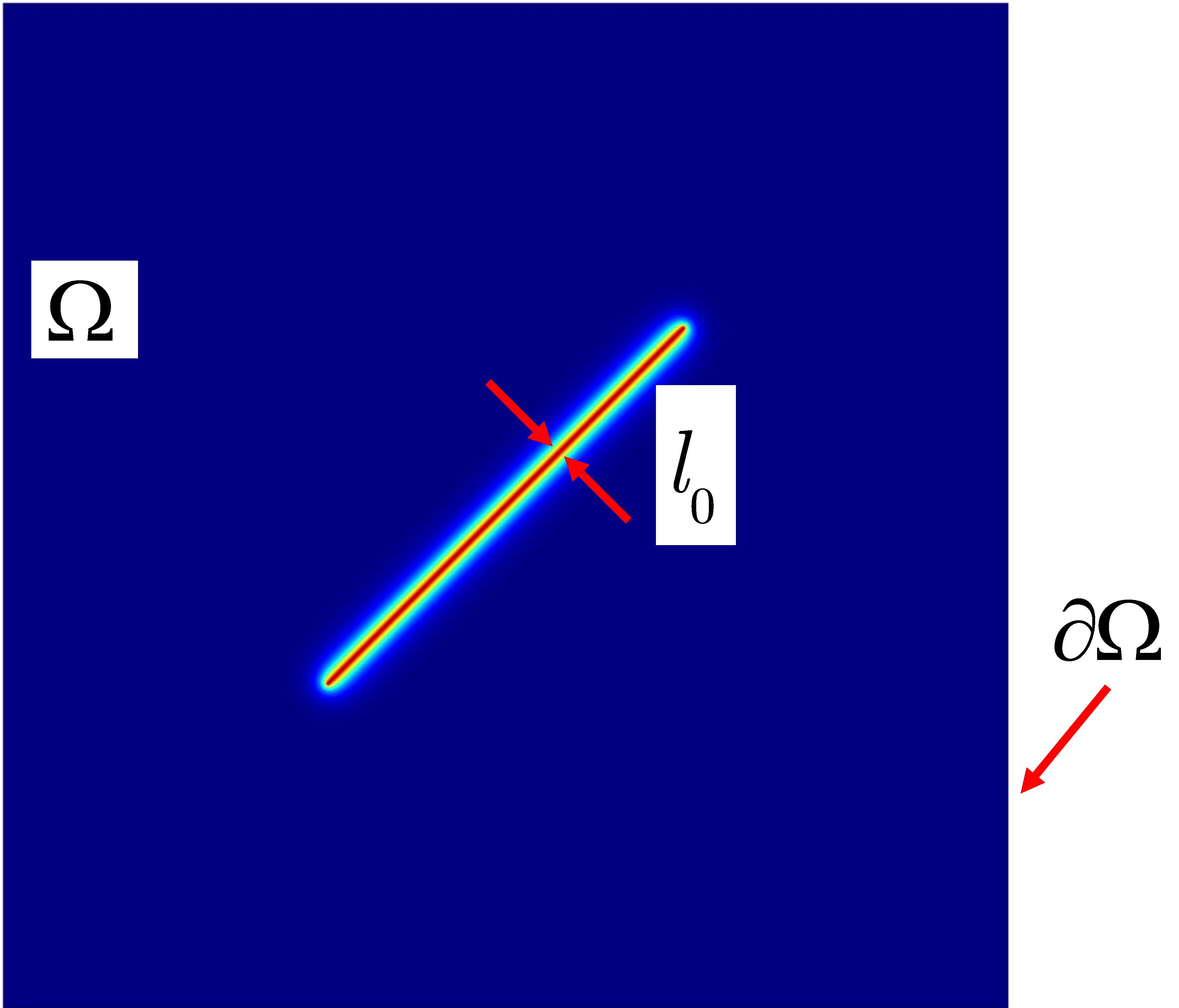}}
	\caption{Sharp and diffusive fractures in the poro-elastic medium}
	\label{Sharp and diffusive fractures in the poro-elastic medium}
	\end{figure}

Let us consider a poro-elastic domain $\Omega\subset \mathbb R^d$ ($d\in \{2,3\} $) in Fig. \ref{Sharp and diffusive fractures in the poro-elastic medium} where the external and internal discontinuity boundaries are denoted as $\partial \Omega$ and $\Gamma$, respectively. The porous medium is assumed homogeneous and isotropic with compressible and viscous fluid in the pores. Let
$T>0$ be the computational time interval and $\bm u(\bm x,t)\subset \mathbb R^d$ be the displacement field at time $t\in[0,T]$ and the position $\bm x$. In addition, the displacement field must satisfy the time-dependent Dirichlet boundary conditions, $u_i(\bm x,t)=g_i(\bm x,t)$, on $\partial \Omega_{g_i} \in \partial\Omega$, and the Neumann condition on $\partial \Omega_{h}$.

The stress, displacement, and fluid pressure of the porous domain are shown in Fig. \ref{Stress, displacement and fluid pressure in geological environment}. Because of the long-term consolidation or other geological effects, the domain has formed an initial stress field $\bm\sigma_0$, initial displacement field $\bm u_0$, and fluid pressure $p_0$, as shown in Fig. \ref{Stress, displacement and fluid pressure in geological environment}a. On the other hand, in Fig. \ref{Stress, displacement and fluid pressure in geological environment}b, after the domain is excavated or fractured by fluid injection, the stress, displacement, and fluid pressure are $\bm \sigma$, $\bm u_0+\bm u$, and $p_0+p$, respectively. Here, the displacement $\bm u$ and pressure $p$ are the induced relative displacement and fluid pressure by engineering activities such as HF. In this work, our presented method is only for calculating $\bm \sigma$, $\bm u$, $p$, and the hydraulic fracture pattern in the case that $\bm \sigma_0$, $\bm u_0$, and $p_0$ are known in advance, while how to obtain these initial fields is beyond the scope of this paper.

It should be noted that for the porous medium in the geological environment, the displacement $\bm u_0$ resulting from long-term geo-stress is always ignored in the stability analysis for underground engineering \citep{zhou2015analytical, zhou2018long} and only the displacement $\bm u$ caused by fracture formation or engineering excavation is calculated. That is, the porous medium $\Omega$ is subjected to an initial stress field $\bm\sigma_0$ and if the stress in $\Omega$ is equal to $\bm\sigma_0$, the displacement field must be 0. In addition, the initial fluid pressure $p_0$ is set as 0 for the purpose of simplicity and because the fluid pressure $p$ results only from the relative displacement $\bm u$ and fluid injection \citep{zhou2018phase2}.

	\begin{figure}[htbp]
	\centering
	\subfigure[Initial state]{\includegraphics[height = 5cm]{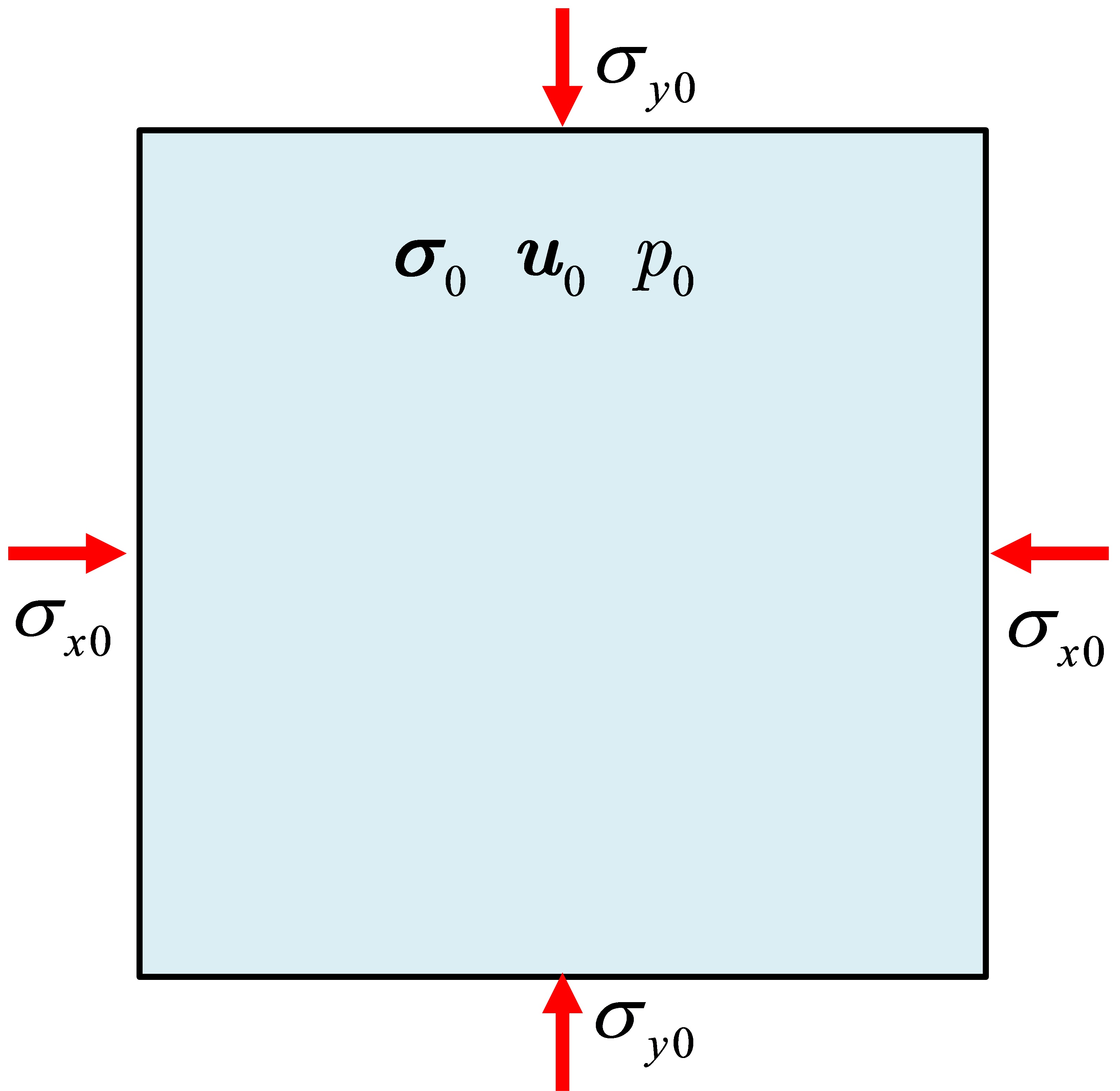}}
	\subfigure[After excavation or HF]{\includegraphics[height = 5cm]{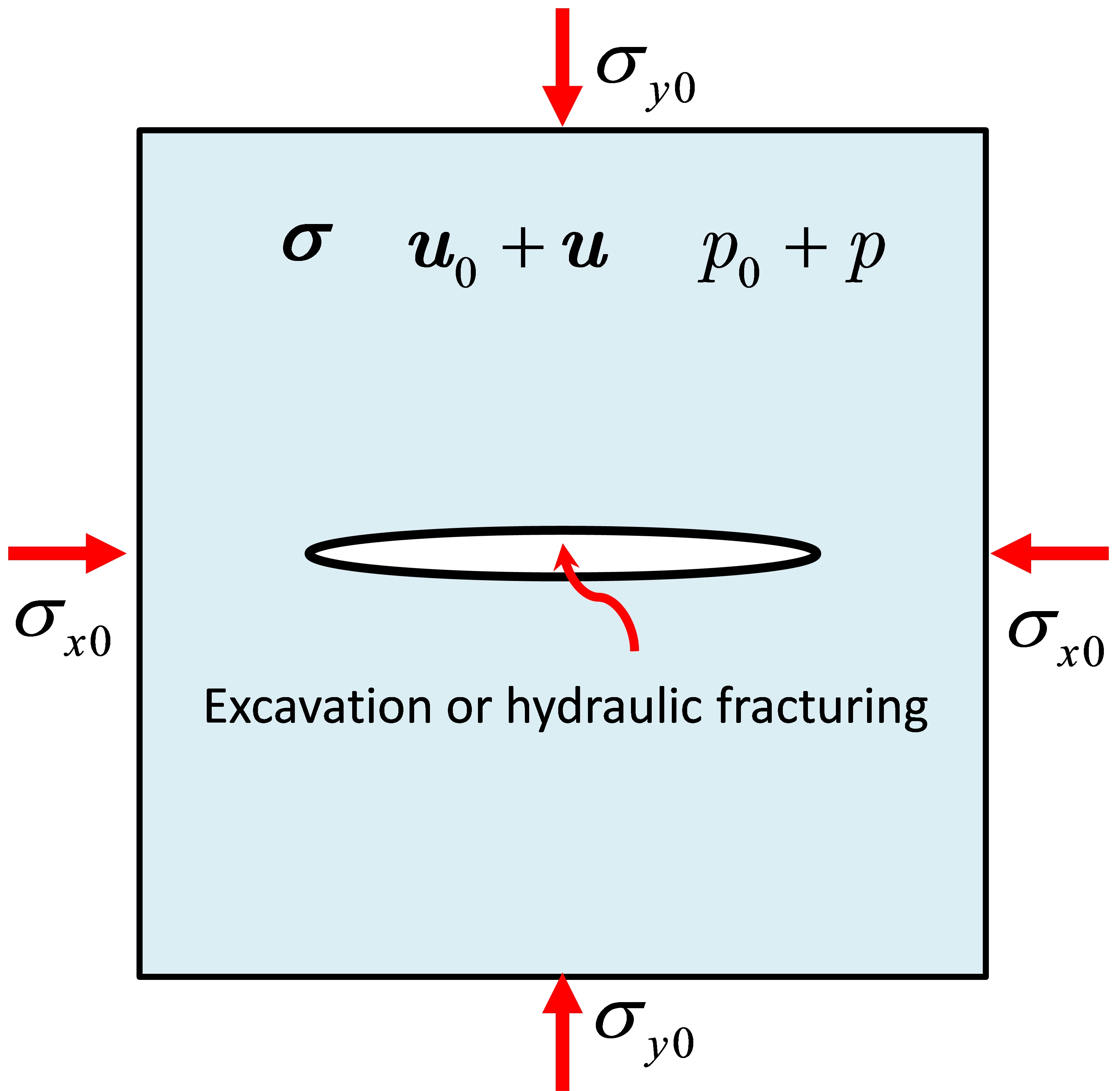}}
	\caption{Stress, displacement and fluid pressure in geological environment}
	\label{Stress, displacement and fluid pressure in geological environment}
	\end{figure}

If the body force is ignored, the basic idea of the previous phase field models for porous-elastic media \citep{zhou2018phase2} is to construct an energy functional $\Psi$ composed of the elastic energy $\Psi_{\varepsilon}(\bm \varepsilon)$, fracture energy $\Psi_f$, external work $W_{ext}$, and the energy contribution of fluid pressure $p$:
	\begin{equation}
	\Psi(\bm u,p,\Gamma) = \underbrace{\int_{\Omega \backslash \Gamma}\psi_{\varepsilon}(\bm \varepsilon) \mathrm{d}{\Omega}}_{\Psi_{\varepsilon}}\underbrace{-\int_{\Omega}\alpha p \cdot (\nabla \cdot \bm u) \mathrm{d}{\Omega}}_{\text{pressure-related term}}+\underbrace{\int_{\Gamma}G_{c} \mathrm{d}\Gamma}_{\Psi_f}\underbrace{- \int_{\partial\Omega_{t}} \bm {f}_t\cdot{\bm u}\mathrm{d}S}_{W_{ext}}
	\label{Original energy functional}
	\end{equation}

\noindent where $\psi_{\varepsilon}(\bm \varepsilon)$ is the elastic energy density; $\alpha\in(0,1]$ is the Biot coefficient; $G_{c}$ is the critical energy release rate; $\bm f_t$ is the traction on the Neumann boundary on $\mathrm{d}S$. In addition, the linear strain tensor $\bm \varepsilon$ is given by
	\begin{equation}
	\varepsilon_{ij}=\frac 1 2 \left(\frac{\partial u_i}{\partial x_j}+\frac{\partial u_j}{\partial x_i}\right)
	\end{equation}

\noindent and for an isotropic linear elastic medium, the elastic energy density $\psi_{\varepsilon}(\bm \varepsilon)$ reads \citep{miehe2010phase}
	\begin{equation}
	\psi_{\varepsilon}(\bm \varepsilon) = \frac{1}{2}\lambda\varepsilon_{ii}\varepsilon_{jj}+\mu\varepsilon_{ij}\varepsilon_{ij}
	\end{equation}

\noindent where $\lambda,\mu>0$ are the Lam\'e constants and
	\begin{equation}
	\left\{
	\begin{aligned}
	\lambda&=\frac{E\nu}{(1+\nu)(1-2\nu)}
	\\ 	\mu&=\frac{E}{2(1+\nu)}
	\end{aligned}\right.
	\end{equation}

\noindent with $E$ and $\nu$ being Young's modulus and Poisson's ratio, respectively.

However, as seen in Eq.\eqref{Original energy functional}, the previously used energy functional  cannot account for the effect of the initial stress field and it is therefore not suitable for modeling fracture propagation in a geomaterial under stress boundary condition. An evidence for this can be observed in \citet{shiozawaeffect} where a stress boundary has a relatively large deformation when the fluid pressure is 0. Hence, to deal with the stress boundary and account for the effect of the initial stress field, we establish a new energy functional $L$ for the phase field modeling:
	\begin{equation}
	L(\bm u,p,\Gamma) = \underbrace{\int_{\Omega \backslash \Gamma}\psi_{\varepsilon}(\bm \varepsilon) \mathrm{d}{\Omega}}_{\Psi_{\varepsilon}}+\underbrace{\int_{\Omega \backslash \Gamma}\bm\sigma_0:\bm\varepsilon \mathrm{d}{\Omega}}_{\text{Initial stress induced}}	
	\underbrace{-\int_{\Omega}\alpha p \cdot (\nabla \cdot \bm u) \mathrm{d}{\Omega}}_{\text{pressure-related term}}+\underbrace{\int_{\Gamma}G_{c} \mathrm{d}\Gamma}_{\Psi_f}\underbrace{- \int_{\partial\Omega_{h}} \bm {f}_t\cdot{\bm u}\mathrm{d}S}_{W_{ext}}
	\label{New energy functional}
	\end{equation}

\noindent where the sum of the first and second terms means the incremental elastic energy from the state of $\bm u_0$ to the state of $\bm u_0+\bm u$. It will be shown in the following sections that this incremental elastic energy drives the fracture propagation in the porous medium $\Omega$. In addition, it should be noted that by using the new energy functional \eqref{New energy functional}, the fracture deflection phenomenon due to stress contrast can be also well captured by the phase field model, as shown in the numerical examples presented in this paper.

\subsection{Phase field description}

It can be seen from Eqs. \eqref{Original energy functional} and \eqref{New energy functional} that the energy functional contains a sharp internal surface $\Gamma$, which increases the difficulty in minimizing the energy functional when the variational method is used. Therefore, to simplify the numerical implementation, a phase field $\phi(\bm x,t)\in[0,1]$ is used to smear the sharp fracture as shown in Fig. \ref{Sharp and diffusive fractures in the poro-elastic medium}b where $\phi=0$ and $\phi=1$ represent an intact material and a fully broken material, respectively. 

For a fracture in a 1D bar, the solution for the phase field is an inverse exponential function $\phi(x)$ \citep{miehe2010phase}:
	\begin{equation}
	\phi(x) = \mathrm{exp}\left(-\frac{|x-a|}{l_0}\right)
	\end{equation}

\noindent where $x=a$ is the fracture location and $l_0$ denotes the intrinsic length scale parameter. The length scale $l_0$ is also required for 2D and 3D problems where the crack surface density per unit volume is used in terms of the phase field and its gradient \citep{miehe2010phase}:
	\begin{equation}
	\gamma(\phi,\bigtriangledown\phi)=\frac{\phi^2}{2l_0}+\frac{l_0}2\nabla\phi\cdot\nabla\phi
	\label{phase field approximation}
	\end{equation}

Note that the length scale parameter controls the width of a diffused fracture and a larger $l_0$ shows a lower nominal tensile strength in the phase field modeling \citep{zhou2018phase}. Note that the sharp fracture can be recovered if $l_0$ tends to zero, which is known as $\Gamma$-convergence \citep{miehe2010phase}. In addition, the length scale is assumed much larger than the pore size of the domain $\Omega$ in this study and therefore the fracture energy in Eq. \eqref{New energy functional} is rewritten as
	\begin{equation}
	\Psi_f = \int_{\Gamma}G_c \mathrm{d}\Gamma \approx \int_{\Omega}G_c \gamma \mathrm{d}\Omega =\int_{\Omega}G_c\left(\frac{\phi^2}{2l_0}+\frac{l_0}2|\nabla\phi|^2 \right) \mathrm{d}\Omega
	\label{psi_f}
	\end{equation}

The elastic energy $\Psi_\varepsilon$ must be decomposed into tensile and compressive parts to ensure cracks only under tension \citep{borden2012phase}. Therefore, we follow the strain spectral decomposition of \citet{miehe2010phase}:
	\begin{equation}
	\bm\varepsilon_\pm=\sum_{a=1}^d \langle\varepsilon_a\rangle_\pm\bm n_a\otimes\bm n_a 
	\end{equation}

\noindent where $\bm\varepsilon_\pm$ are the tensile and compressive strain tensors, respectively; $\varepsilon_a$ and $\bm n_a$ are the principal strain and its direction; the operators $\langle*\rangle_\pm$ are defined as \citep{miehe2010phase}: $\langle*\rangle_{\pm}=(*\pm|*|)/2$.

Applying the decomposed strain tensor, the tensile and compressive parts of the elastic energy density are written as
	\begin{equation}
	\psi_{\varepsilon}^\pm(\bm \varepsilon) = \frac{\lambda}{2}\langle \mathrm{tr}(\bm\varepsilon)\rangle_\pm^2+\mu \mathrm{tr} \left(\bm\varepsilon_\pm^2\right) 
	\end{equation}

We follow \citet{borden2012phase} and the compressive part of the elastic energy density is assumed not to affect the fracture propagation. Therefore, the elastic energy is rewritten as
	\begin{equation}
	\Psi_{\varepsilon} = \int_{\Omega \backslash \Gamma}\psi_{\varepsilon}(\bm \varepsilon) \mathrm{d}{\Omega} = \int_{\Omega} \left[g(\phi) \psi_{\varepsilon}^+(\bm \varepsilon)+\psi_{\varepsilon}^-(\bm \varepsilon)\right] \mathrm{d}{\Omega}
	\label{Elastic energy}
	\end{equation}
	
\noindent where $g(\phi)$ is a degradation function, and $g(0)=1$, $g(1)=0$, $g'(1)=0$ must be satisfied \citep{ambati2015review}. Although there are many forms for $g(\phi)$, a quadratic form of $g(\phi)=(1-k)(1-\phi)^2+k$ is applied in this work with $k=10^{-9}$ being a stability parameter to prevent numerical singularity when $\phi=0$.
	
The degradation function is also applied to the energy variation due to the initial stress field:
	\begin{equation}
\int_{\Omega \backslash \Gamma}\bm\sigma_0:\bm\varepsilon \mathrm{d}{\Omega}\approx
\int_{\Omega}g(\phi)\bm\sigma_0:\bm\varepsilon \mathrm{d}{\Omega}
	\label{Initial stress field}
	\end{equation}

\noindent which means the initial stress field does not contribute to the energy functional in a fully broken region with $\phi=0$.	
	
\subsection{Governing equations for the displacement}

Now, substituting the fracture energy \eqref{psi_f}, the elastic energy \eqref{Elastic energy}, and Eq. \eqref{Initial stress field} into Eq. \eqref{New energy functional}, the energy functional can be rewritten as
	%\begin{equation}
	\begin{multline}
	L=\int_{\Omega}\left[g(\phi)\psi_{\varepsilon}^+(\bm \varepsilon)+\psi_{\varepsilon}^-(\bm \varepsilon)\right]\mathrm{d}{\Omega}
	+\int_{\Omega}g(\phi)\bm\sigma_0:\bm\varepsilon\mathrm{d}{\Omega}
	-\int_{\Omega}\alpha p \cdot (\nabla \cdot \bm u) \mathrm{d}{\Omega}+
	\\ \int_{\Omega}G_c\left[\frac{\phi^2}{2l_0}+\frac{l_0}2\nabla\phi\cdot\nabla\phi\right]\mathrm{d}{\Omega}-\int_{\partial\Omega_{h}} \bm f_t\cdot\bm u\mathrm{d}S
	\label{final functional}
	\end{multline}
	%\end{equation}

We then apply the variational approach \citep{francfort1998revisiting} where fracture initiation and propagation at  time $t\in[0,T]$ is a process to minimize the energy functional $L$. Therefore, the first variation of the energy functional $L$ is set as 0, yielding
	\begin{multline}
	\delta L=\underbrace{\int_{\partial \Omega_{h}}\left[\left(\sigma_{ij}^e+g(\phi)\sigma_{0ij}-\alpha p\delta_{ij}\right)m_j-f_{ti} \right] \delta u_i \mathrm{d}{S}}_{\textcircled{1}}
	-\underbrace{\int_{\Omega}\left(\sigma_{ij}^e+g(\phi)\sigma_{0ij}-\alpha p \delta_{ij}\right)_{,j}\delta  u_i \mathrm {d}{\Omega}}_{\textcircled{2}}
	-\\ \underbrace{\int_{\Omega} \left[ g'(\phi)(\psi_{\varepsilon}^+ +\sigma_{0ij}\varepsilon_{ij}) + \frac {G_c \phi}{l_0}-G_c l_0\frac{\partial^2\phi}{\partial x_i^2} \right]\delta\phi \mathrm{d}{\Omega}}_{\textcircled{3}} + \underbrace{\int_{\partial\Omega}\left( \frac{\partial\phi}{\partial x_i}m_i\right)\delta\phi \mathrm{d}S}_{\textcircled{4}}=0
	\label{first variation of the functional}
	\end{multline}

\noindent where $g'(\phi)=\mathrm{d}g(\phi)/\mathrm{d}\phi=2(\phi-1)(1-k)$ and $m_j$ is the component of the outward-pointing normal vector of the boundary. $\sigma_{ij}^e$ is the component of the effective stress tensor $\bm \sigma(\bm\varepsilon)^e$ induced by the displacement $\bm u$:	
	\begin{equation}
	\begin{aligned}
	\bm\sigma^e&=g(\phi)\frac {\partial{\psi_\varepsilon^+}}{\partial {\bm\varepsilon}}+\frac {\partial{\psi_\varepsilon^-}}{\partial\bm\varepsilon}\\
	&=\left [(1-k)(1-\phi)^2+k \right]\left[\lambda \langle tr(\bm\varepsilon)\rangle_+ \bm I+ 2\mu \bm\varepsilon_+ \right]+\lambda \langle tr(\bm\varepsilon)\rangle_- \bm I+ 2\mu \bm\varepsilon_-
	\end{aligned}
	\label{sigma_e}
	\end{equation}
	
\noindent where $\bm I$ is the identity tensor $\in \mathbb R^{d\times d}$.

Now, we define the total stress tensor $\bm\sigma$ as
	\begin{equation}
	\bm \sigma(\bm\varepsilon)=\bm \sigma^e(\bm\varepsilon)+g(\phi)\bm\sigma_0-\alpha p \bm I,\hspace{0.5cm} \mathrm{in} \hspace{0.1cm} \Omega\times(0,T]
	\label{sigma}
	\end{equation}
	
Combining Eqs. \eqref{sigma_e} and \eqref{sigma}, it can be seen that in a pure tension state without fluid, if the phase field $\phi=1$, the total stress $\bm\sigma$ is 0 in the fully broken region. In addition, because arbitrary admissible $\delta u$ must satisfy Eq. \eqref{first variation of the functional}, Eq. \eqref{first variation of the functional}$\textcircled{2}$ gives rise to the governing equation:
	\begin{equation}
	\frac {\partial {\sigma_{ij}}}{\partial x_j}=0, \hspace{0.5cm} \mathrm{in} \hspace{0.1cm} \Omega\times(0,T]
	\label{governing equation 1}
	\end{equation}
	
\noindent subjected to the Dirichlet boundary and the Neumann boundary given by
	\begin{equation}
	\sigma_{ij}m_j=f_{ti}, \hspace{1cm} \mathrm{on}\hspace{0.5cm} \partial\Omega_{h}\times(0,T]
	\end{equation}
	
\noindent which can be derived from Eq. \eqref{first variation of the functional}$\textcircled{1}$.

\subsection{Governing equations for the phase field}

Because of the arbitrariness of $\delta \phi$, Eq. \eqref{first variation of the functional}$\textcircled{3}$ results in the original governing equation for the phase field:
	\begin{equation}
	 \left[\frac{2l_0(1-k)(\psi_{\varepsilon}^++\sigma_{0ij}\varepsilon_{ij})}{G_c}+1\right]\phi-l_0^2\frac{\partial^2 \phi}{\partial {x_i^2}}=\frac{2l_0(1-k)(\psi_{\varepsilon}^++\sigma_{0ij}\varepsilon_{ij})}{G_c}, \hspace{0.5cm} \mathrm{in} \hspace{0.1cm} \Omega\times(0,T]
	\label{governing equations 2_1}
	\end{equation}
	
However, Eq. \eqref{governing equations 2_1} cannot ensure the irreversibility condition $\Gamma(\bm x,s)\in\Gamma(\bm x,t)(s<t)$. Therefore, a history reference field $H$ is constructed to form a monotonically increasing phase field:
	\begin{equation}
	H(\bm x,t) = \max \limits_{s\in[0,t]}\left[\psi_\varepsilon^+\left(\bm\varepsilon(\bm x,s)\right)+\bm\sigma_0:\bm\varepsilon(\bm x,s)\right], \hspace{0.5cm} \mathrm{in} \hspace{0.1cm} \Omega\times(0,T]
	\end{equation}

Replacing $\psi_\varepsilon^++\sigma_{0ij}\varepsilon_{ij}$ by $H(\bm x,t)$  in Eq. \eqref{governing equations 2_1}, the strong form for the phase field is given by
	\begin{equation}
	\left[\frac{2l_0(1-k)H}{G_c}+1\right]\phi-l_0^2\frac{\partial^2 \phi}{\partial {x_i^2}}=\frac{2l_0(1-k)H}{G_c}, \hspace{0.5cm} \mathrm{in} \hspace{0.1cm} \Omega\times(0,T]
	\label{governing equation 2_2}
	\end{equation}
	
\noindent which is subjected to the Neumann condition:
	\begin{equation}
	\frac{\partial \phi}{\partial x_i} m_i = 0, \hspace{0.5cm} \mathrm{on}\hspace{0.5cm} \partial\Omega\times(0,T]
	\end{equation}
	
Therefore, the evolution equation of the phase field \eqref{governing equation 2_2} indicates that the fracture initiation and propagation is driven by the history reference field $H(x,t)$, which represents the historic high incremental elastic energy from the $\bm u_0$ state to the $\bm u_0+\bm u$ state during the period $(0,t]$.

\subsection{Fluid flow in the porous medium}

As described in Subsection \ref{subsection New energy functional}, the fluid in poro-elastic medium is assumed compressible and viscous. Then, we calculate the fluid flow in three subdomains, namely the unbroken domain (reservoir domain) $\Omega_r(t)$, fractured domain $\Omega_f(t)$ and transition domain $\Omega_t(t)$, respectively. These subdomains are distinguished according to the phase field in Table \ref{Subdomain definition} where $c_1$ and $c_2$ are two phase field thresholds. 

	\begin{table}[htbp]
		\caption{Subdomain definition}
		\label{Subdomain definition}
		\centering
		\begin{tabular}{ll}
			\toprule[1pt]
			Subdomain&Phase field\\
			Reservoir domain & $\phi\le c_1$\\
			Transition domain & $c_1<\phi<c_2$\\
			Fractured domain & $\phi\ge c_2$\\
			\bottomrule[1pt] 
		\end{tabular}
	\end{table}

In this study, we calculate the hydraulic parameters in the transition domain by using a linear interpolation from the reservoir and fractured domains. Therefore, two indicator functions $\chi_r$ and $\chi_f$ \citep{shiozawaeffect} are established. $\chi_r=1$ in the reservoir domain and $\chi_r=0$ in the fractured domain. In contrast, $\chi_f=0$ in the reservoir domain and $\chi_f=1$ in the fractured domain. In addition, the indicator functions satisfy the following:
	\begin{equation}
	\left\{\begin{aligned}
	&\chi_r(\cdot,\phi)=\frac{c_2-\phi}{c_2-c_1} \hspace{1cm}&c_1<\phi<c_2\\
	&\chi_f(\cdot,\phi)=\frac{\phi-c_1}{c_2-c_1} \hspace{1cm}&c_1<\phi<c_2
	\end{aligned}
	\right.
	\label{function2}
	\end{equation}

Darcy's law is used to model the fluid flow and the mass conservation equation in \citet{zhou2018phase2} is applied to the whole calculation domain:
	\begin{equation}
	\rho S \frac{\partial p}{\partial t}+\nabla\cdot(\rho\bm v)=q_m-\rho\alpha\chi_r\frac{\partial \varepsilon_{vol}}{\partial t}
	\label{mass conservation of the whole domain}
	\end{equation}

\noindent where $\rho$, $S$, $\bm v$, $\varepsilon_{vol}=\nabla\cdot\bm u$, and $q_m$ represent the fluid density, storage coefficient, flow velocity, volumetric strain, and fluid source term, respectively. In Eq. \eqref{mass conservation of the whole domain}, $\rho=\rho_r\chi_r+\rho_f\chi_f$ and $\alpha=\alpha_r\chi_r+\alpha_f\chi_f$ where $\rho_r$ and $\rho_f$ are the fluid densities in the reservoir and fracture domains while $\alpha_r$ and $\alpha_f$ are the Biot coefficients of the reservoir and fractured domains. Naturally, $\alpha_f=1$ is set for the fractured domain and therefore $\alpha=\alpha_{r}\chi_r+\chi_f$.

In Eq. \eqref{mass conservation of the whole domain}, the storage coefficient $S$ is expressed as \citep{zhou2018phase2}  
	\begin{equation}
	S=\varepsilon_pc+\frac{(\alpha-\varepsilon_p)(1-\alpha)}{K_{Vr}}
	\end{equation}

\noindent where $\varepsilon_p$, $c$, and $K_{Vr}$ are the porosity, fluid compressibility, and bulk modulus of the calculation domain, respectively. Denoting $c_r$ and $c_f$ as the fluid compressibility in the reservoir and fractured domains, we have $c=c_r\chi_r+c_f\chi_f$. 

The gravity is not considered in this study; therefore the Darcy's velocity $\bm v$ is calculated as
	\begin{equation}
	\bm v=-\frac{K_{eff}}{\mu_{eff}}\nabla p
	\label{velocity of the whole domain}
	\end{equation}
	
\noindent where $\mu_{eff}$ denotes fluid viscosity. Similarly, $\mu_{eff}=\mu_{r}\chi_r+\mu_f \chi_f$ with $\mu_{r}$ and $\mu_{f}$ being the fluid viscosity in the reservoir and fractured domains, respectively. $ K_{eff}$ is the effective permeability and $K_{eff}=k_{r}\chi_r+k_f \chi_f$ where $k_{r}$ and $k_{f}$ are the permeabilities of the reservoir and fractured domains. Replacing Eq. \eqref{velocity of the whole domain} into Eq. \eqref{mass conservation of the whole domain}, the governing equation for fluid flow is expressed in terms of the fluid pressure $p$:
	\begin{equation}
	\rho S \frac{\partial p}{\partial t}-\nabla\cdot \frac{\rho K_{eff}}{\mu_{eff}}\nabla p=q_m-\rho\alpha\chi_r\frac{\partial \varepsilon_{vol}}{\partial t}
	\label{governing equation of the whole domain}
	\end{equation}
	
\noindent which is subjected to the Dirichlet condition on $\partial\Omega_D$ and Neumann condition on $\partial\Omega_N$ \citep{zhou2018phase2}:
	\begin{equation}
	\left\{
	\begin{aligned}
	&p=p_D \hspace{2cm}&\mathrm{on}\quad\partial\Omega_D\\
	&-\bm n \cdot \rho\bm v=M_N  \hspace{2cm}&\mathrm{on}\quad\partial\Omega_N
	\end{aligned}\right.
	\end{equation}

\noindent where $p_D$ and $M_N$ are the prescribed pressure and mass flux.

\section {Numerical algorithm}\label {Numerical algorithm}

\begin{figure}[htbp]
	 	\centering
	 	\includegraphics[width = 15cm]{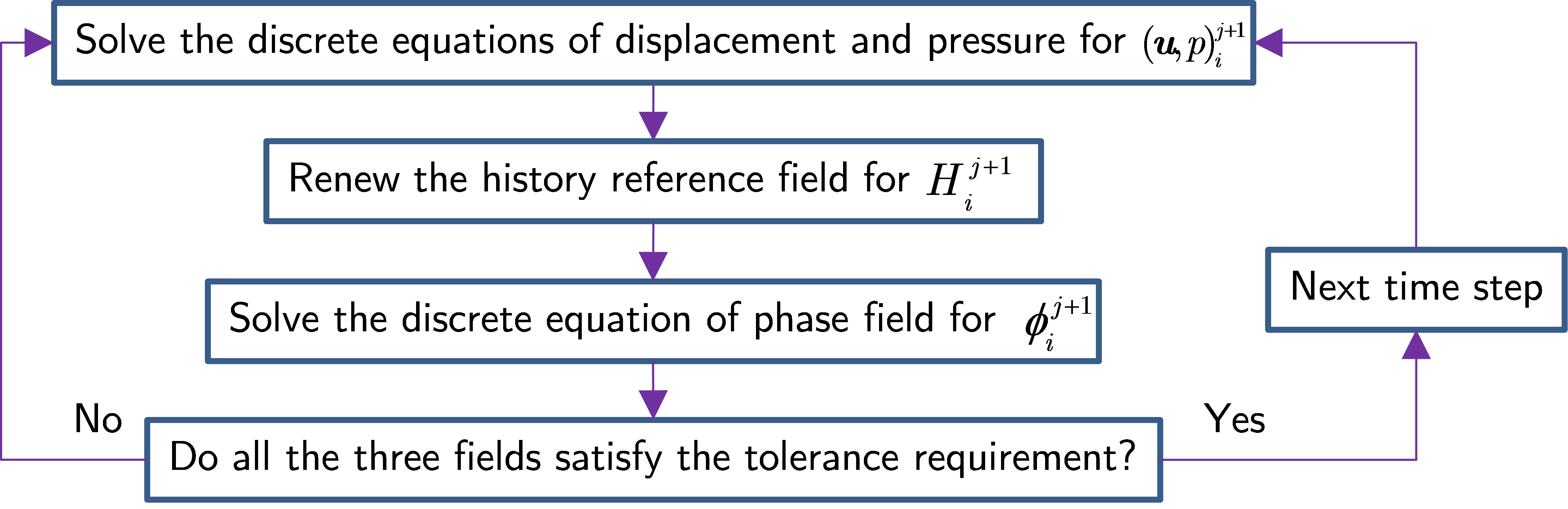}
	 	\caption{Numerical scheme in each time step}
	 	\label{Numerical scheme in each time step}
	 \end{figure}

 We implement the proposed phase field model in the framework of finite element method (FEM) and the FE discretization can be seen in Appendix. In addition, a finite difference method is applied for the time discretization. The overall procedure of our numerical algorithm for solving the three fields is shown in Fig. \ref{Numerical scheme in each time step} where a staggered scheme is used. That is, the three fields are solved sequentially and independently in each time step.
 
To reduce the computational effort, we implement the numerical method in the commercial software, COMSOL Multiphysics, which has been proven to have high capability for multi-field problems. In all simulations, the maximum iteration number is set as 150 due to high nonlinearity resulting from the derivative  $\partial\bm\sigma^e/\partial\bm\varepsilon$ and the phase field which degrades the stiffness matrix of the displacement field. In addition, a stabilization and convergence acceleration method--the Anderson acceleration technique is applied. For more detail on the COMSOL implementation, the readers can be referred to the previous study \citep{zhou2019phase}.

\section{2D examples}\label{2D examples}

In this section, 2D examples of specimens subjected to internal fluid injection are
presented to prove the capability of the proposed phase field method. All the pre-existing cracks are established by introducing an initial history field with a relatively high $H$ \citep{borden2012phase} and the source term in the pre-existing notches is set as $q_F = 10$ kg/(m$^3\cdot$s) in all examples.

\subsection{Fractures from a horizontal notch}\label{Fractures from a horizontal notch}

\begin{figure}[htbp]
		\centering
		\includegraphics[height = 6cm]{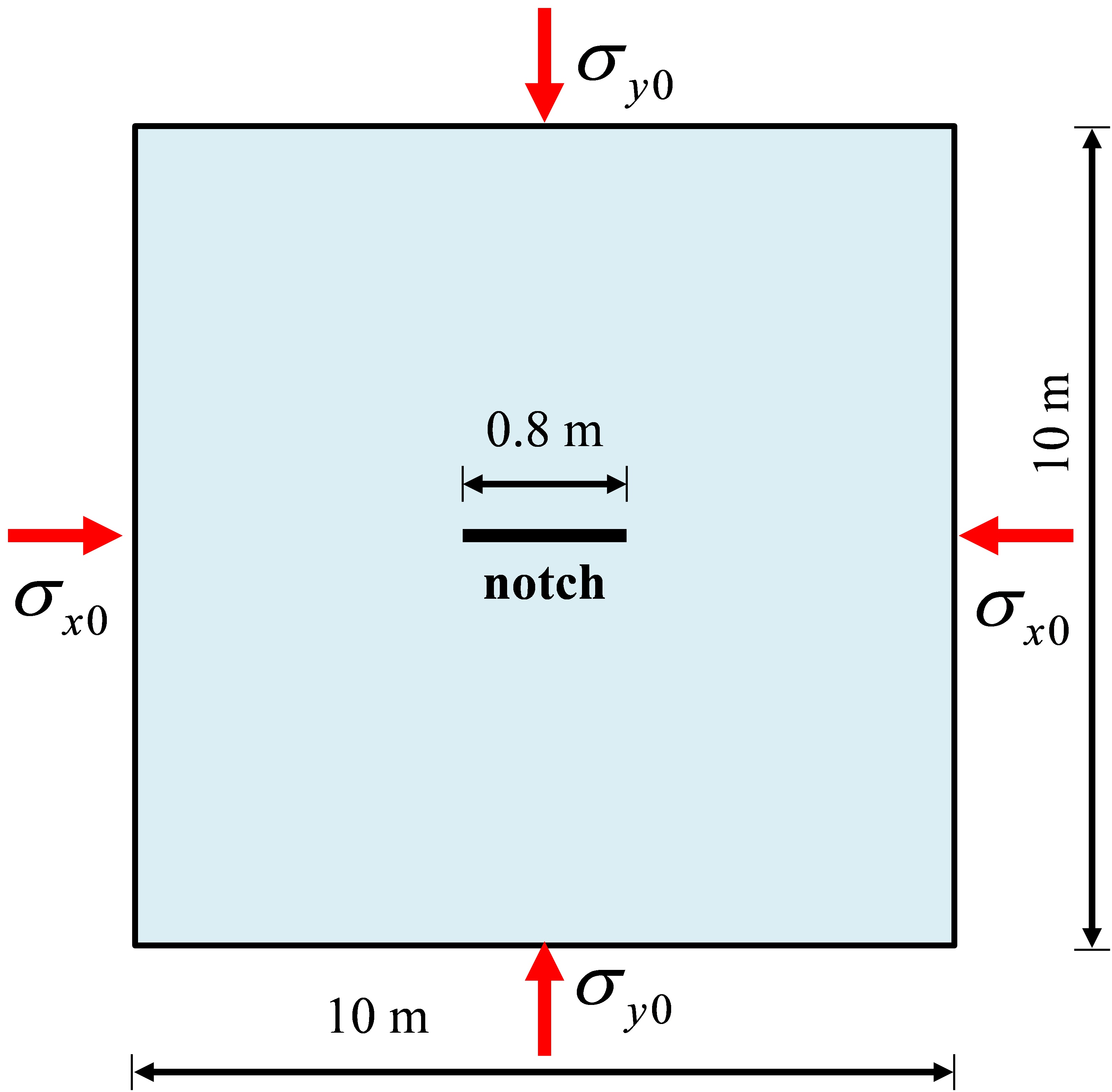}
		\caption{Geometry and boundary conditions of the calculation domain}
		\label{Geometry and boundary conditions of the calculation domain}
	\end{figure}

The first example tests the hydraulic fracture propagation from a horizontal notch in a square domain. The geometry and boundary conditions of this example are shown in Fig. \ref{Geometry and boundary conditions of the calculation domain} with increasing fluid volume in the initial notch. All the outer boundaries of the calculation domain has a fluid pressure of $p=0$ and a zero tangential displacement in order to remove the effect of rigid body displacement.  In addition, the parameters for calculation are listed in Table \ref{Basic calculation parameters}.

	\begin{table}[htbp]
	\small
	\caption{Basic calculation parameters}
	\label{Basic calculation parameters}
	\centering
	\begin{tabular}{lll||lll}
		\hline
		Parameter & Value &Unit & Parameter & Value &Unit\\
		\hline
		$\mu$ & 23.08 & GPa & $\lambda$ &34.62 & GPa \\
		$G_c$ & 500 & N/m & $k$ &$1\times10^{-9}$ & --\\
		$l_0$ & 0.1 & m & $c_1$ & 0.4 & -- \\
		$c_2$ & 1.0 & --& $\varepsilon_{pR}$ & 0.05 & --\\
		$\rho_{R}$,  $\rho_{F}$& $1.0\times10^{3}$& kg/m$^3$  & $\alpha_R$ &0.05&-- \\
		$q_R$ & 0 & kg/(m$^3\cdot$s)& $q_F$ & 0& kg/(m$^3\cdot$s)\\
		$k_R$ & $1\times10^{-15}$ &m$^2$ & $k_F$ & $8.333\times10^{-4}$ &m$^2$\\
		$c_R$ & $1\times10^{-8}$& 1/Pa & $c_F$ & $1\times10^{-8}$ &1/Pa \\
		$\mu_R$ & $1\times10^{-3}$& Pa$\cdot$s & $\mu_F$ & $1\times10^{-3}$ &Pa$\cdot$s\\
		\hline
		\normalsize
	\end{tabular}
	\end{table}

We discretize the calculation domain with unstructured triangular elements with the maximum element size $h=0.05$ m. Therefore, the linear shape functions are used for all the three physical fields. In addition, the time step $\Delta t = 0.05$s is adopted for the simulation. The initial stress state of $\sigma_{x0}=0.5$ MPa, $\sigma_{y0}=0.5$ MPa and $\sigma_{z0}=\nu(\sigma_{x0}+\sigma_{y0})$ is applied with $\nu$ being Poisson's ratio, which can be evaluated through the well-known elasticity relationship from $\lambda$ and $\mu$. As a comparison, we also model the fracture propagation by using the previously developed method \citep{zhou2018phase2} without considering the effect of initial stress field.

\begin{figure}[htbp]
	\centering
	\subfigure[$t=2.4$ s]{\includegraphics[width = 5.5cm]{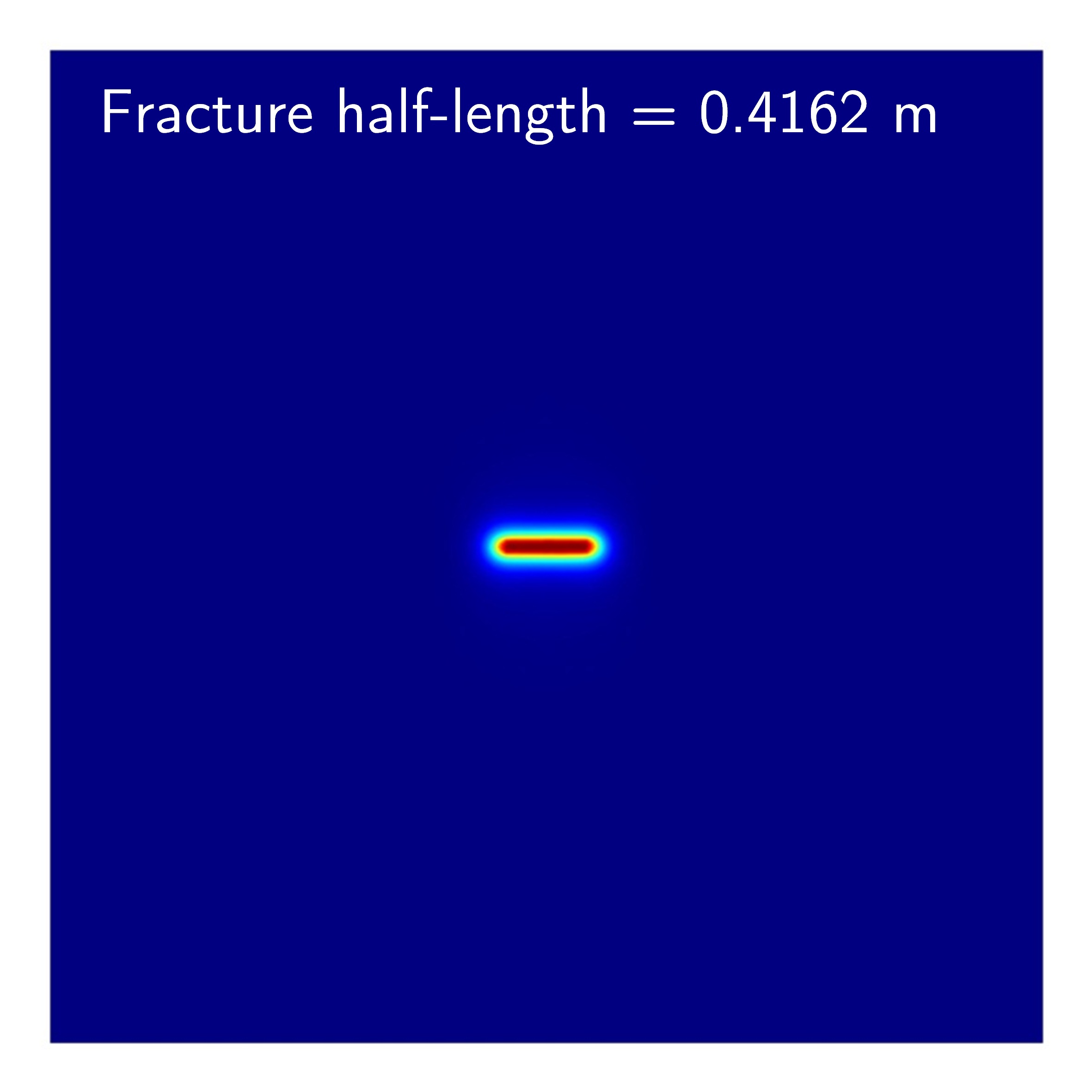}}
	\subfigure[$t=3.6$ s]{\includegraphics[width = 5.5cm]{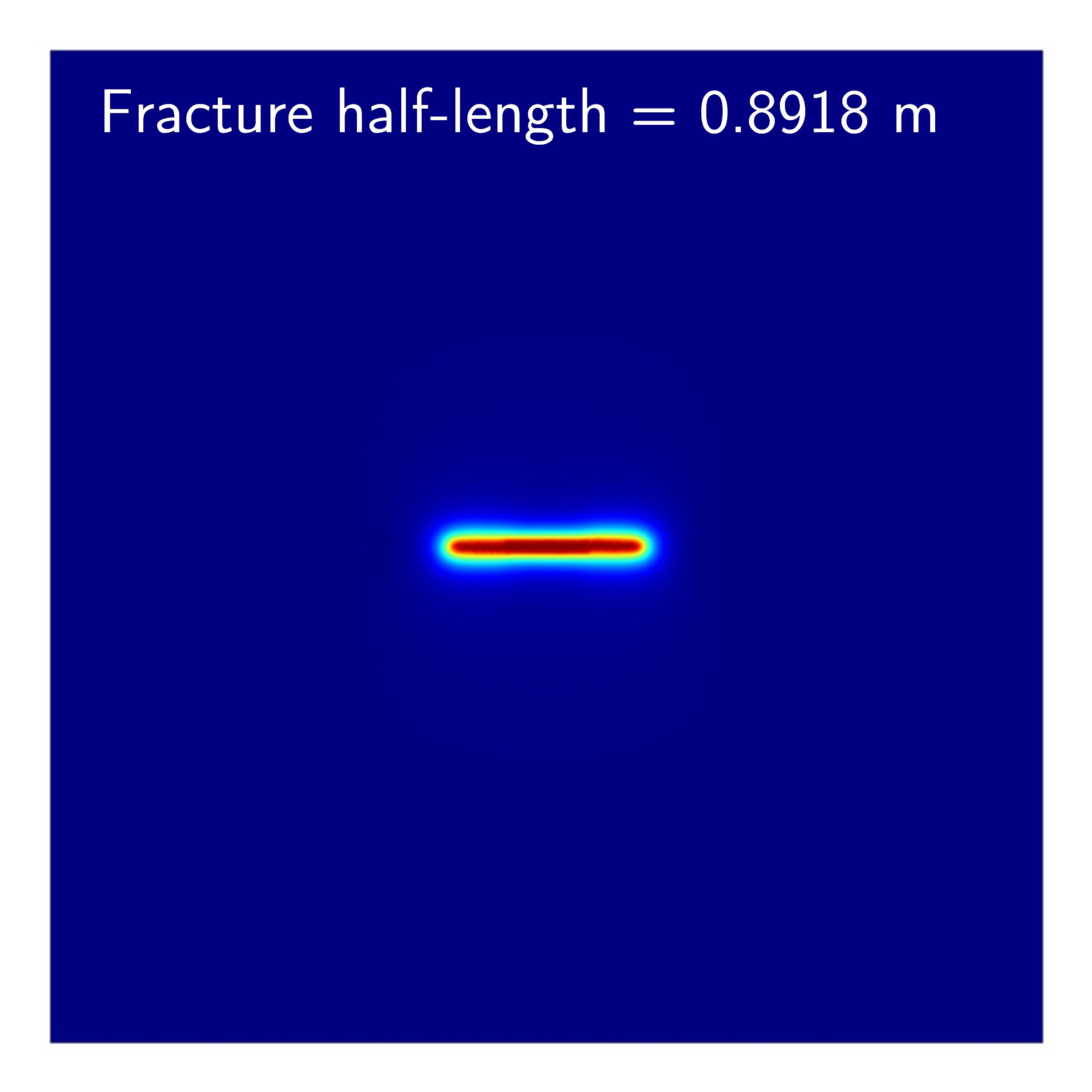}}
	\subfigure[$t=4.8$ s]{\includegraphics[width = 5.5cm]{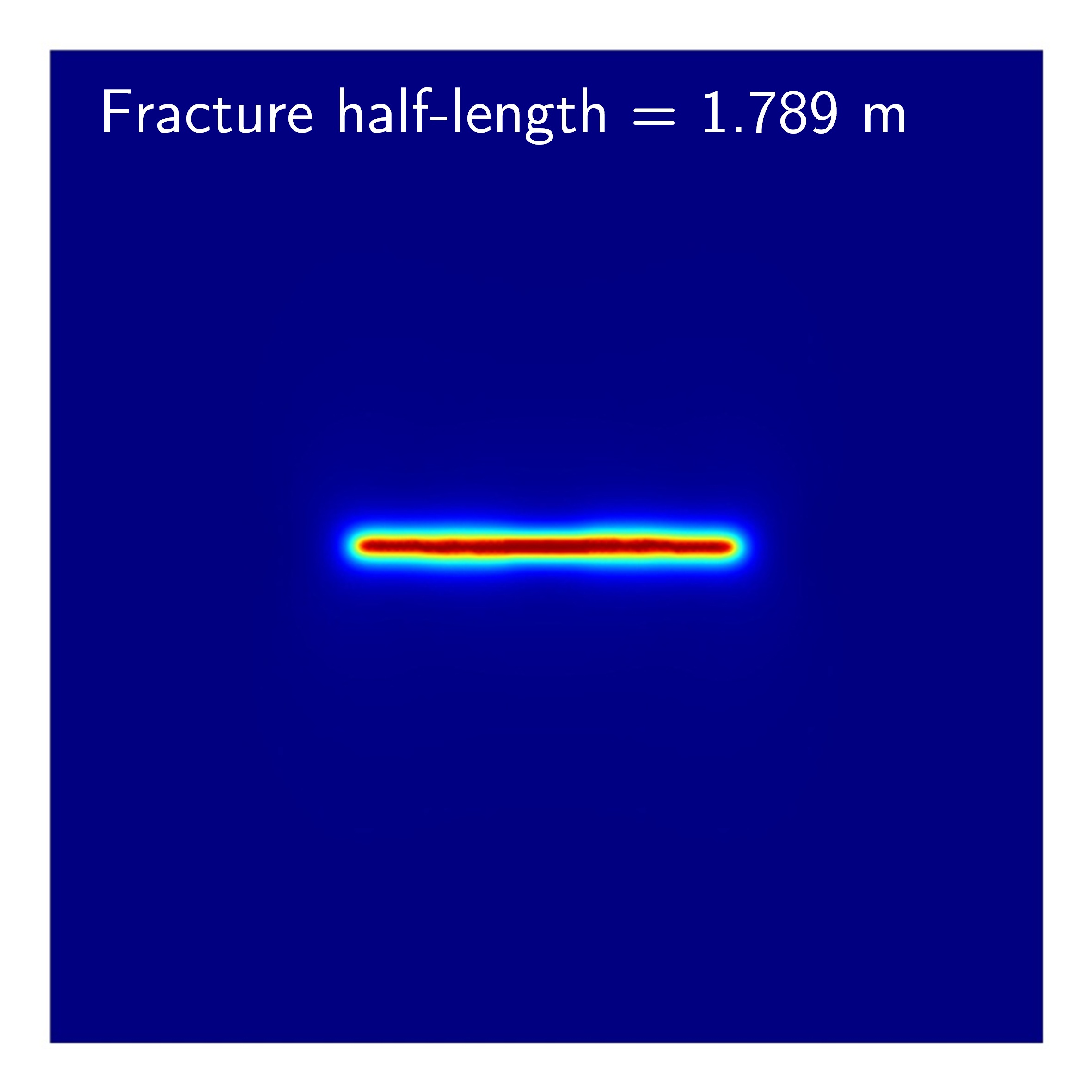}}
	\caption{Fracture propagation pattern by using the proposed PFM}
	\label{Fracture propagation pattern by using the proposed PFM}
	\end{figure}

\begin{figure}[htbp]
	\centering
	\subfigure[$t=2.4$ s]{\includegraphics[width = 5.5cm]{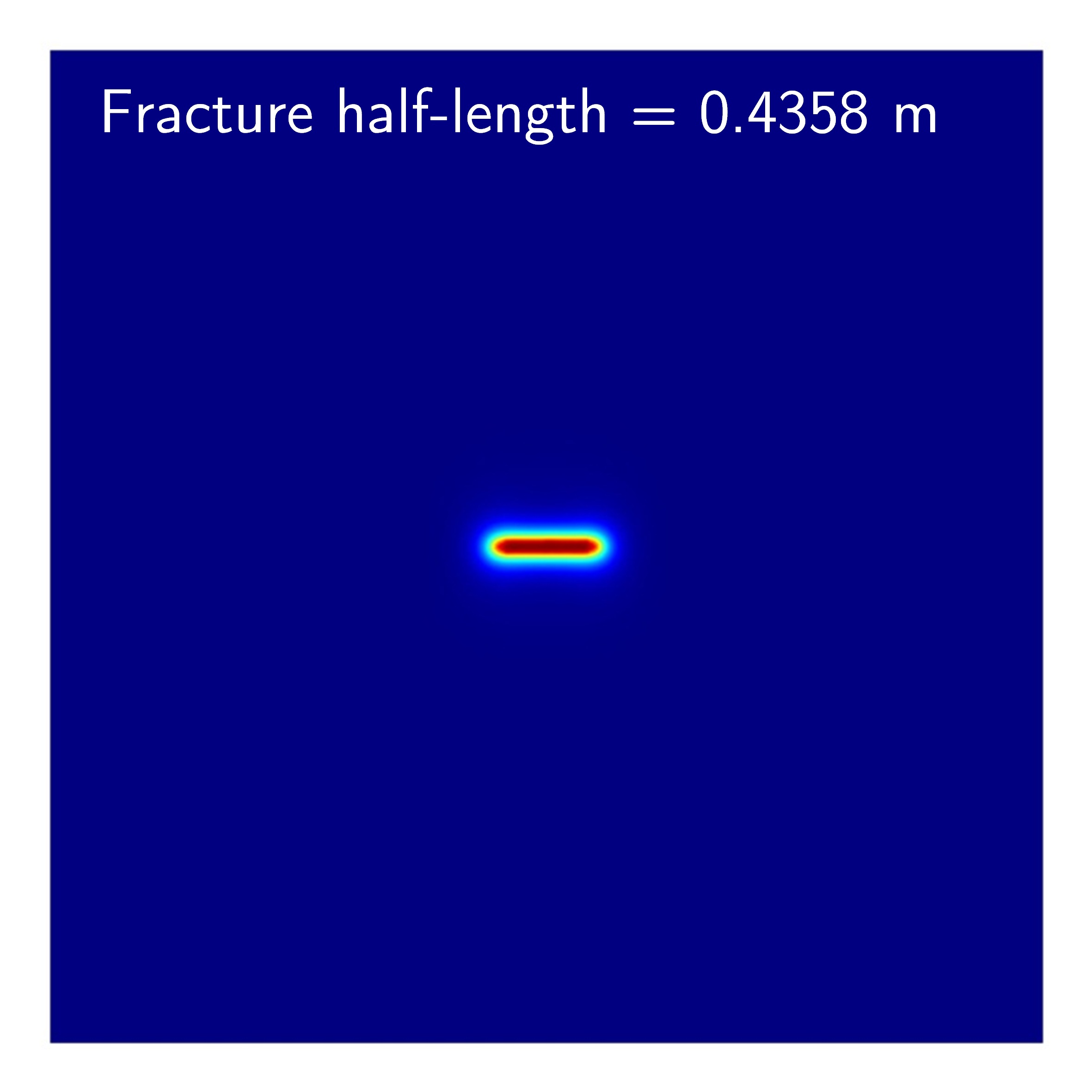}}
	\subfigure[$t=3.6$ s]{\includegraphics[width = 5.5cm]{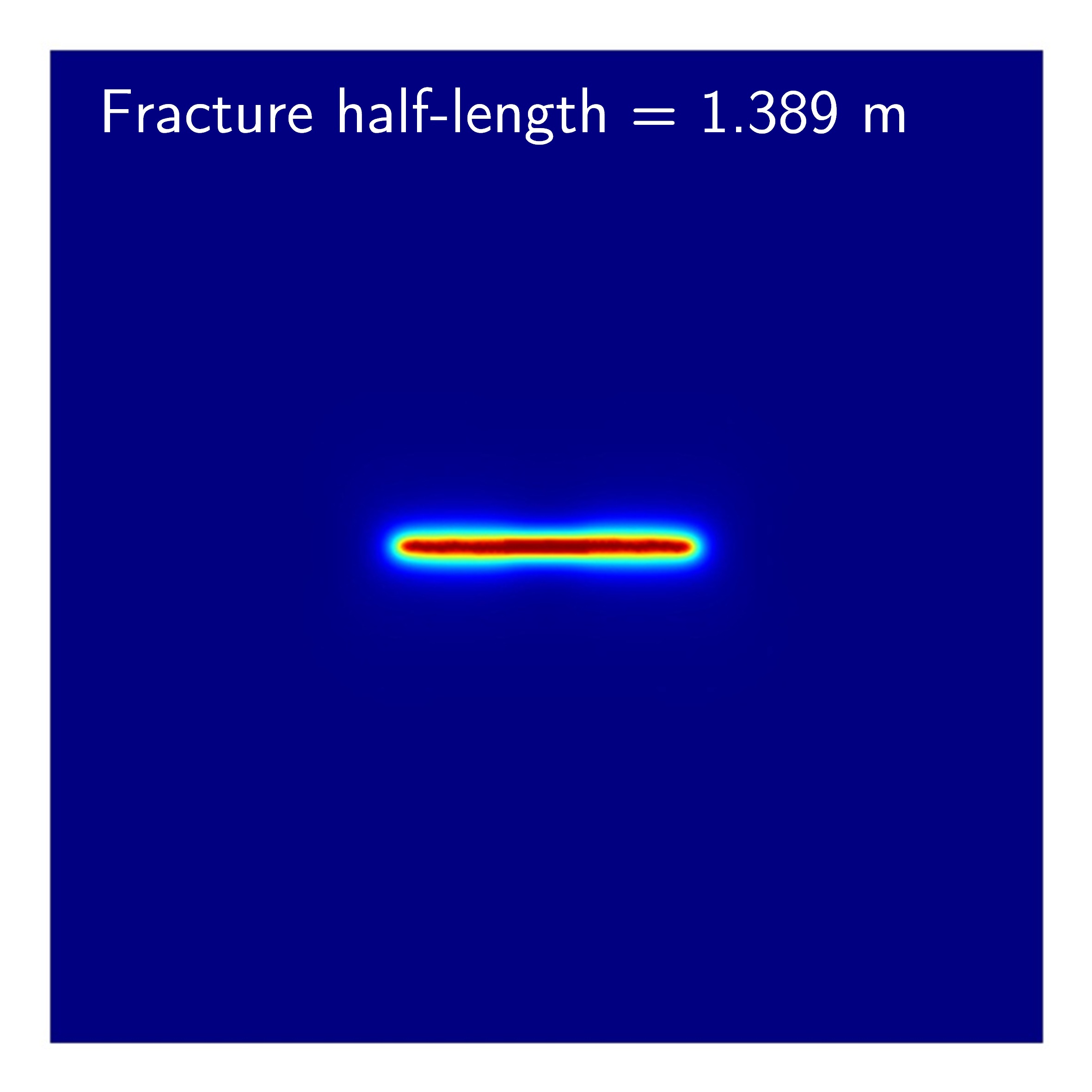}}
	\subfigure[$t=4.8$ s]{\includegraphics[width = 5.5cm]{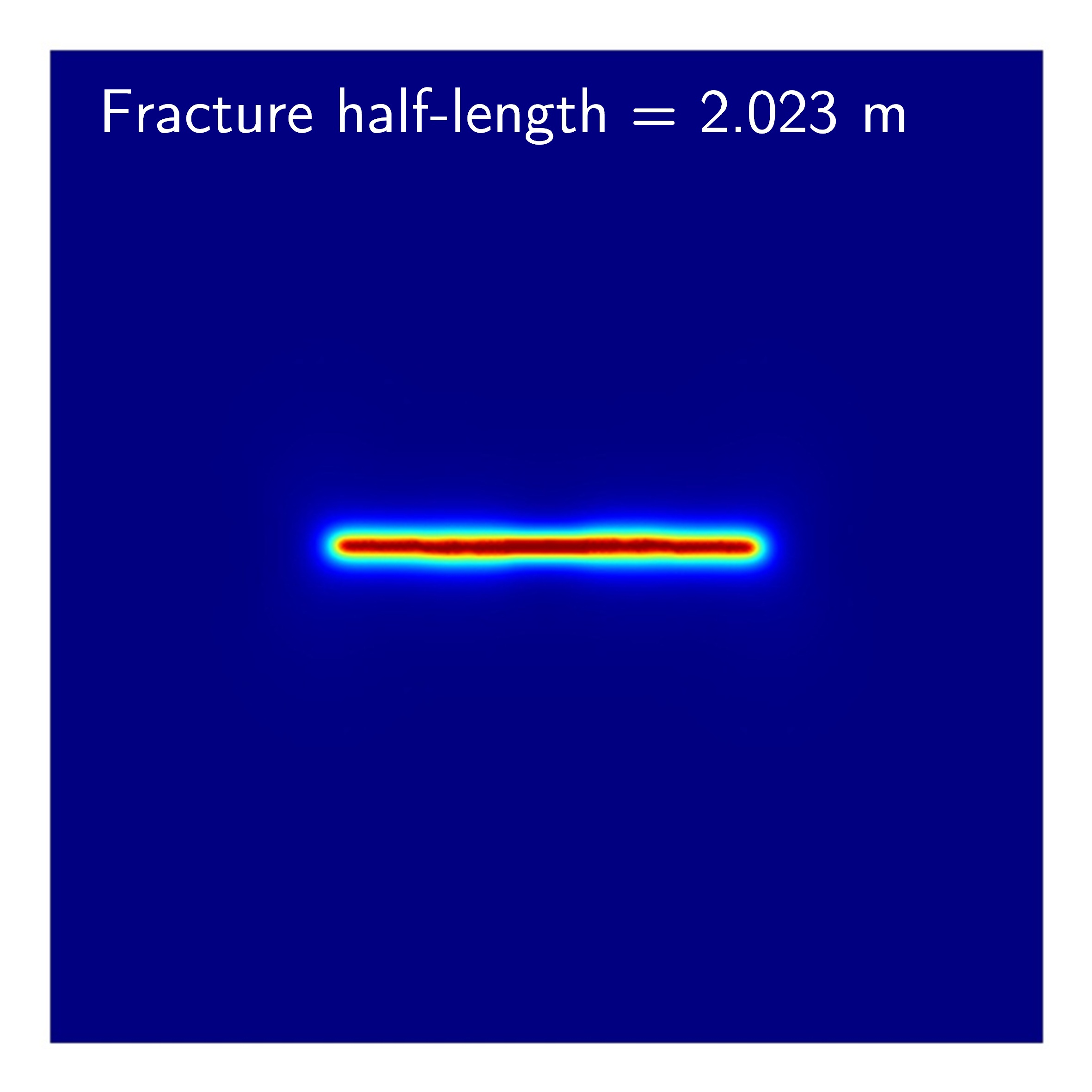}}
	\caption{Fracture propagation pattern without considering the effect of initial stress field}
	\label{Fracture propagation pattern without considering the effect of initial stress field}
	\end{figure}

Figures \ref{Fracture propagation pattern by using the proposed PFM} and \ref{Fracture propagation pattern without considering the effect of initial stress field} show the fracture propagation obtained by using the proposed phase field method and that without considering the effect of initial stress field. The figures indicate that discarding the effect of initial stress does not affect the fracture pattern for the first example; in both situation the fracture propagates along the horizontal direction. However, the time for fracture initiation and propagation is different and the fracture length is smaller if the initial stress field is considered in the constitutive model.

\begin{figure}[htbp]
	\centering
	\subfigure[The proposed PFM]{\includegraphics[width = 5.5cm]{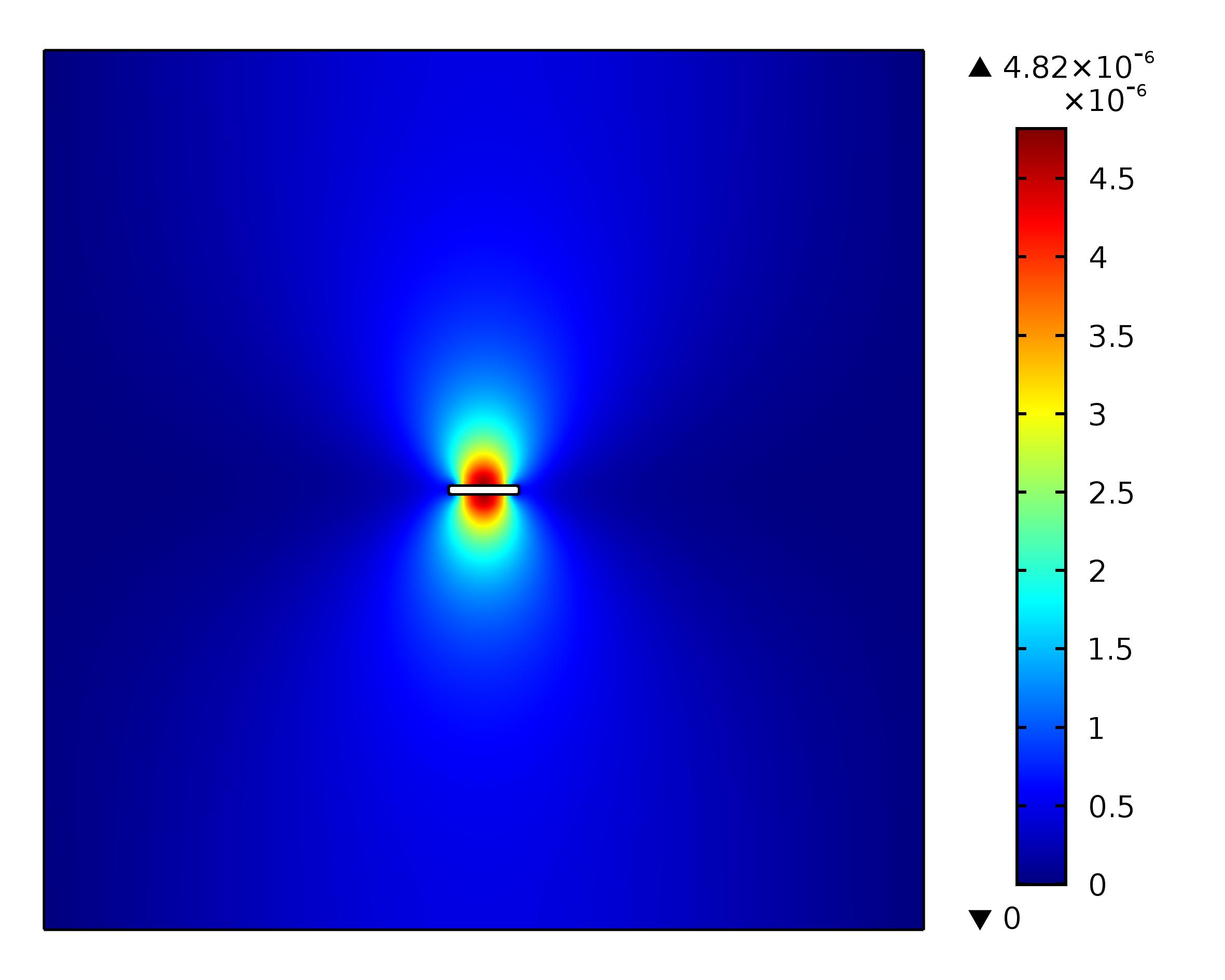}}
	\subfigure[Without considering the initial stress field]{\includegraphics[width = 5.5cm]{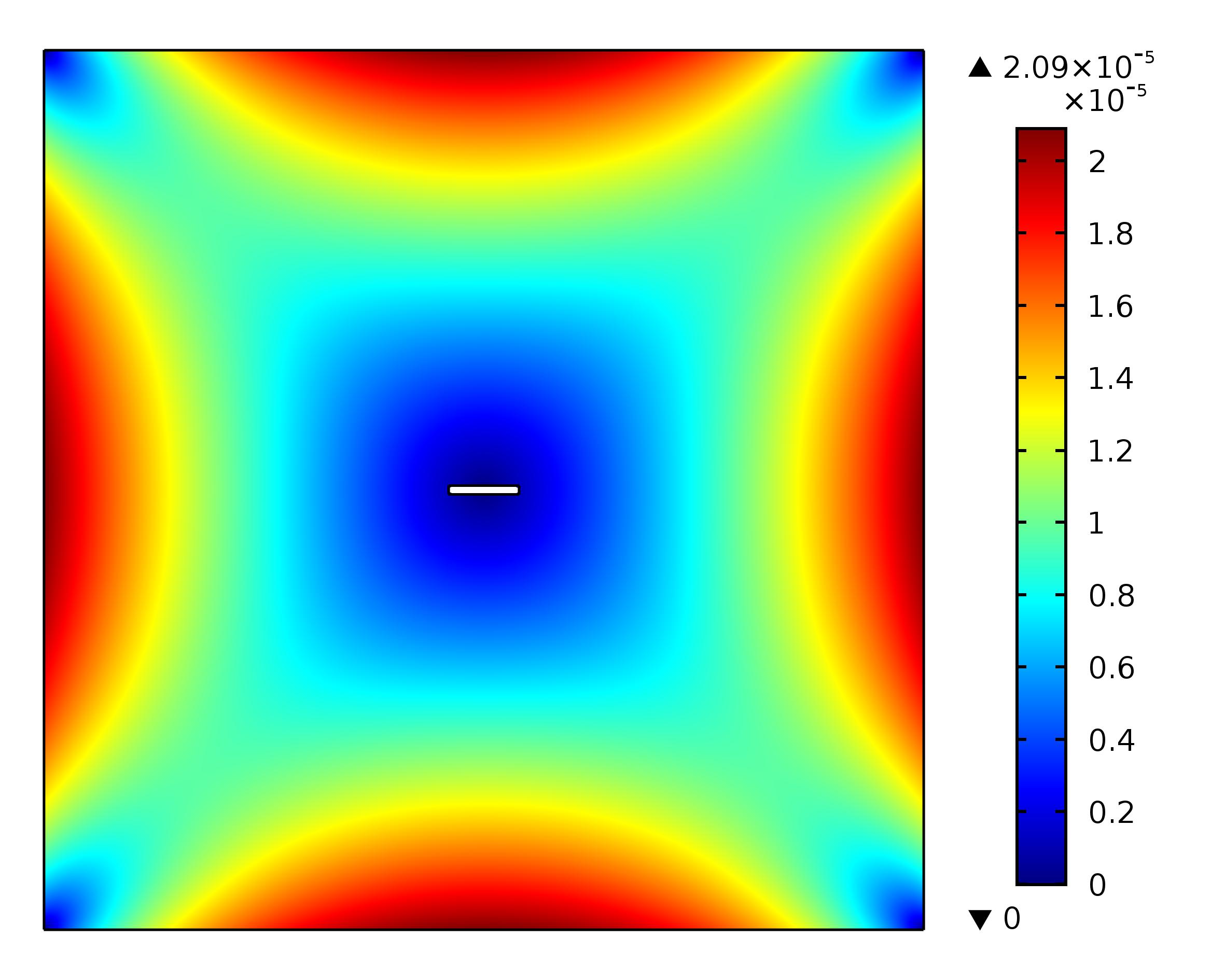}}
	\caption{Comparison of the displacement field at $t=0$ s for the proposed PFM and \citet{zhou2018phase2} (Unit: m)}
	\label{Comparison of the displacement field at t=0 s}
	\end{figure}

\begin{figure}[htbp]
	\centering
	\subfigure[Top boundary]{\includegraphics[width = 7.5cm]{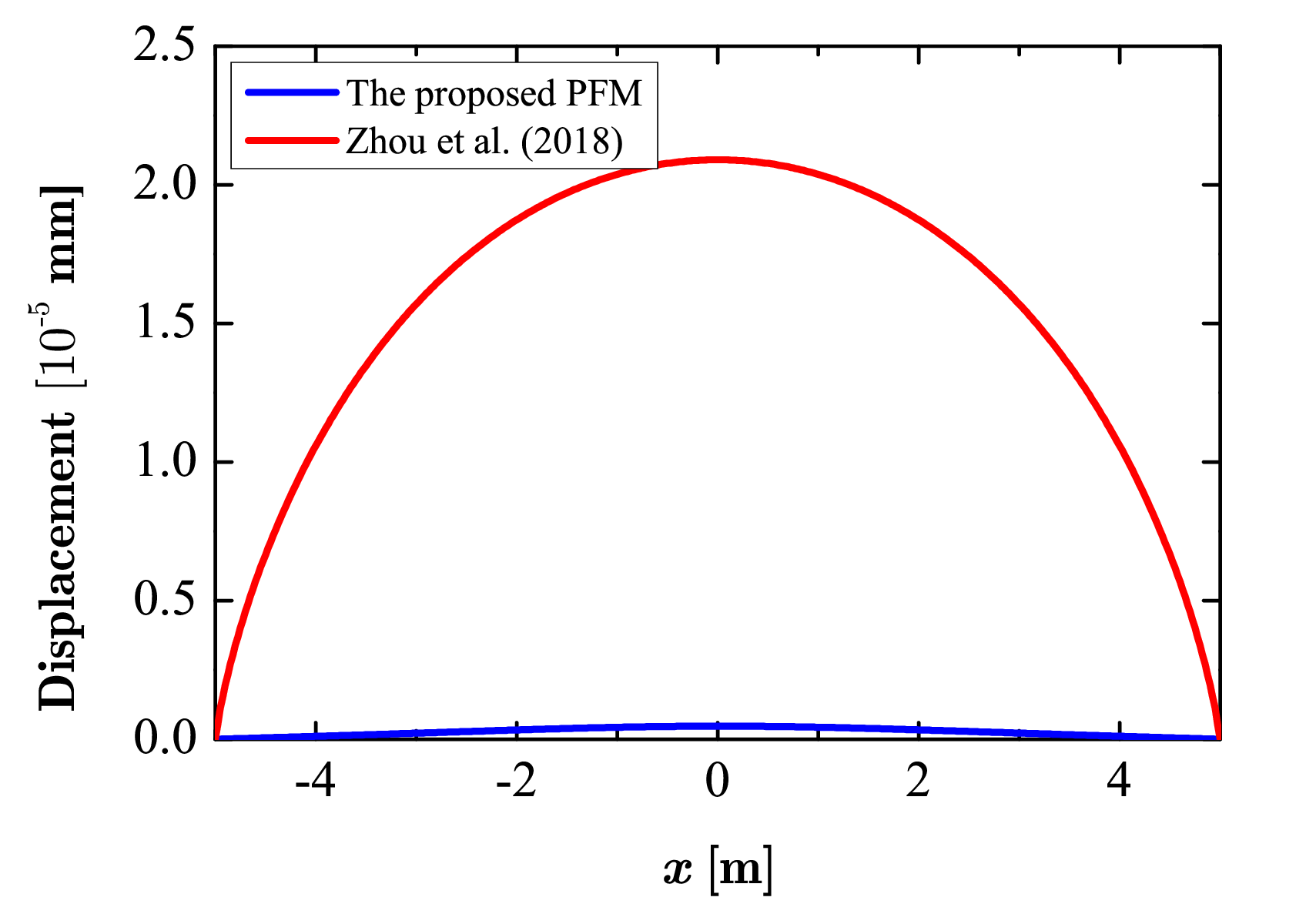}}
	\subfigure[Left boundary]{\includegraphics[width = 7.5cm]{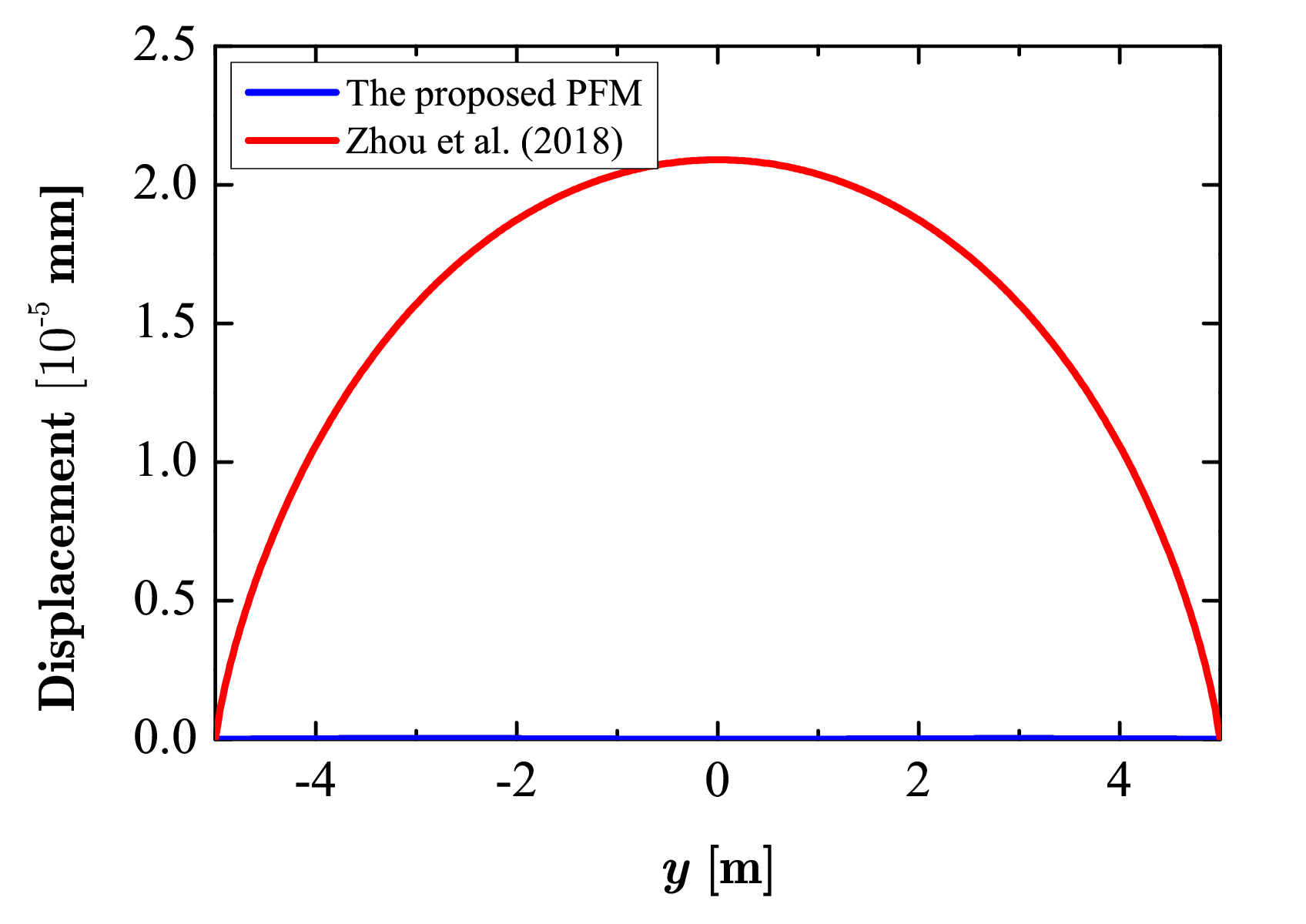}}
	\caption{Comparison of the displacement along the top and left boundaries at $t=0$ s for the proposed PFM and \citet{zhou2018phase2}}
	\label{Comparison of the displacement along the top and left boundaries at t=0 s}
	\end{figure}

The displacement field at time $t=0$ s is shown in Fig. \ref{Comparison of the displacement field at t=0 s} where the region of $\phi\ge0.95$ is removed to reflect the shape of the fully broken domain. It can be observed that the proposed PFM and the method of \citet{zhou2018phase2} achieve different initial displacement field for the problem of a poro-elastic domain subjected to stress boundary condition. Fig. \ref{Comparison of the displacement field at t=0 s}a indicates that only a small displacement appear around the center of the notch for the proposed PFM because the ``excavation" or stiffness degradation of the initial notch produces displacement towards the broken domain. However, if the PFM does not consider the effect of initial stress field, all the outer boundaries have rather large displacements as shown in Fig. Fig. \ref{Comparison of the displacement field at t=0 s}b. In addition, Fig. \ref{Comparison of the displacement along the top and left boundaries at t=0 s} compares the displacements along the top and left boundaries at time $t=0$ s. As observed, for the method of \citet{zhou2018phase2}, the stress boundary condition produces large initial displacements along the left and top boundaries while the initial displacement on the boundaries is negligible by using the proposed PFM. In summary, comparisons in Figs. \ref{Comparison of the displacement field at t=0 s} and \ref{Comparison of the displacement along the top and left boundaries at t=0 s} indicate the proposed PFM will achieve better displacement distribution compared with those PFMs without considering the effect of initial stress field, especially for the poro-elastic medium subjected to stress boundary condition in the geological environment.

\begin{figure}[htbp]
	\centering
	\includegraphics[width = 10cm]{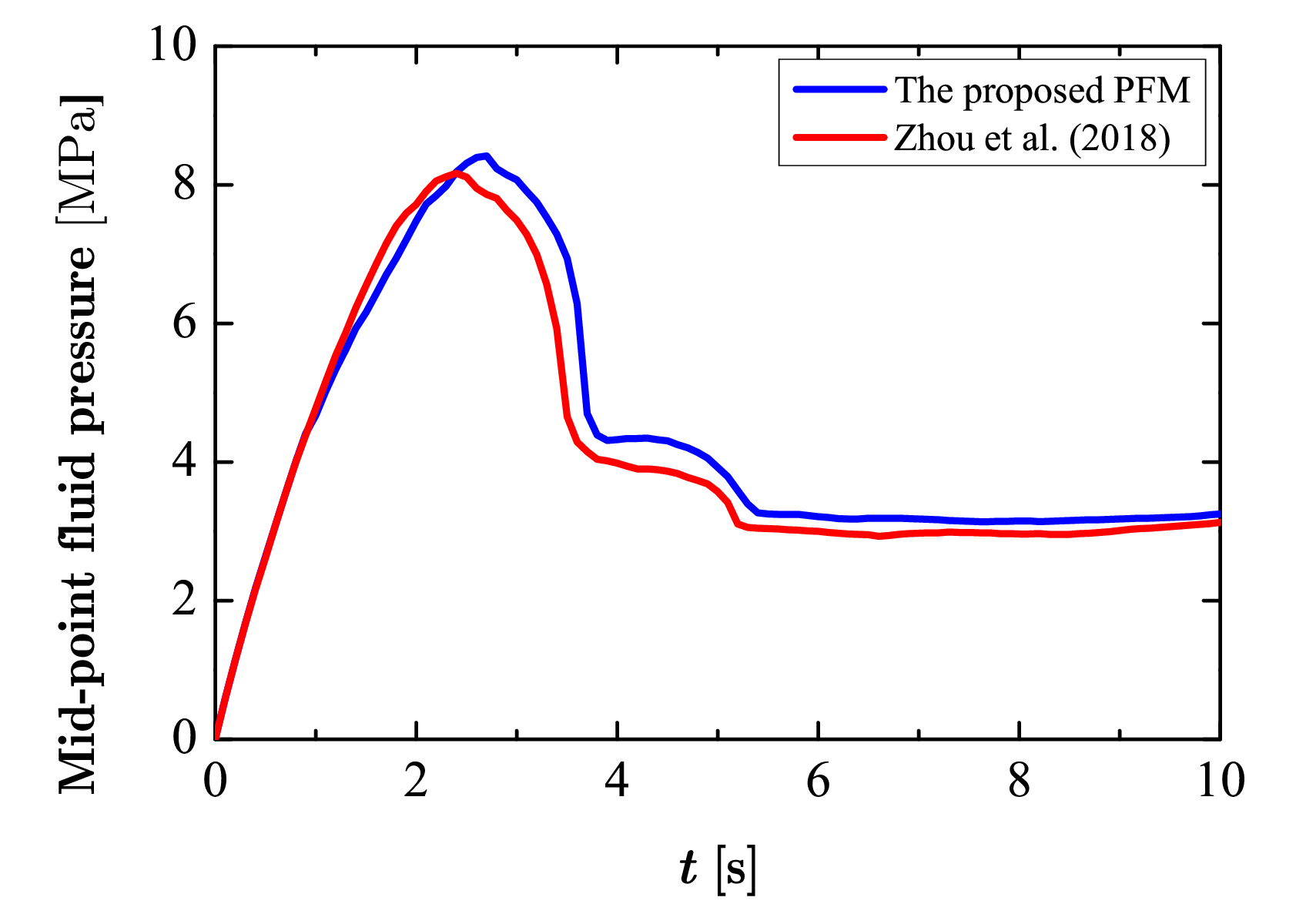}
	\caption{Comparison of the fluid pressure-time curve for the proposed PFM and \citet{zhou2018phase2}}
	\label{Comparison of the fluid pressure-time curve for the proposed PFM}
	\end{figure}

Figure \ref{Comparison of the fluid pressure-time curve for the proposed PFM} shows the effect of initial stress field on the fluid pressure-time curve. Note that the data at the center of the initial notch are selected. Similar to the fracture pattern, Fig. \ref{Comparison of the fluid pressure-time curve for the proposed PFM} indicates the fluid pressure is only slightly affected by the initial stress field. In the current example, the time for fracture initiation is reduced if the initial stress field is not considered; therefore the fluid pressure-time curve has a earlier drop stage and a lower maximum pressure compared with those obtained by the proposed PFM as shown in Fig. \ref{Comparison of the fluid pressure-time curve for the proposed PFM}.

\begin{figure}[htbp]
	\centering
	\subfigure[$t=2.4$ s]{\includegraphics[width = 5.5cm]{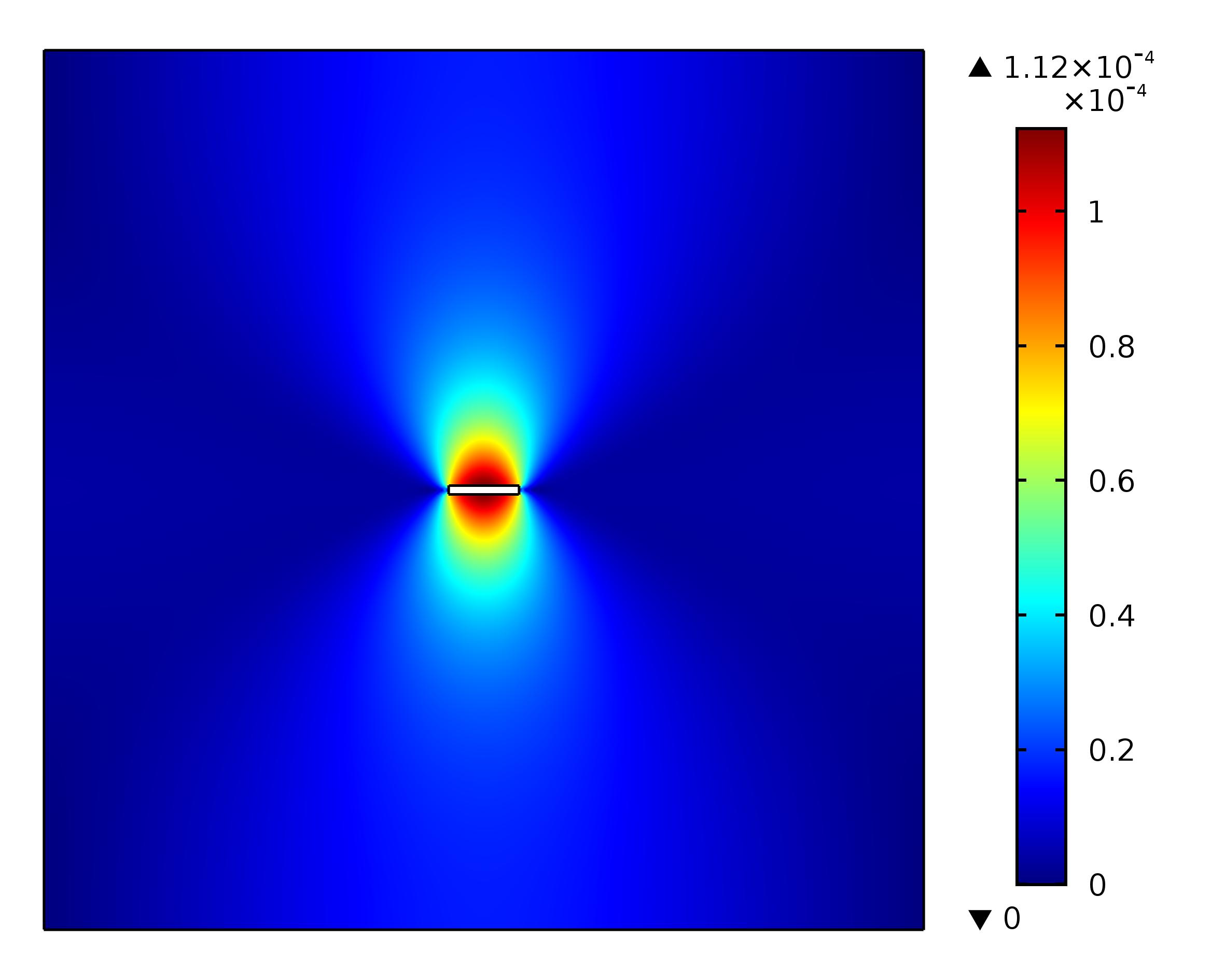}}
	\subfigure[$t=3.6$ s]{\includegraphics[width = 5.5cm]{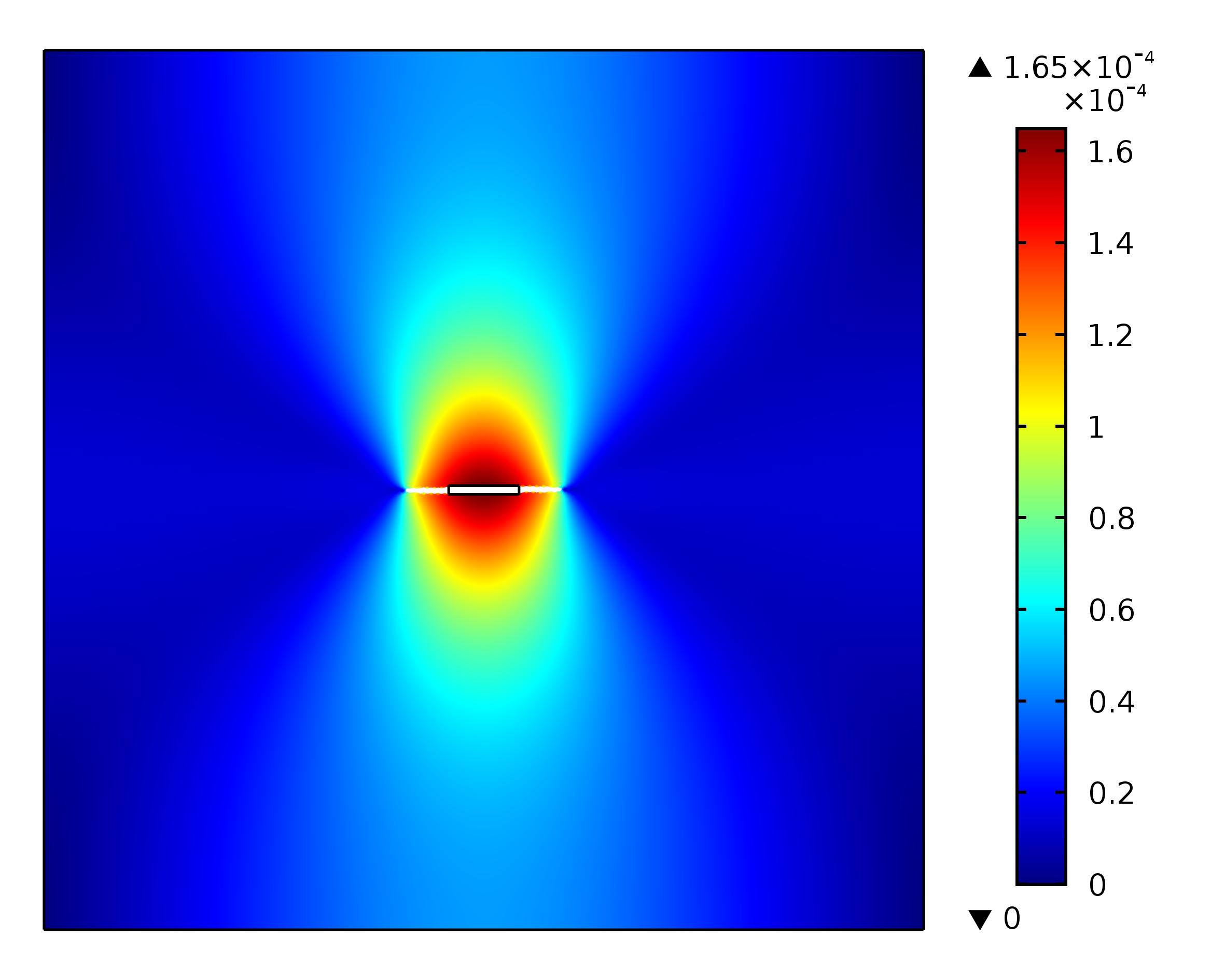}}
	\subfigure[$t=4.8$ s]{\includegraphics[width = 5.5cm]{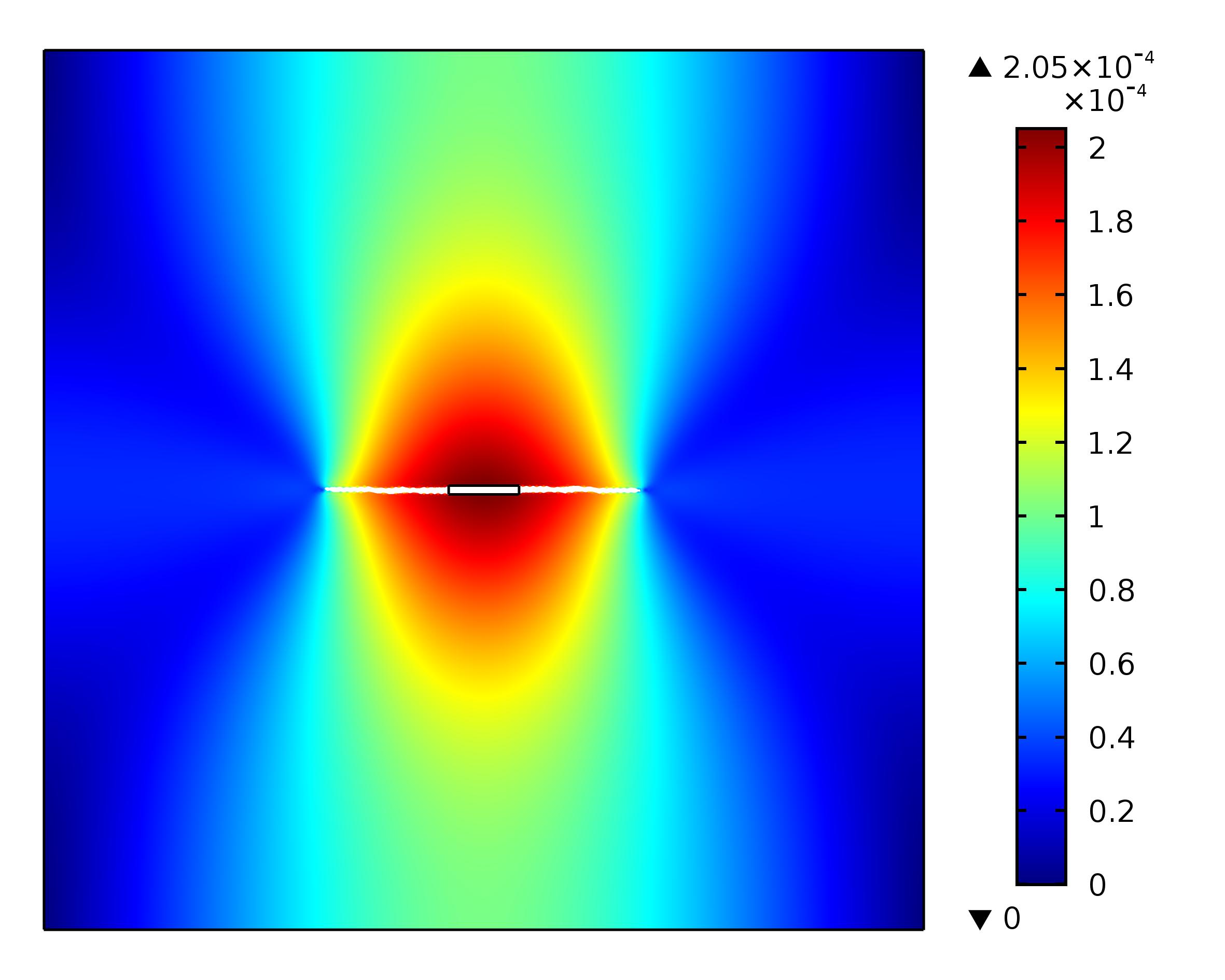}}
	\caption{Evolution of the displacement field by using the proposed PFM (Unit: m)}
	\label{Evolution of the displacement field by using the proposed PFM}
	\end{figure}

The evolution of the displacement field obtained by the proposed PFM is shown in Fig. \ref{Evolution of the displacement field by using the proposed PFM}. As expected, the maximum displacement occurs in the center of the fracture and it increases with the increasing time. This phenomenon is also consistent with those observations in a porous medium with fixed displacement boundaries \citep{mikelic2013phase,mikelic2015phase,zhou2018phase2}, which indirectly reflects the applicability of the proposed PFM in this study.

\begin{figure}[htbp]
	\centering
	\subfigure[$\sigma_{y0}/\sigma_{x0}=0.5$]{\includegraphics[width = 5.5cm]{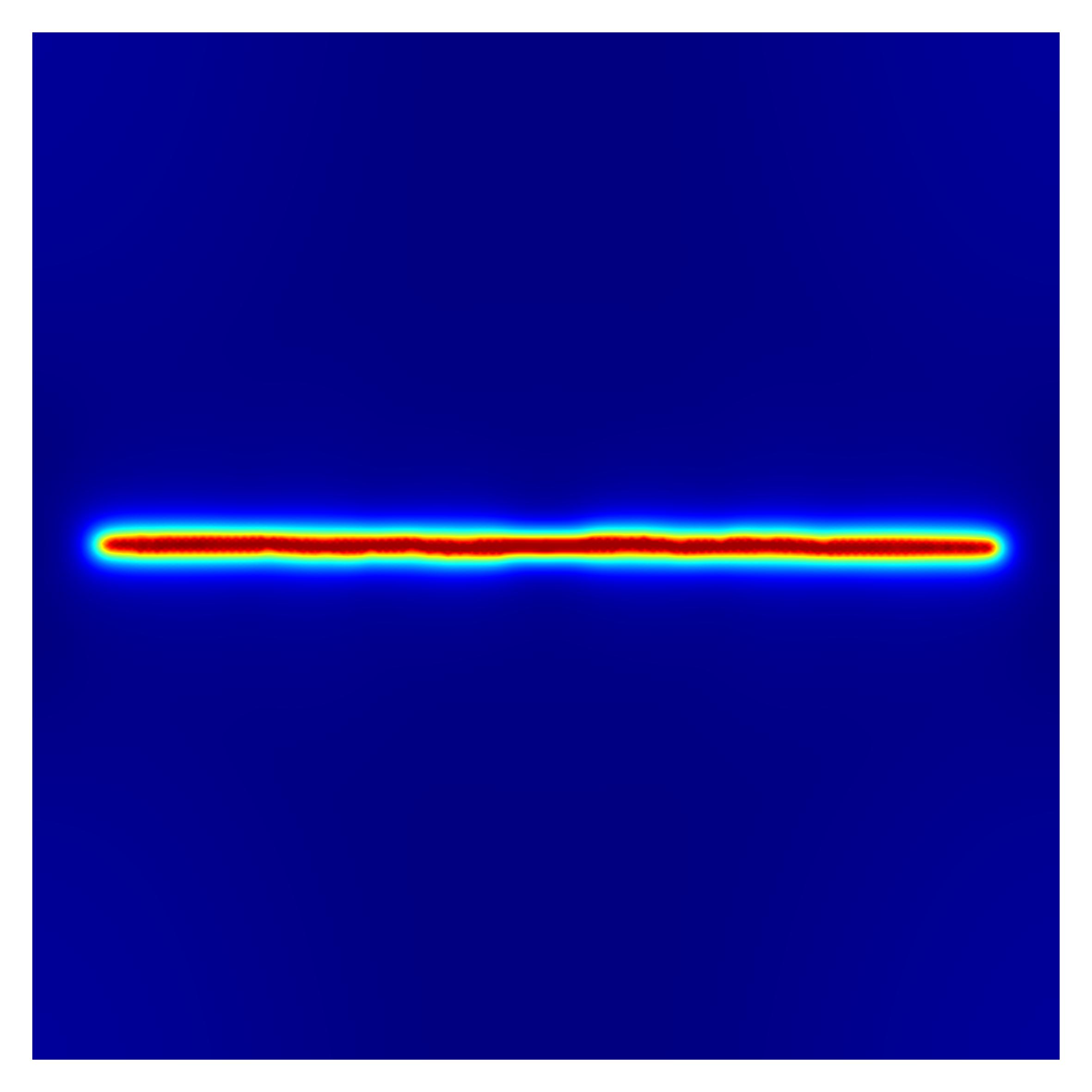}}
	\subfigure[$\sigma_{y0}/\sigma_{x0}=1$]{\includegraphics[width = 5.5cm]{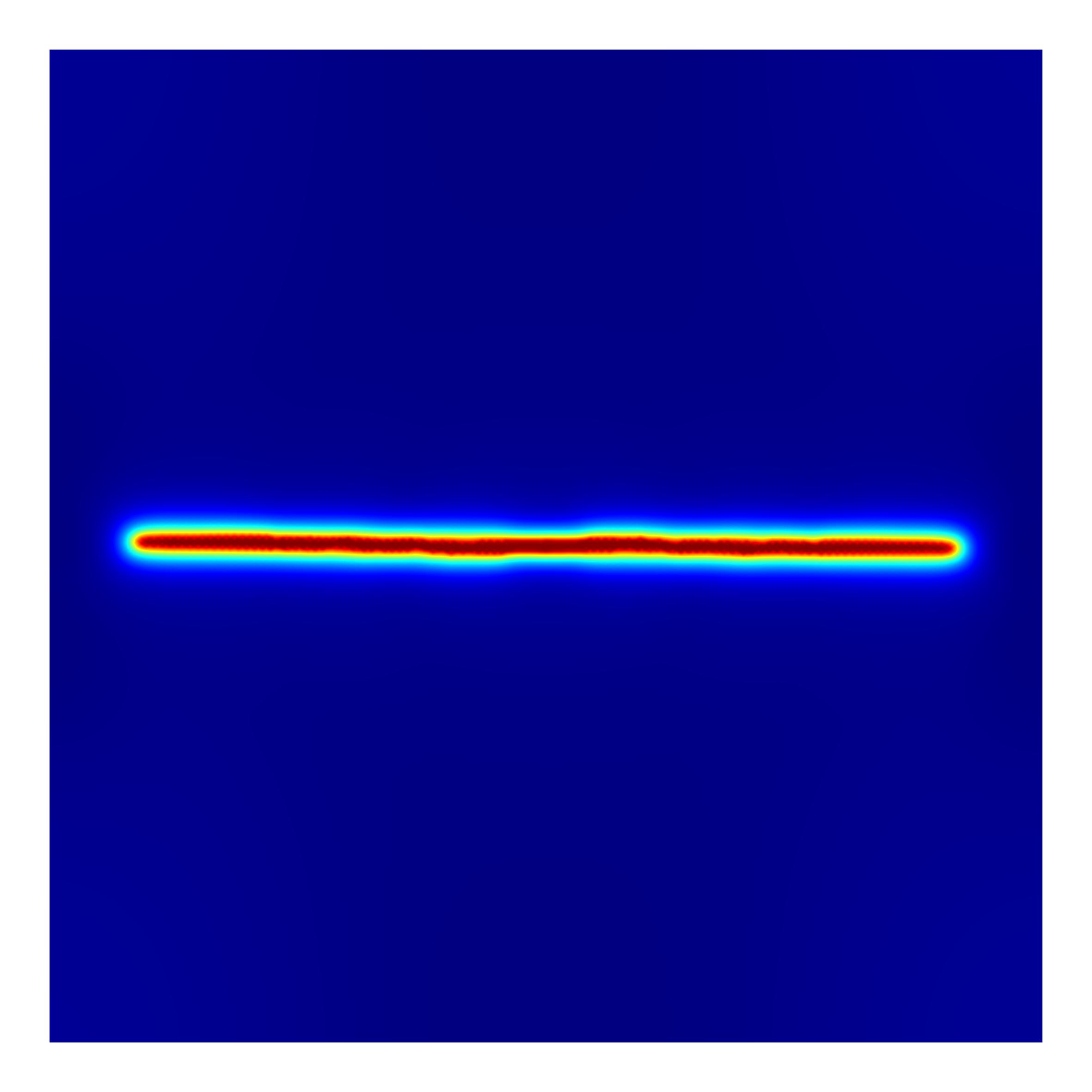}}\\
	\subfigure[$\sigma_{y0}/\sigma_{x0}=2$]{\includegraphics[width = 5.5cm]{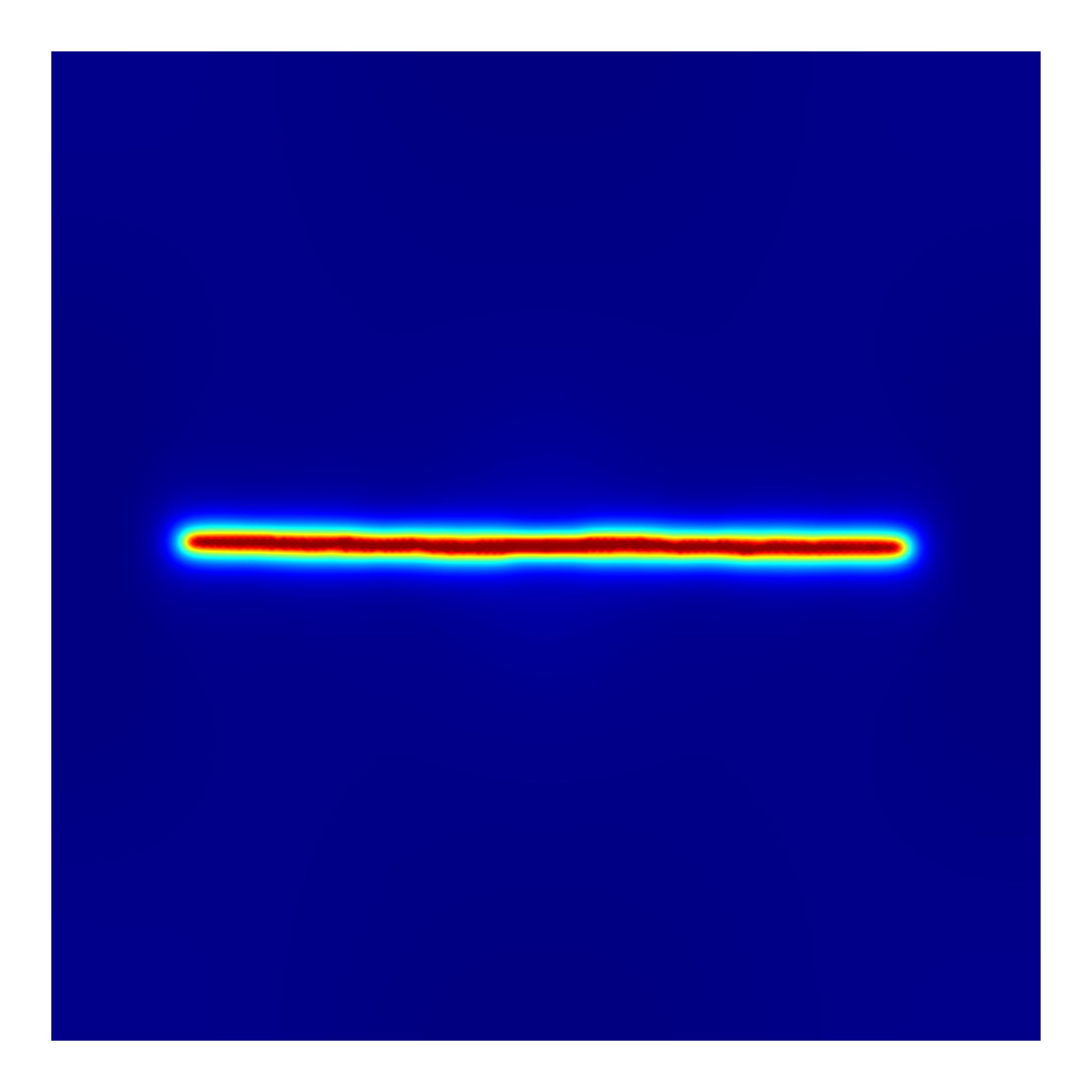}}
	\subfigure[$\sigma_{y0}/\sigma_{x0}=10$]{\includegraphics[width = 5.5cm]{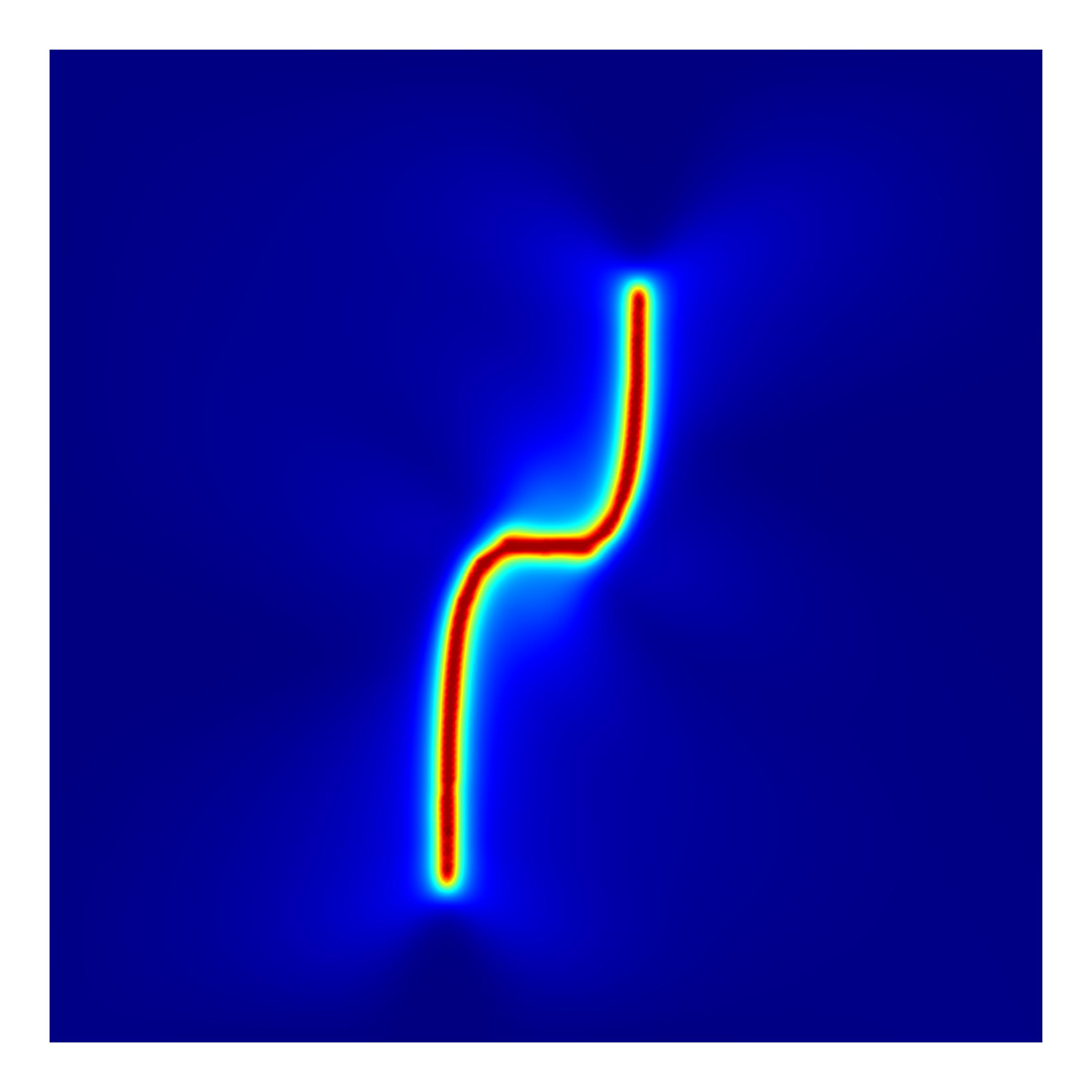}}
	\caption{Fracture propagation from a horizontal initial notch at time $t=10$ s under different $\sigma_{y0}/\sigma_{x0}$}
	\label{Fracture propagation from a horizontal initial notch at time t=10 s under different}
	\end{figure}

\begin{figure}[htbp]
	\centering
	\includegraphics[width = 10cm]{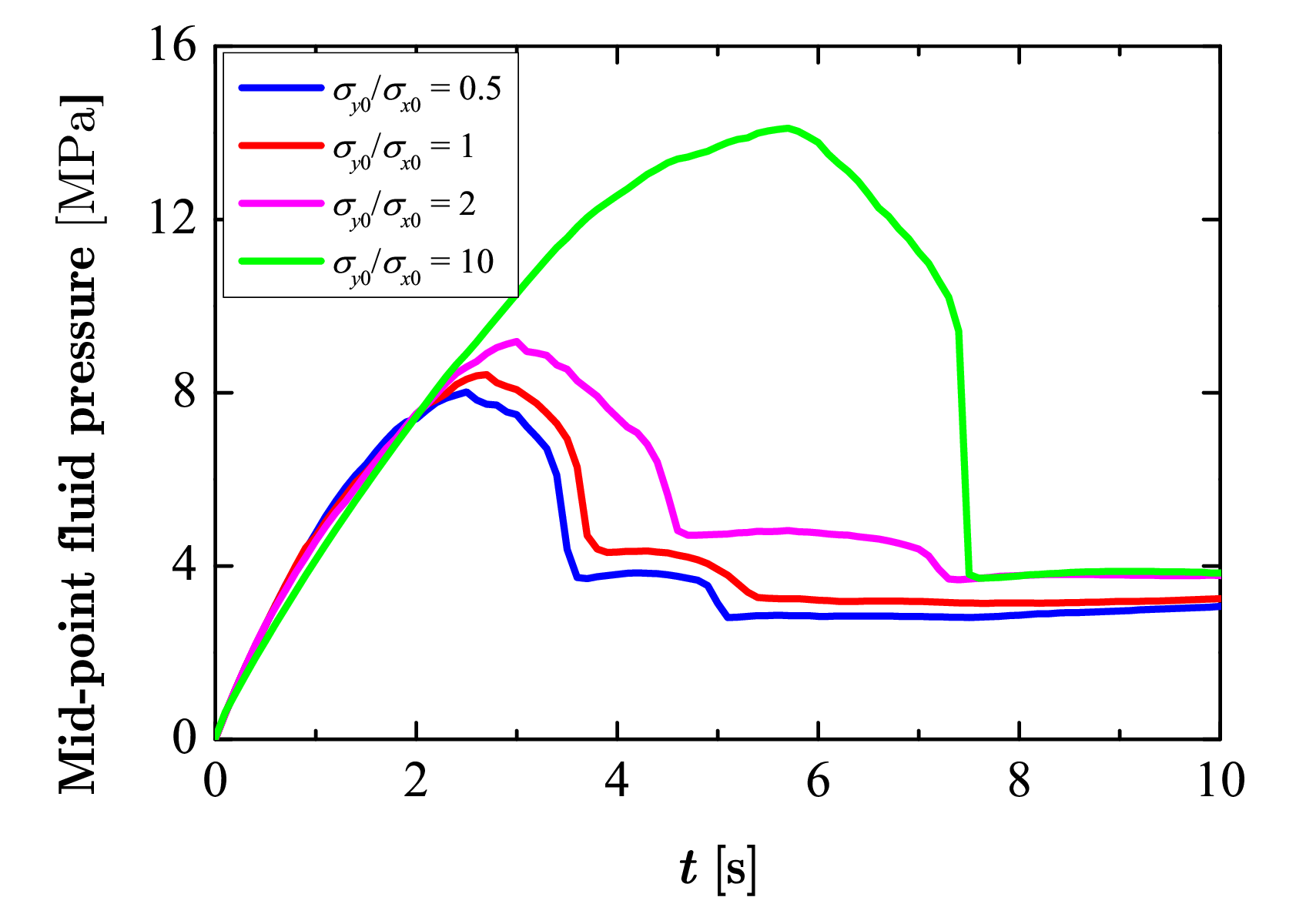}
	\caption{Effect of the ratio of $\sigma_{y0}/\sigma_{x0}$ on fluid pressure-time curve for a horizontal initial notch}
	\label{Effect of the ratio of sysx on fluid pressure-time curve for a horizontal initial notch}
	\end{figure}

It is well-known that the hydraulic fracture pattern is highly affected by the stress contrast acting on the outer boundaries of the calculation domain. Therefore, in this example, we change the ratio of $\sigma_{y0}/\sigma_{x0}$ to $[0.5,1,2,10]$ with $\sigma_{x0}=0.5$ MPa unchanged to demonstrate the effect of stress contrast. By using the proposed PFM, the calculated fracture paths at time $t=10$ s are shown in Fig. \ref{Fracture propagation from a horizontal initial notch at time t=10 s under different}. It can be observed that for $\sigma_{y0}/\sigma_{x0}=$ 0.5, 1, and 2, the fracture from the initial notch propagates horizontally and the fracture length decreases as the ratio of $\sigma_{y0}/\sigma_{x0}$ increases. However, when $\sigma_{y0}/\sigma_{x0}=10$, the fracture deflects and propagates along the direction of the maximum in-situ stress $S_{max}$, which is consistent with the engineering observations in hydraulic fracturing. In addition, the effect of the ratio of $\sigma_{y0}/\sigma_{x0}$ on the fluid pressure-time curve is shown in Fig. \ref{Effect of the ratio of sysx on fluid pressure-time curve for a horizontal initial notch}. The maximum fluid pressure at the mid-point of the initial notch is observed to increase with the increasing $\sigma_{y0}/\sigma_{x0}$.

\begin{figure}[htbp]
	\centering
	\includegraphics[width = 10cm]{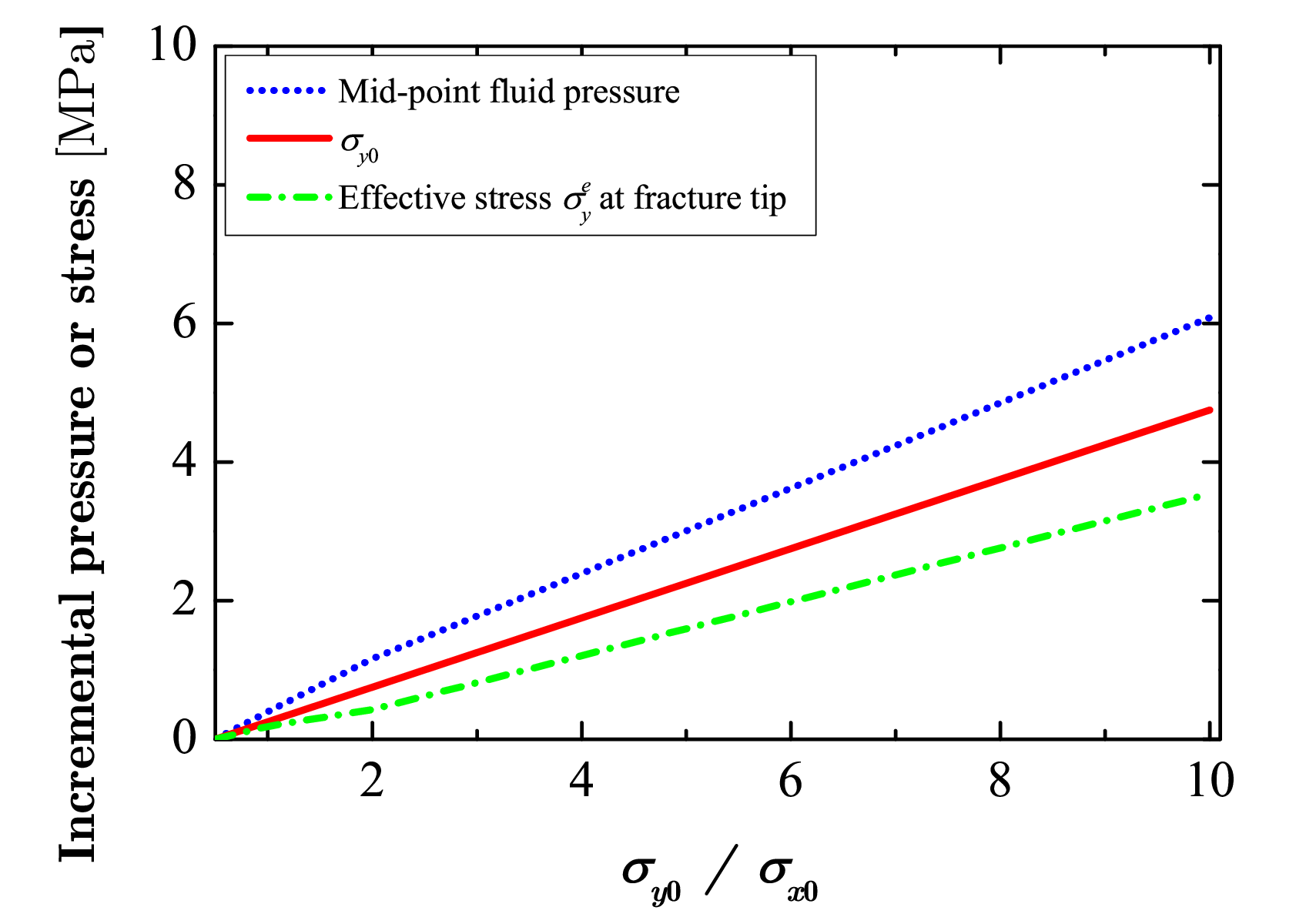}
	\caption{Relative incremental fluid pressure and effective stress at fracture tip under different $\sigma_{y0}/\sigma_{x0}$ in the case of a horizontal initial notch}
	\label{Relative incremental fluid pressure and effective stress at fracture tip under different sigmay0/sigmax0 in the case of a horizontal initial notch}
	\end{figure}

Figure \ref{Relative incremental fluid pressure and effective stress at fracture tip under different sigmay0/sigmax0 in the case of a horizontal initial notch} describes the relative incremental mid-point fluid pressure and effective stress at fracture tip under different $\sigma_{y0}/\sigma_{x0}$ in the case of a horizontal initial notch. Note that the effective stress along the $y$ direction is accordingly used to show the effect of $\sigma_{y0}$, and the case of $\sigma_{y0}/\sigma_{x0}=0.5$ is used as the reference. As shown in Fig. \ref{Relative incremental fluid pressure and effective stress at fracture tip under different sigmay0/sigmax0 in the case of a horizontal initial notch}, when the vertical stress $\sigma_{y0}$ increases linearly, the mid-point fluid pressure and effective stress at the fracture tip also increase linearly. This phenomena indirectly verifies the feasibility and practicability of the proposed PFM. Furthermore, the incremental fluid pressure is slightly larger than the vertical stress variation while the incremental effective stress at fracture tip is slightly lower.

\subsection{Fracture from an inclined notch}\label{Fracture from an inclined notch}

We now consider an inclined initial notch in the calculation domain in Fig. \ref{Geometry and boundary conditions of the calculation domain}. The notch has an length of 0.8 m and an inclination angle $\theta=45^\circ$ while the other simulation settings are the same as those in Subsection \ref{Fractures from a horizontal notch}. Stress contrast $\sigma_{x0}/\sigma_{y0}=[1,2,4,6,8,10]$ is applied with the vertical stress $\sigma_{y0}=0.5$ MPa unchanged. Therefore, there are six simulations performed to show the effect of stress contrast on the fracture propagation from the inclined notch.

\begin{figure}[htbp]
	\centering
	\subfigure[$\sigma_{x0}/\sigma_{y0}=1$]{\includegraphics[width = 5.5cm]{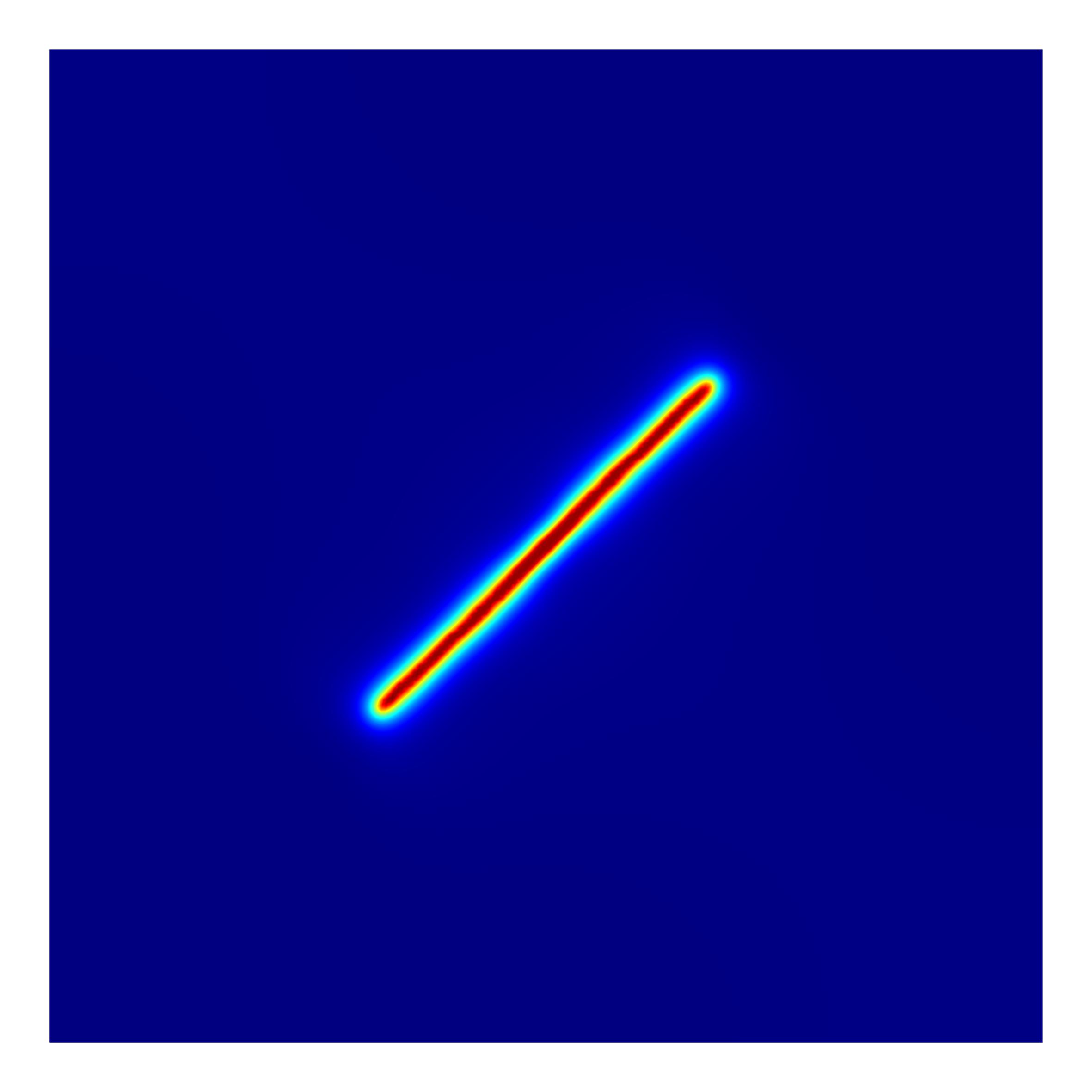}}
	\subfigure[$\sigma_{x0}/\sigma_{y0}=2$]{\includegraphics[width = 5.5cm]{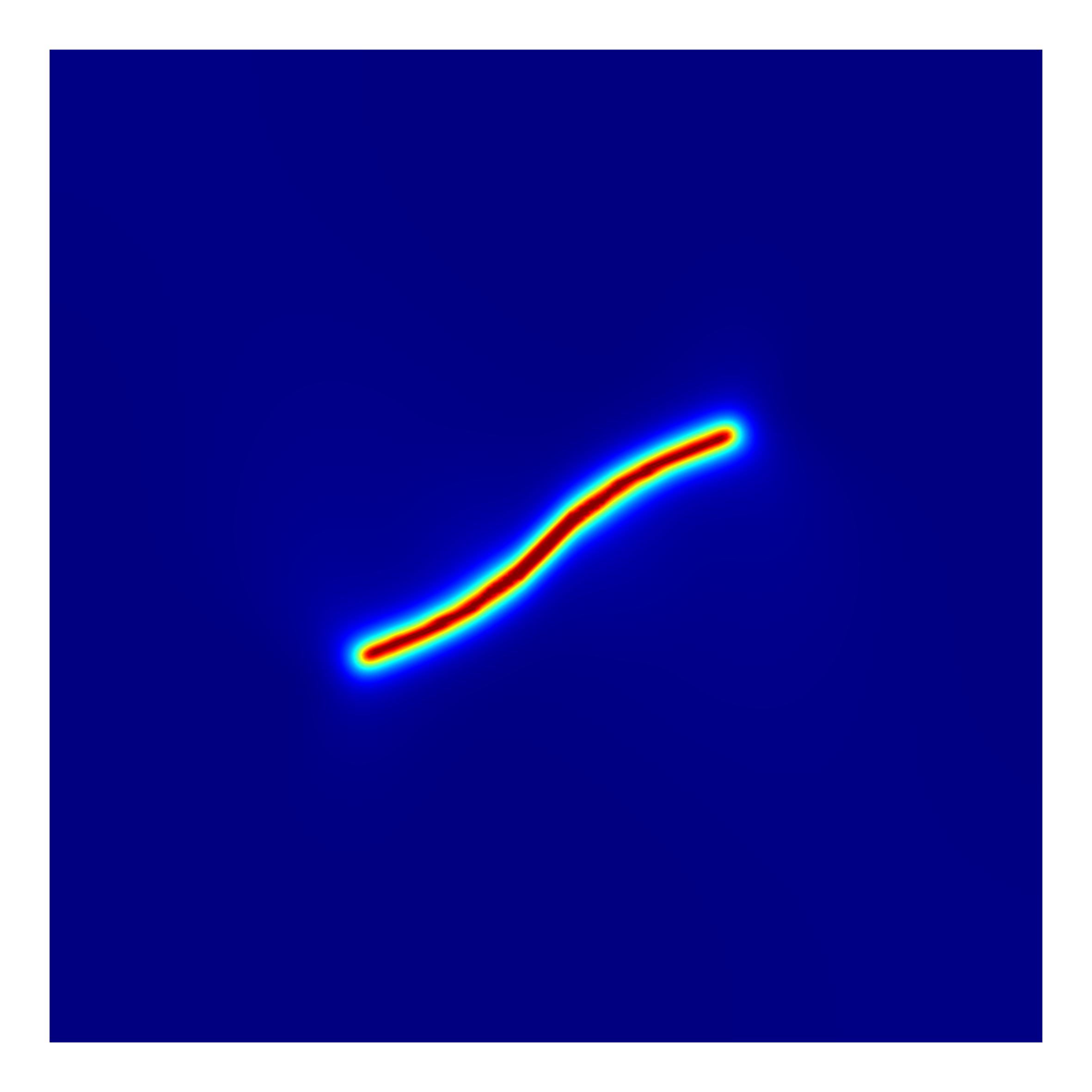}}
	\subfigure[$\sigma_{x0}/\sigma_{y0}=4$]{\includegraphics[width = 5.5cm]{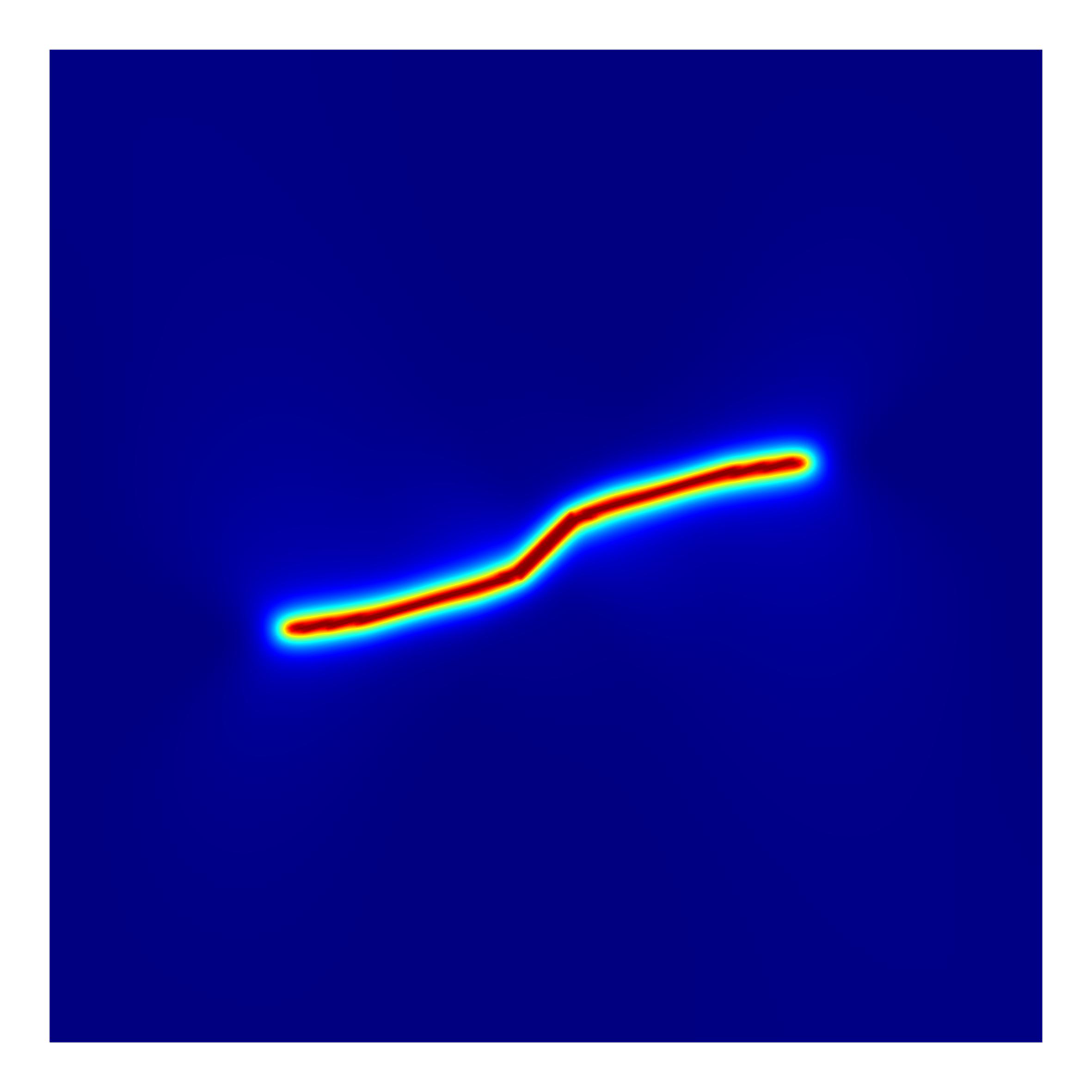}}\\
	\subfigure[$\sigma_{x0}/\sigma_{y0}=6$]{\includegraphics[width = 5.5cm]{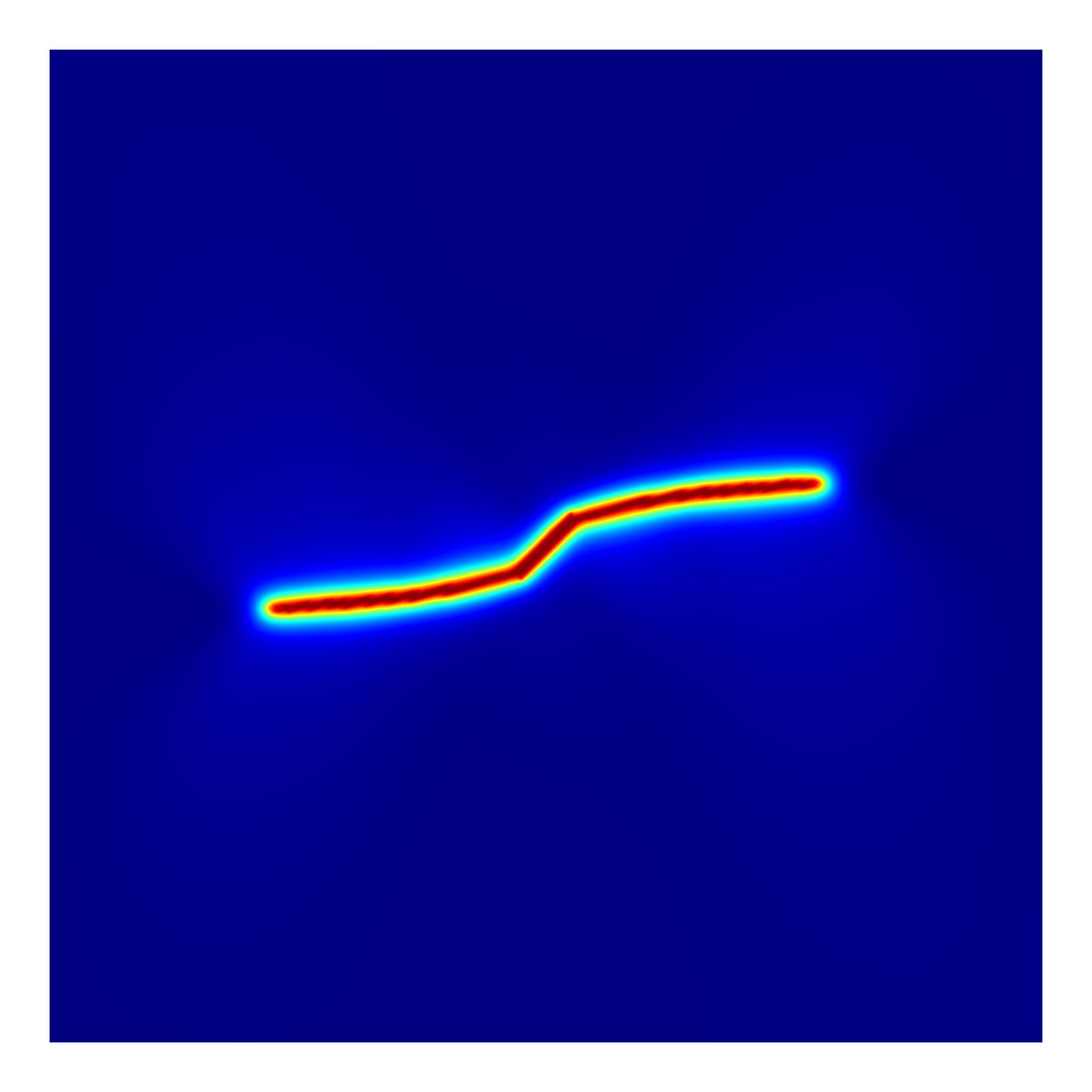}}
	\subfigure[$\sigma_{x0}/\sigma_{y0}=8$]{\includegraphics[width = 5.5cm]{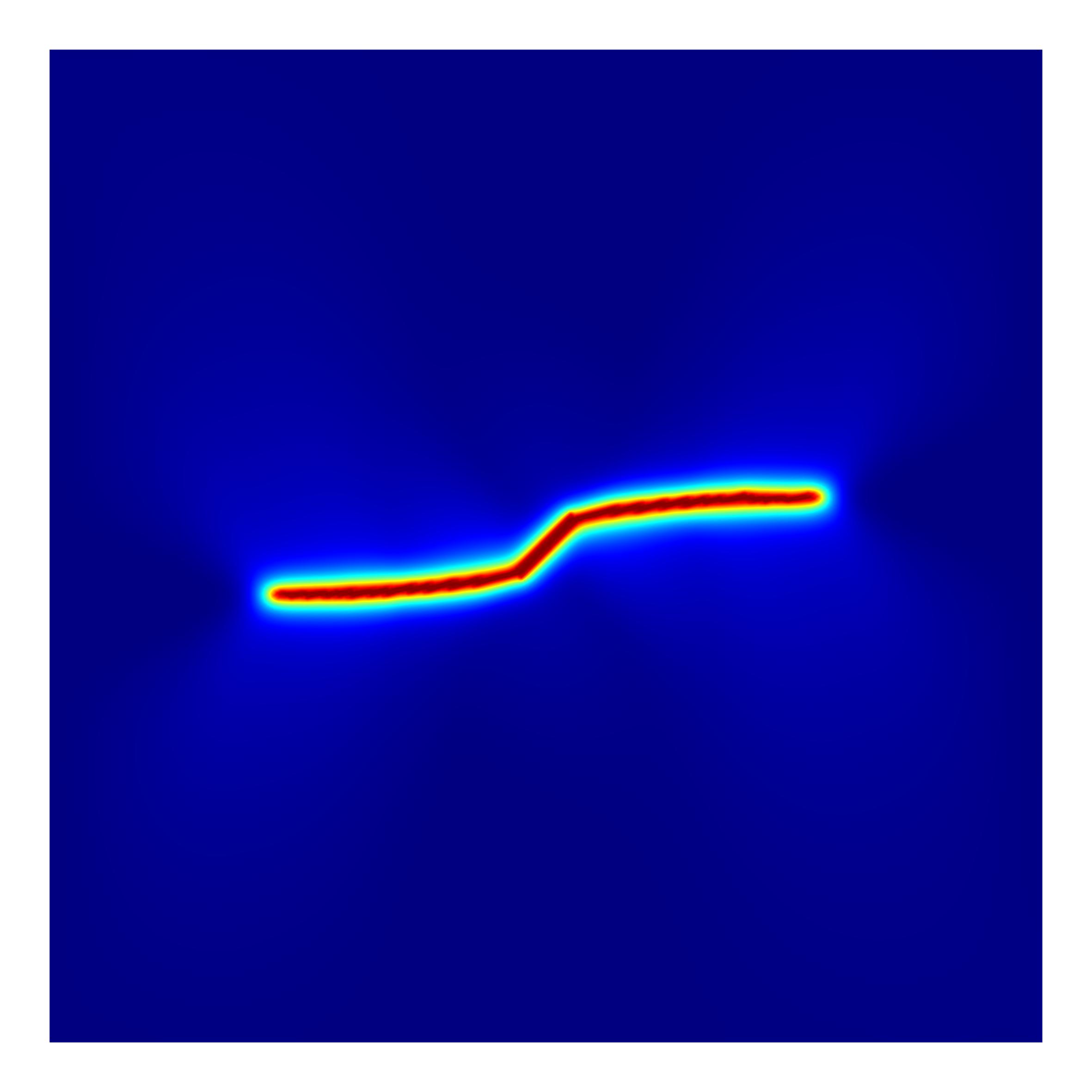}}
	\subfigure[$\sigma_{x0}/\sigma_{y0}=10$]{\includegraphics[width = 5.5cm]{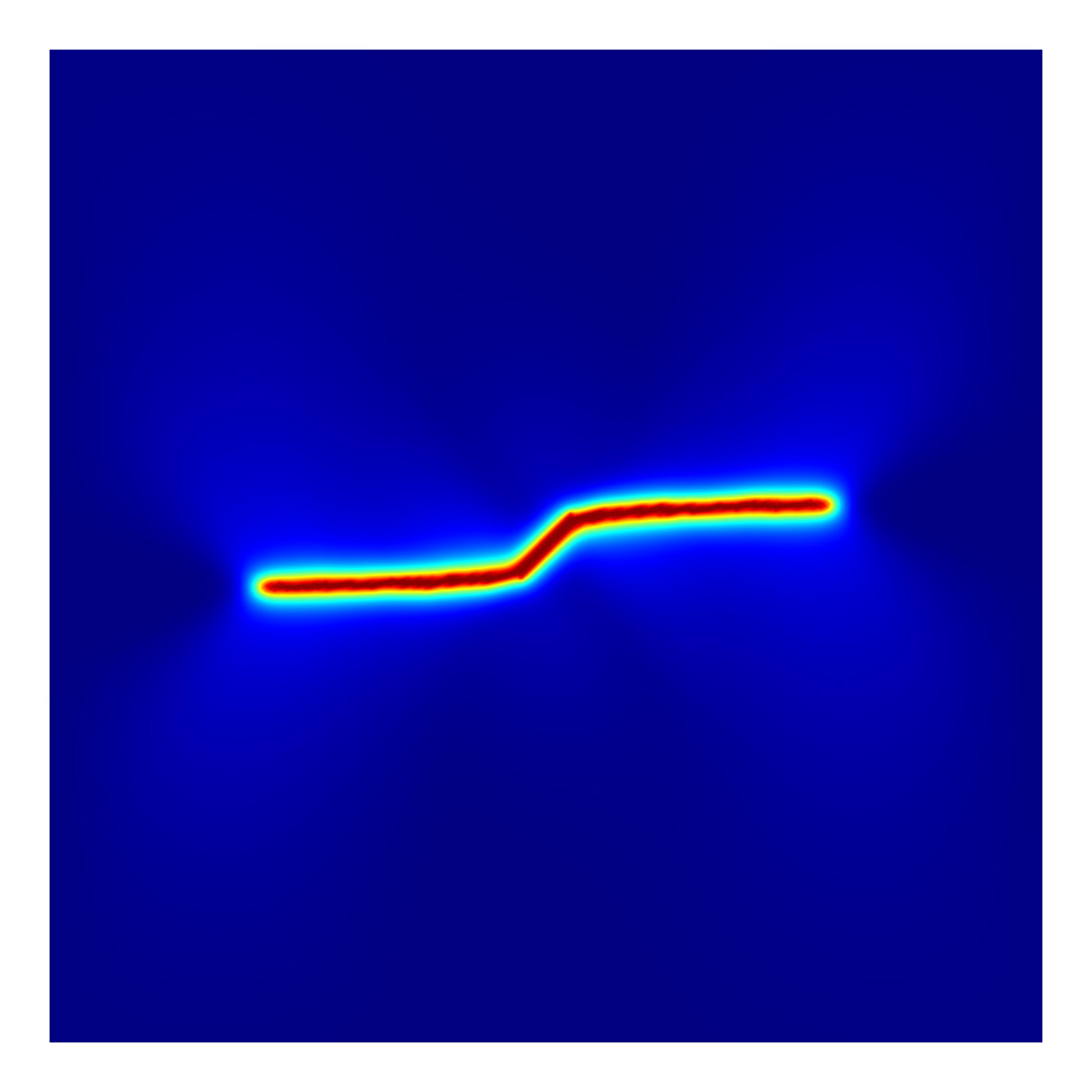}}
	\caption{Fracture propagation from an inclined initial notch at time $t=6$ s under different $\sigma_{x0}/\sigma_{y0}$}
	\label{Fracture propagation from an inclined initial notch at time t=6 s under different}
	\end{figure}

\begin{figure}[htbp]
	\centering
	\includegraphics[width = 10cm]{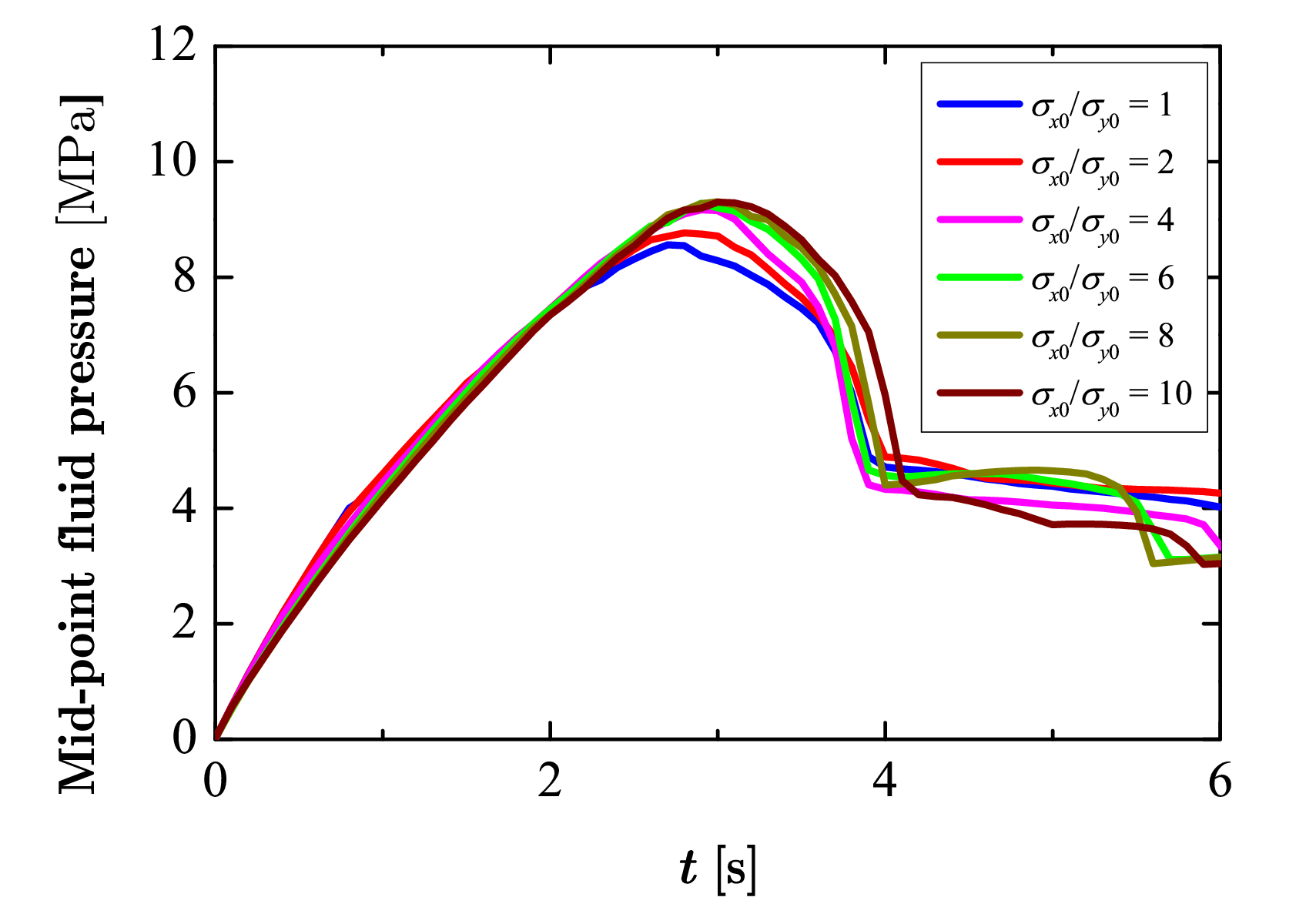}
	\caption{Effect of the ratio of $\sigma_{x0}/\sigma_{y0}$ on fluid pressure-time curve for an inclined initial notch}
	\label{Effect of the ratio of sysx on fluid pressure-time curve for an inclined initial notch}
	\end{figure}

By using the proposed PFM, the phase field distribution for different $\sigma_{x0}/\sigma_{y0}$ at time $t=6$ s is shown in Fig. \ref{Fracture propagation from an inclined initial notch at time t=6 s under different}. As observed, the fracture propagates straight along the direction of the initial notch when $\sigma_{x0}/\sigma_{y0}=1$ at time $t=6$ s; however, if the stress ratio is larger than 1, the fracture deflects from the direction of the notch ($\theta=45^\circ$). All the simulations consistently indicate that the larger the ratio of $\sigma_{x0}/\sigma_{y0}$ is, the smaller angle the deflected fracture has from the direction of the maximum in-situ stress $S_{max}$ ($\sigma_{x0}$) in this example. The fluid pressure-time curve for different $\sigma_{x0}/\sigma_{y0}$ is shown in Fig. \ref{Effect of the ratio of sysx on fluid pressure-time curve for an inclined initial notch}. Because the minimum in-situ stress $\sigma_{y0}$ is kept constant, the fluid pressure at the center of the notch only increases slightly with the increasing stress contrast $\sigma_{x0}/\sigma_{y0}$.
	
\subsection{Fracture from two perpendicularly crossed notches}\label{Fracture from two perpendicularly crossed notches}

We also test the hydraulic fracture propagation from two perpendicularly crossed notches. The notches are located in the center of the calculation domain shown in Fig. \ref{Geometry and boundary conditions of the calculation domain} and both have a length of 0.8 m. The other simulation settings are the same as those in Subsection \ref{Fractures from a horizontal notch}. We only consider four cases of in-situ stress as described in Table \ref{Four cases of boundary and in-situ stresse}. This means that the direction of the maximum stress $S_{max}$ and the minimum stress $S_{min}$ varies in the four cases.

	\begin{table}[htbp]
	\caption{Four cases of boundary and in-situ stresses}
	\label{Four cases of boundary and in-situ stresse}
	\centering
	\begin{tabular}{l||ll}
		\hline
		Case & Horizontal stress & Vertical stress\\
		\hline
		Case 1 & $\sigma_{x0}=1$ MPa & $\sigma_{y0}=0.5$ MPa\\
		Case 2 & $\sigma_{x0}=5$ MPa & $\sigma_{y0}=0.5$ MPa\\
		Case 3 & $\sigma_{x0}=0.5$ MPa & $\sigma_{y0}=1$ MPa\\
		Case 4 & $\sigma_{x0}=0.5$ MPa & $\sigma_{y0}=5$ MPa\\
		\hline
	\end{tabular}
	\end{table}

\begin{figure}[htbp]
	\centering
	\subfigure[Case 1]{\includegraphics[width = 5.5cm]{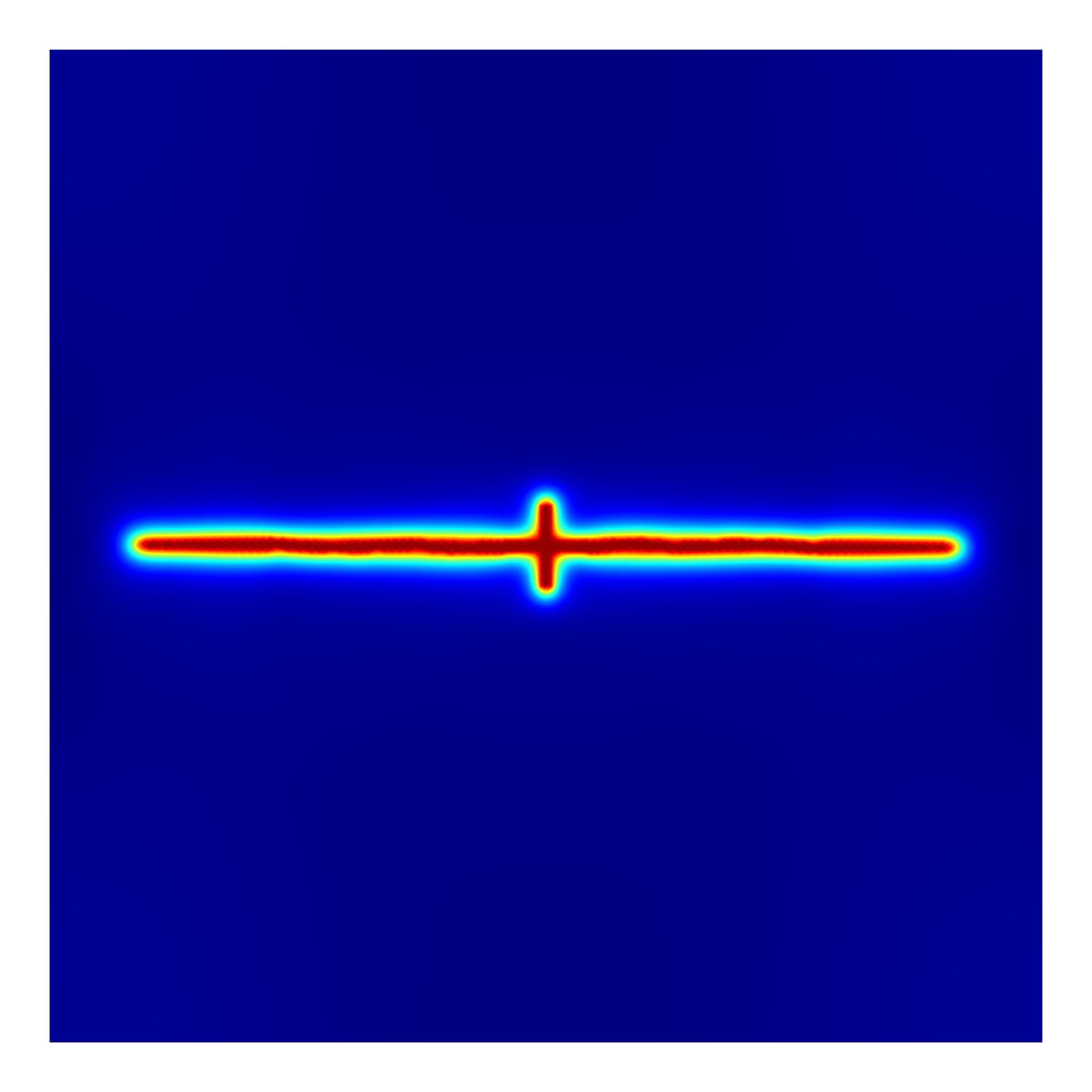}}
	\subfigure[Case 2]{\includegraphics[width = 5.5cm]{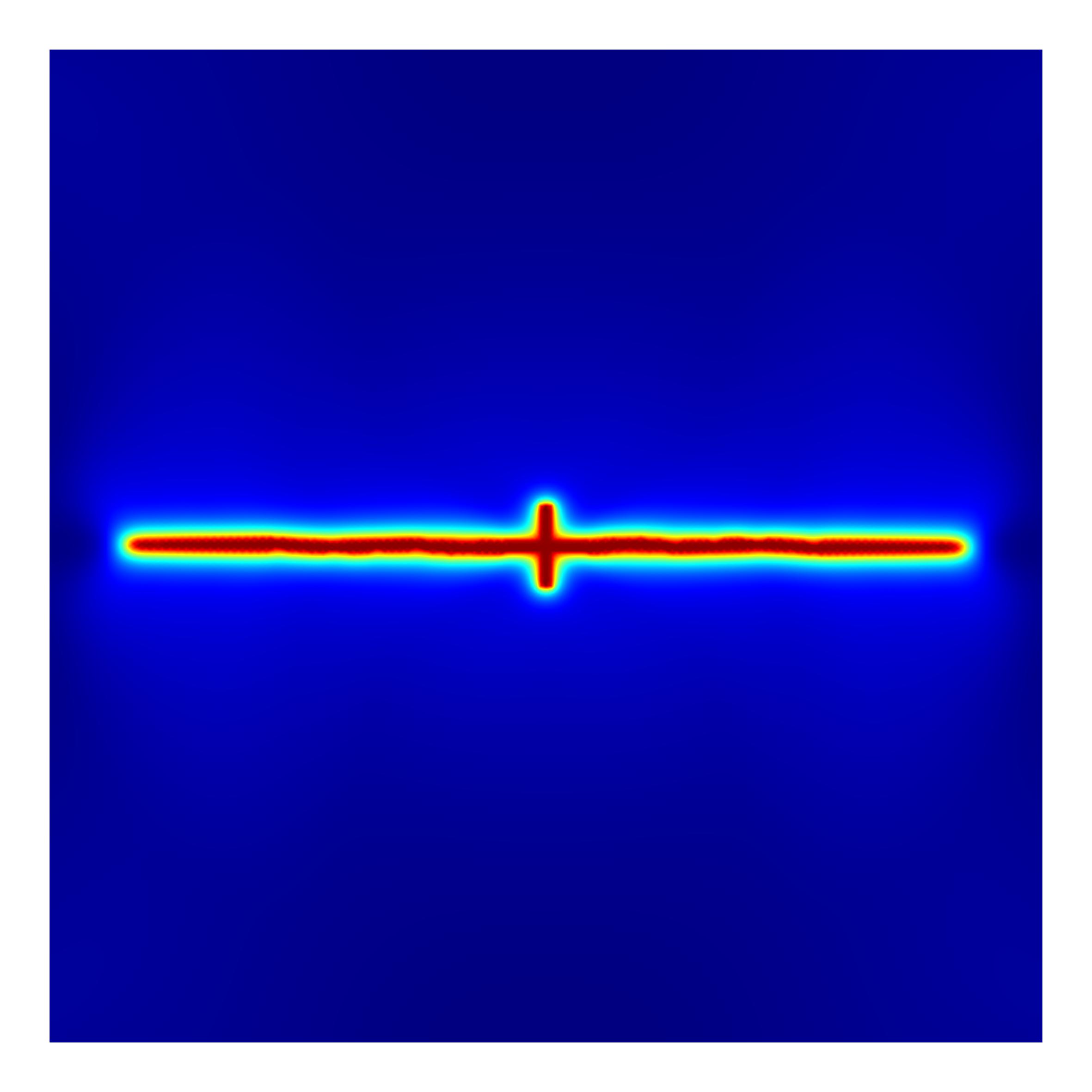}}\\
	\subfigure[Case 3]{\includegraphics[width = 5.5cm]{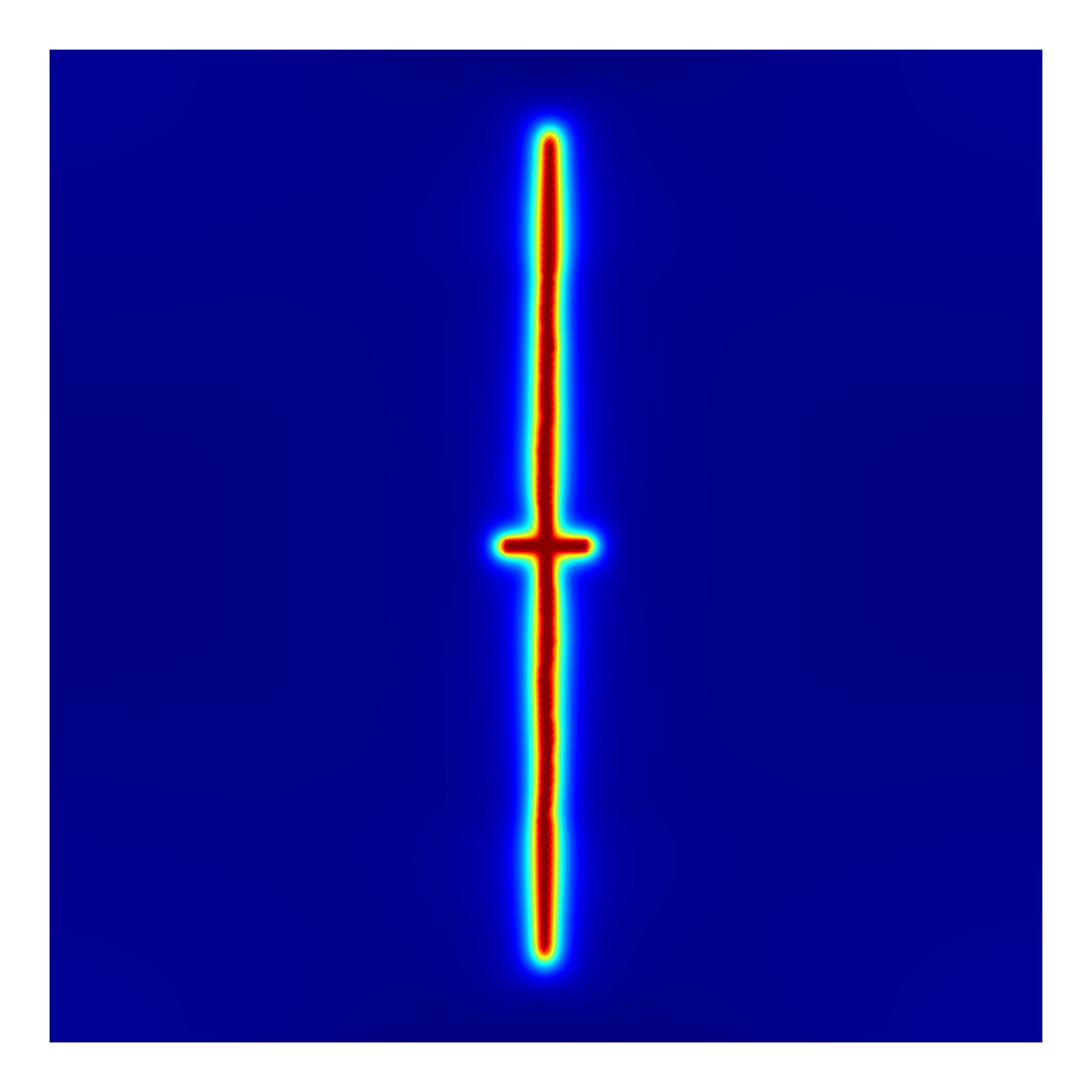}}
	\subfigure[Case 4]{\includegraphics[width = 5.5cm]{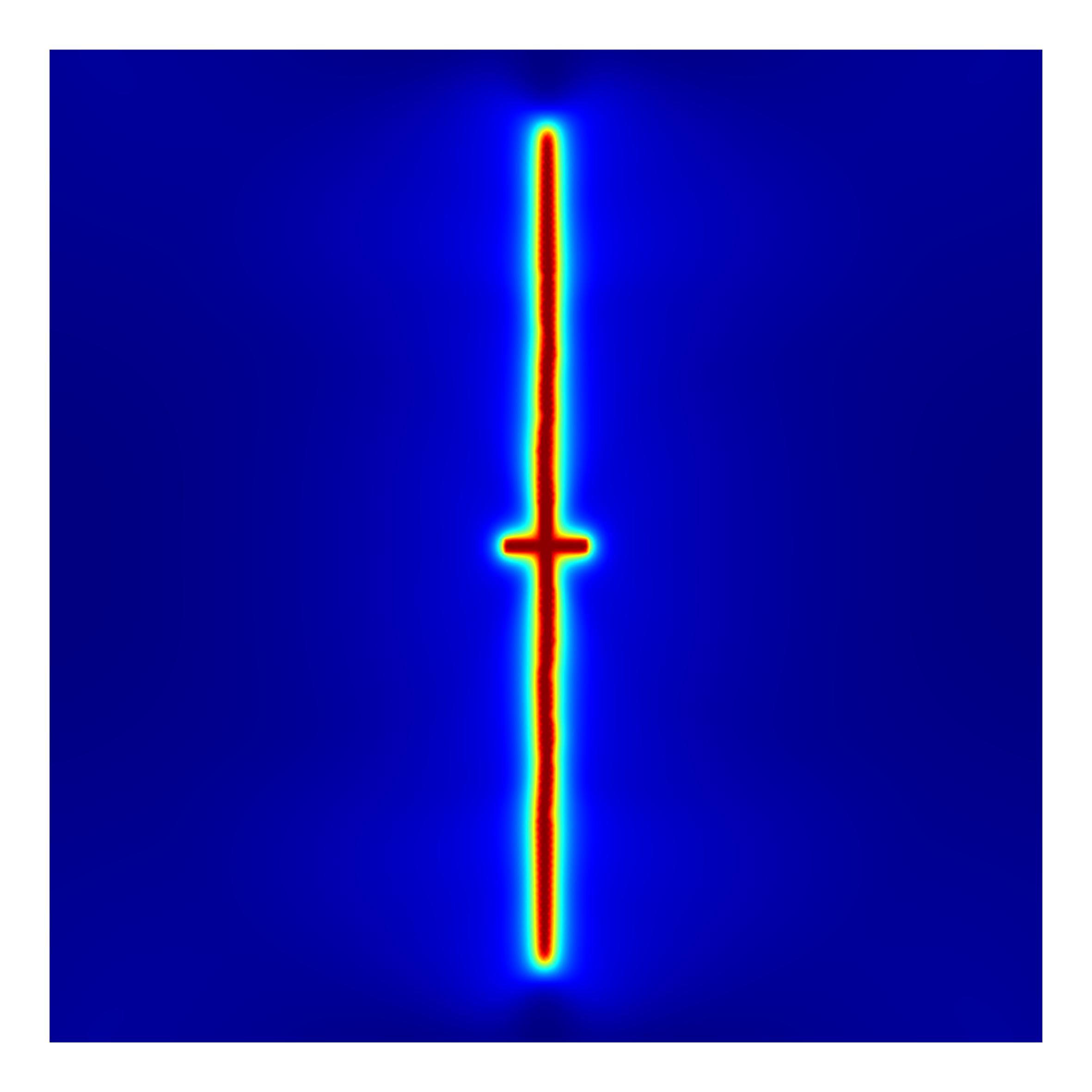}}
	\caption{Fracture propagation from two perpendicularly crossed notches at time $t=5$ s}
	\label{Fracture propagation from two perpendicularly crossed notches at time $t=5$ s}
	\end{figure}

\begin{figure}[htbp]
	\centering
	\includegraphics[width = 10cm]{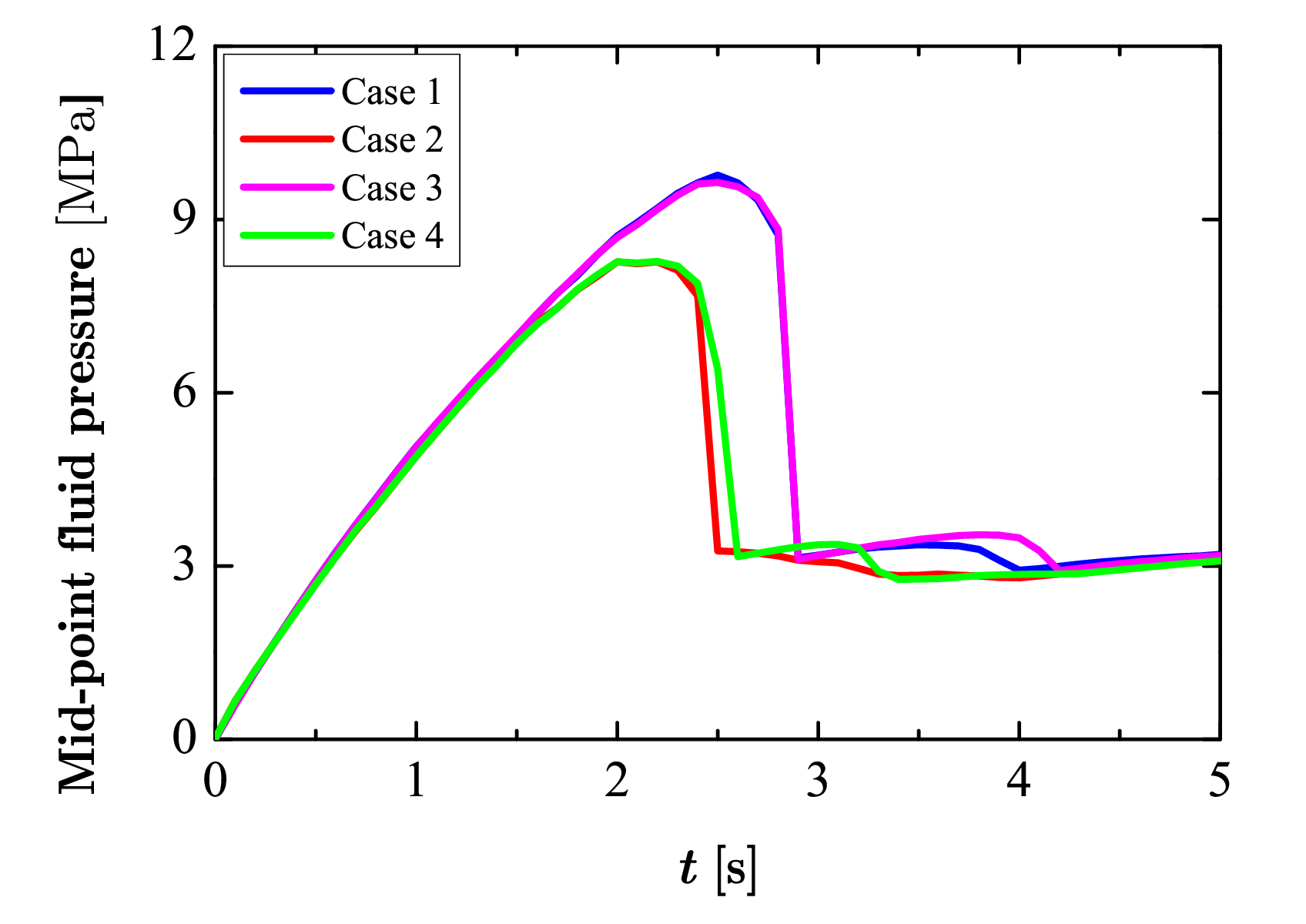}
	\caption{Fluid pressure-time curve for two perpendicularly crossed notch}
	\label{Fluid pressure-time curve for two perpendicularly crossed notch}
	\end{figure}

By using the proposed PFM, the phase field distribution for different cases is shown in Fig. \ref{Fracture propagation from two perpendicularly crossed notches at time $t=5$ s}. As observed, fractures only initiate and propagate from the notch perpendicular to the direction of the minimum stress $S_{min}$ while the other notches do not even grow. Note that the direction of these fractures is consistent with the direction of $S_{min}$. Therefore, Fig. \ref{Fracture propagation from two perpendicularly crossed notches at time $t=5$ s} validates the engineering observation that the fractures perpendicular to the direction of the minimum in-situ stress will initiate and propagate more easily. The fluid pressure-time curves for different cases are shown in Fig. \ref{Fluid pressure-time curve for two perpendicularly crossed notch}. The curves for Cases 1 and 3 and those for Case 2 and 4 are almost the same because the maximum and minimum in-situ stresses are identical. 

\subsection{Fracture from two horizontal notches}\label{Fracture from two horizontal notches}
\begin{figure}[htbp]
	\centering
	\subfigure[Geometry and boundary condition]{\includegraphics[height = 5.5cm]{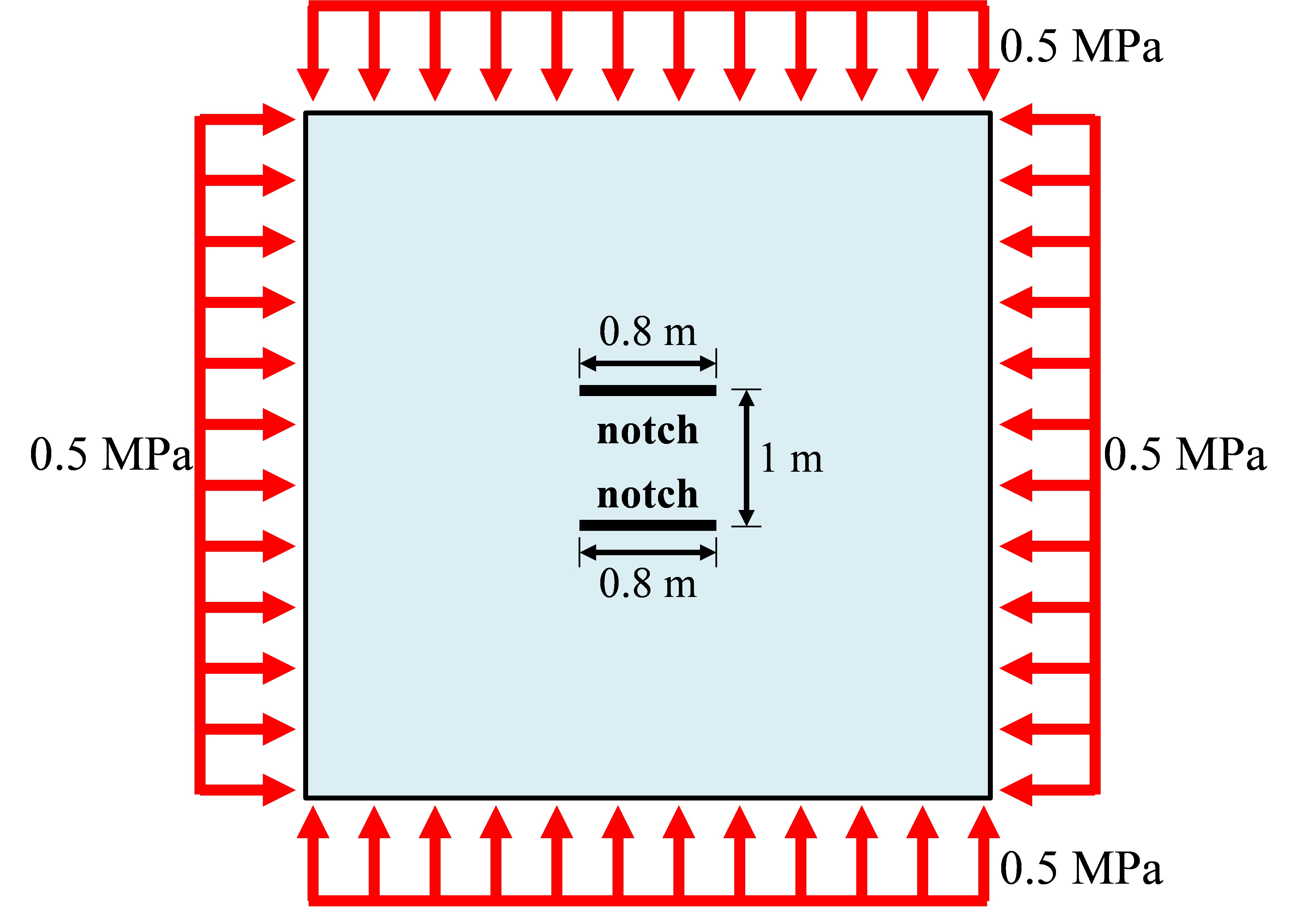}}
	\subfigure[Phase field at $t=2.5$ s]{\includegraphics[height = 5.5cm]{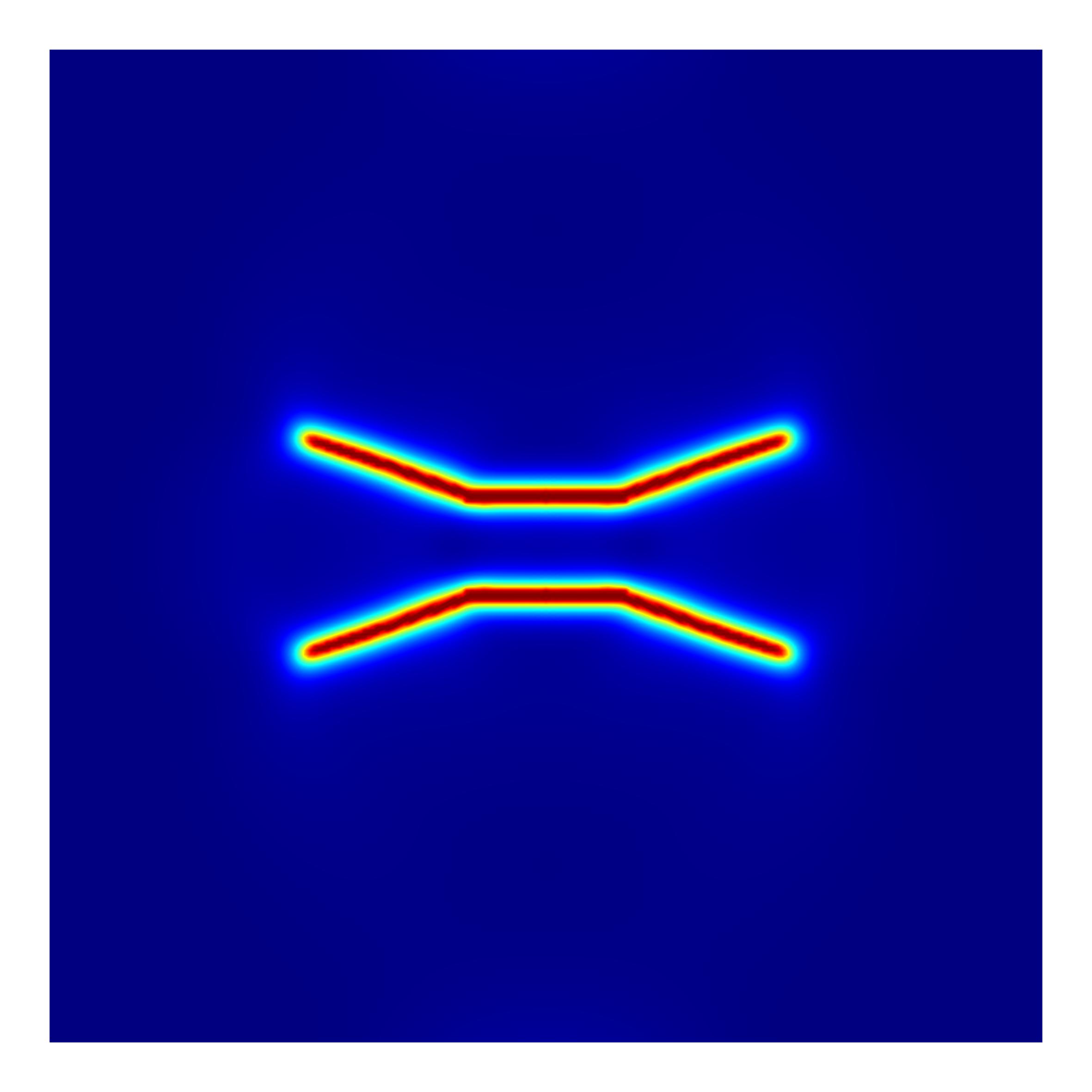}}\\
	\caption{Fracture propagation from two horizontal notches}
	\label{Fracture propagation from two horizontal notches}
	\end{figure}

In this 2D example, hydraulic fracture from two horizontal notches is presented to further validate the proposed PFM. The geometry and boundary stress condition of this example are shown in Fig. \ref{Fracture propagation from two horizontal notches}a where the horizontal and vertical remote stresses are both 0.5 MPa for simplicity. The parameters and numerical settings are the same as Subsection \ref{Fractures from a horizontal notch}. The phase field at $t=2.5$ s is shown in Fig. \ref{Fracture propagation from two horizontal notches}b. A symmetrical fracture pattern is observed and the fracture deflection occurs during fluid injection. This observation is in good agreement with the ``stress shadowing" phenomenon in engineering practice \citep{sobhaniaragh2019computational}.

\subsection{Linearly varying stress}\label{Linearly varying stress}

\begin{figure}[htbp]
	\centering
	\includegraphics[height = 6cm]{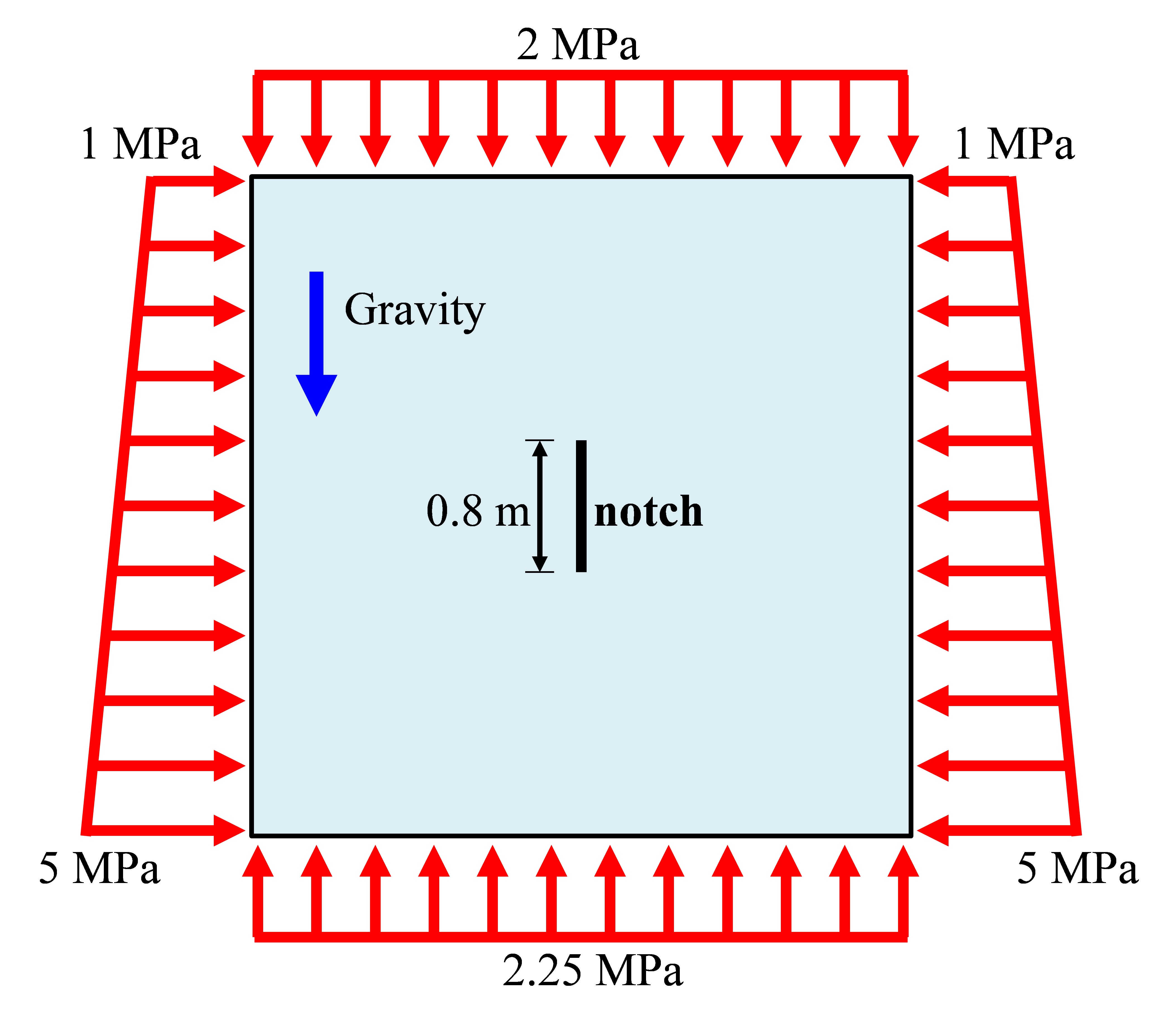}
	\caption{Linearly varying stress boundary condition}
	\label{Linearly varying stress boundary condition}
	\end{figure}

In this final 2D example, we test the effect of linearly varying stress field on hydraulic fracture propagation. The geometry of this example is the same as that in Fig. \ref{Geometry and boundary conditions of the calculation domain} while the boundary stress condition is shown in \ref{Linearly varying stress boundary condition}. Note that in this example, the gravity is applied in the vertical direction, and linearly varying horizontal stress acts in the horizontal direction, which can be considered as a combination of self-weight stress field and tectonic stress field in underground geological environment. The parameters and numerical settings are the same as Subsection \ref{Fractures from a horizontal notch}.

\begin{figure}[htbp]
	\centering
	\subfigure[$t=4$ s]{\includegraphics[width = 5.5cm]{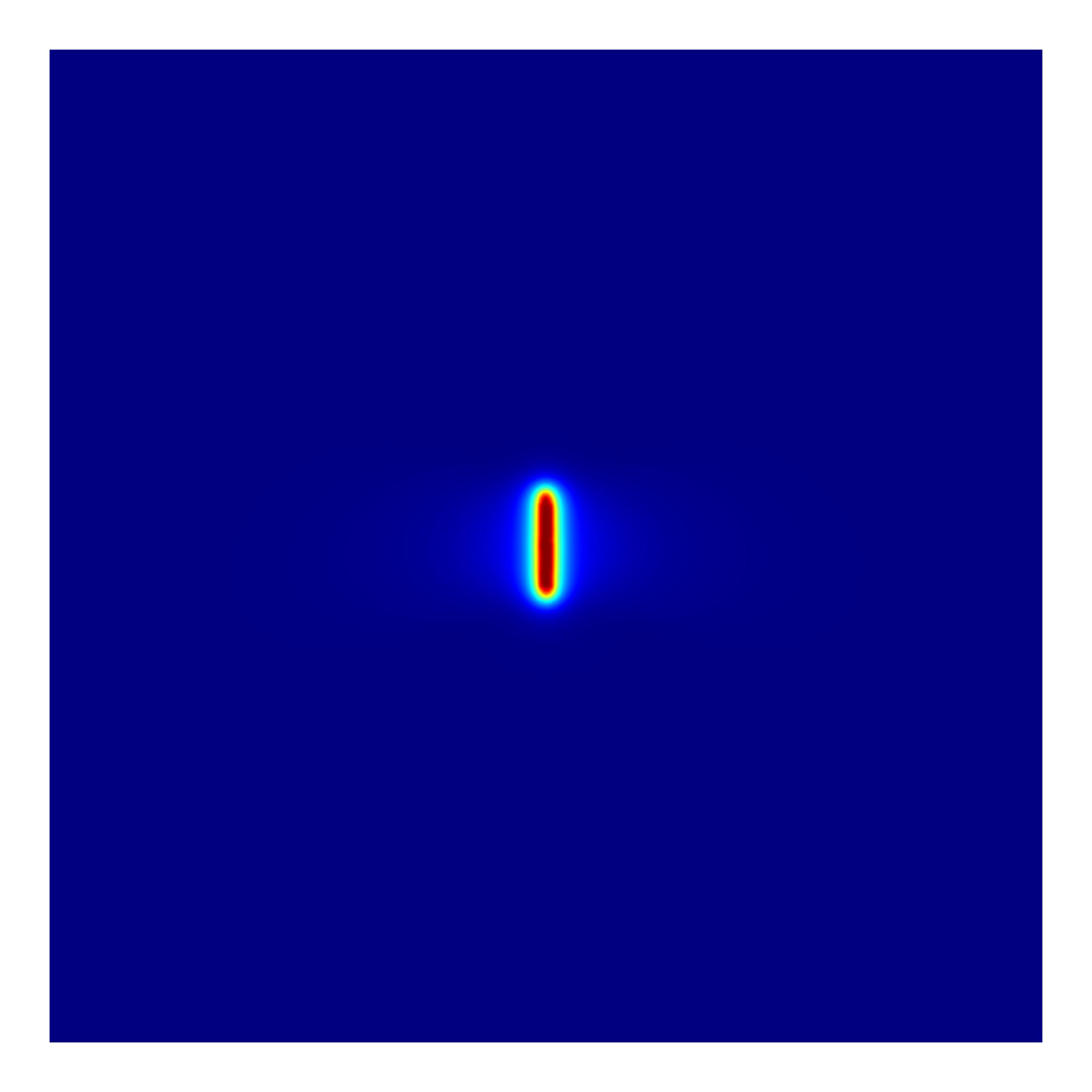}}
	\subfigure[$t=9$ s]{\includegraphics[width = 5.5cm]{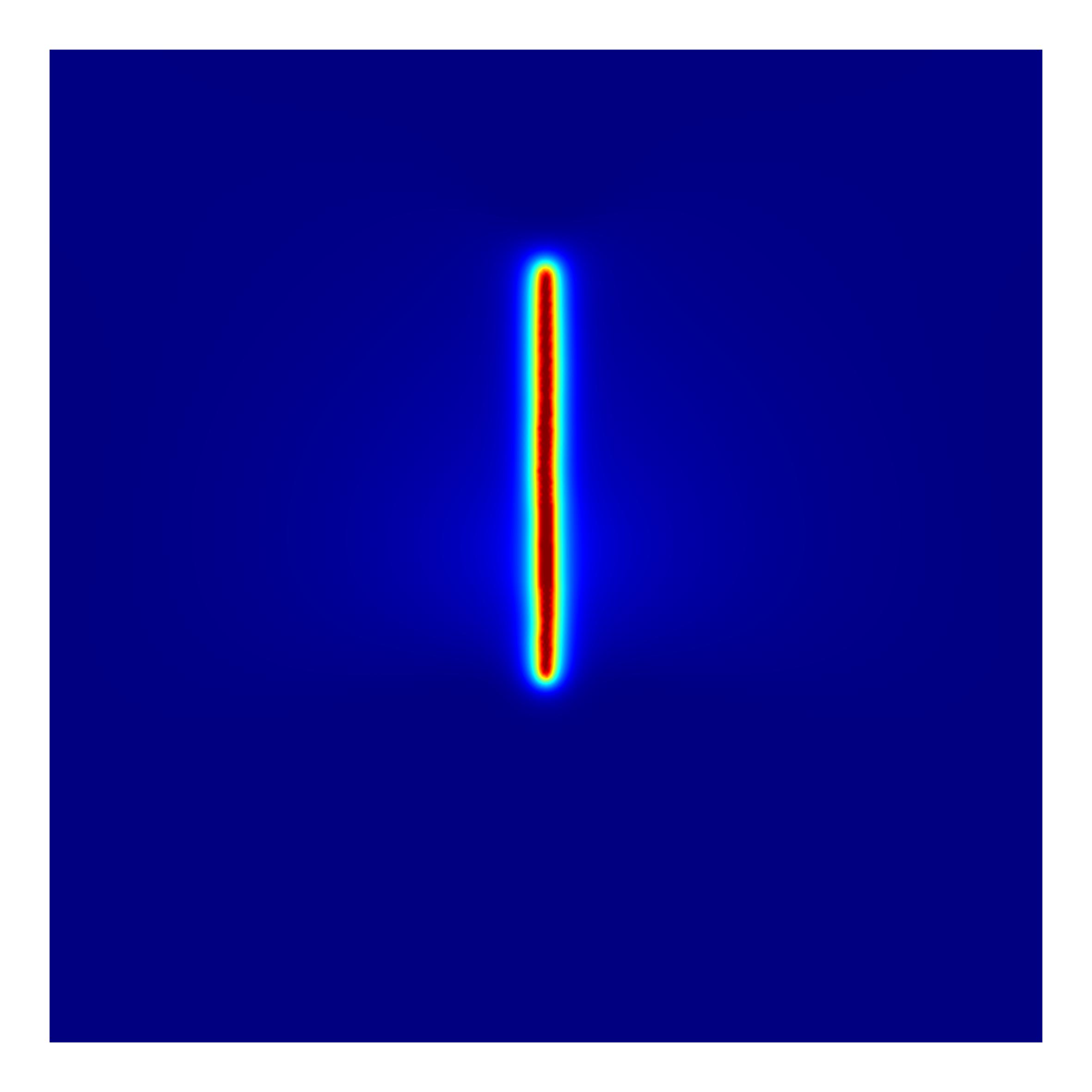}}
	\subfigure[$t=14$ s]{\includegraphics[width = 5.5cm]{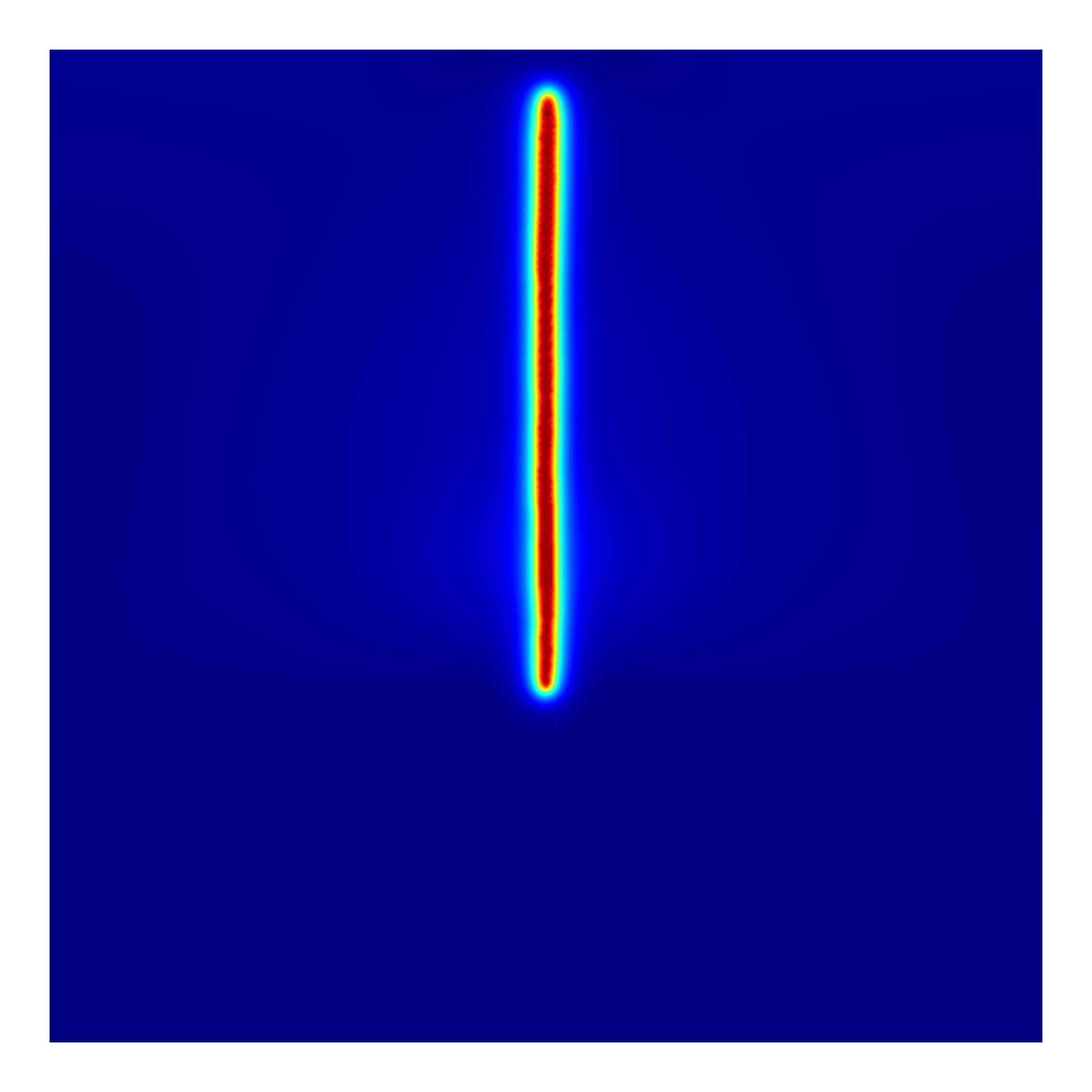}}\\
	\caption{Fracture propagation under varying vertical stress field}
	\label{Fracture propagation under varying vertical stress field}
	\end{figure}

\begin{figure}[htbp]
	\centering
	\includegraphics[width = 10cm]{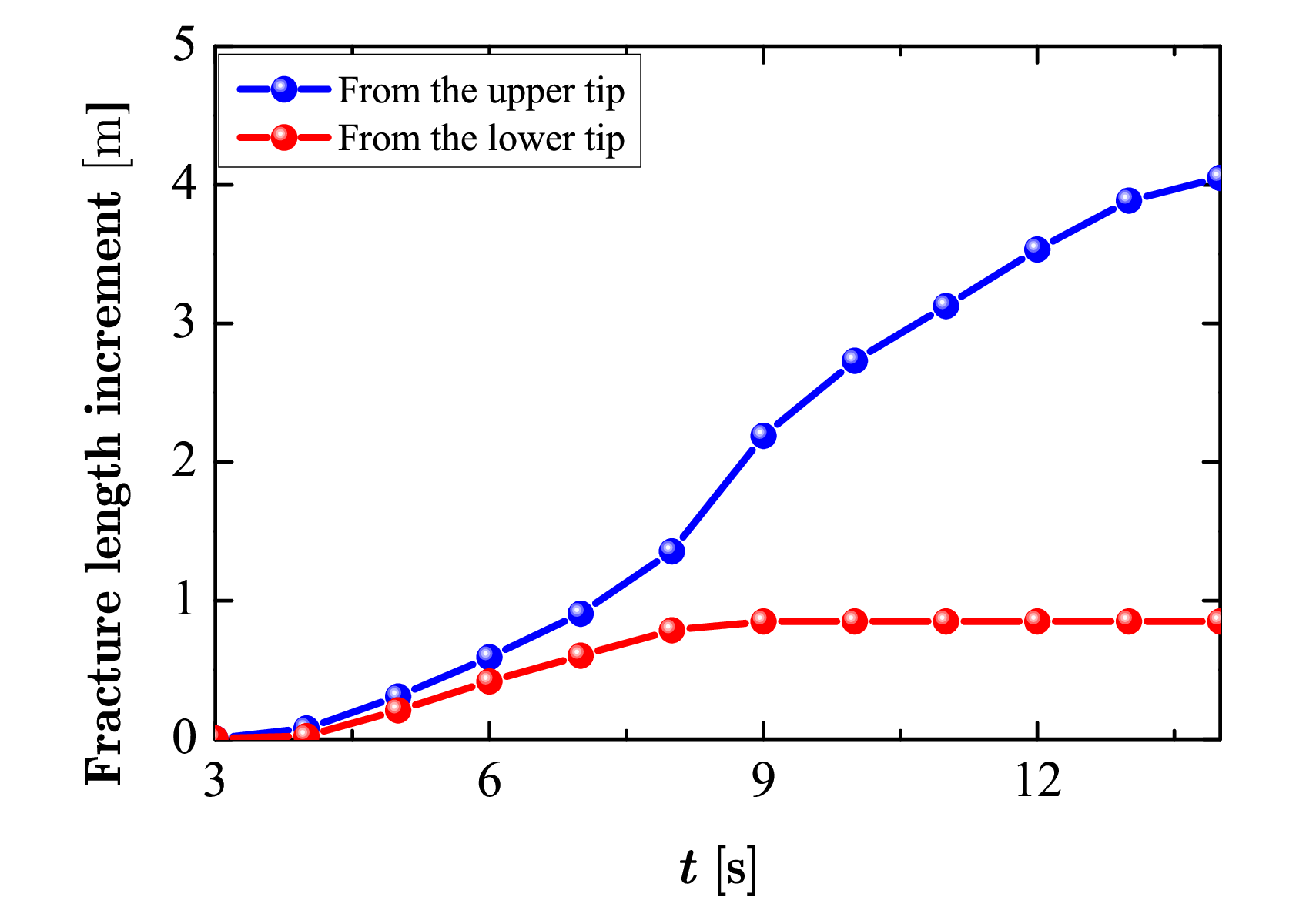}
	\caption{Fracture length increment under nearly varying stress field}
	\label{Fracture length increment under nearly varying stress field}
	\end{figure}

The phase field evolution at different time is shown in Fig. \ref{Fracture propagation under varying vertical stress field}. An asymmetric fracture pattern is observed and the fracture propagation is much easier towards the upper boundary than towards the bottom. This finding can be also verified in Fig. \ref{Fracture length increment under nearly varying stress field}, which depicts the fracture length increment at different time. Figure \ref{Fracture length increment under nearly varying stress field} indicates that the fracture length from the upper tip of the pre-existing notch is much larger than that from the lower tip. The downwards hydraulic fracture is hampered and even cannot propagate after $t=9$ s. Figures \ref{Fracture propagation under varying vertical stress field} and \ref{Fracture length increment under nearly varying stress field} strongly verifies that the hydraulic fracture tends to propagate towards the region with lowest fracture resistance.

In summary, the 2D examples in Subsections \ref{Fractures from a horizontal notch} to \ref{Linearly varying stress} indicate the sensitiveness of the hydraulic fracture propagation to the stress boundary condition. Our proposed model, which involves the effect of initial stress field, is feasible and practicable in capturing the effect of remote stresses on hydraulic fracture propagation and in producing the correct displacement field.

\section{3D example}\label{3D example}

In this section, we test the performance of our method in modeling 3D hydraulic fracture propagation. Here, the last example is a 3D isotropic medium with a penny-shaped initial notch. The fluid source in the notch is set as $10$ kg/(m$^3\cdot$s). The calculation domain is a cube with a dimension of $10\times10\times10$ m$^3$ and has the same center as the initial notch, the height of which is in the $z$ direction. The notch is parallel to the $x$-$y$ plane and the radius and height are 0.8 m and 0.4 m, respectively. 

The parameters for calculation are identical to those in Table \ref{Basic calculation parameters} except the length scale parameter is $l_0 = 0.25$ m and the permeability $k_f$ is $5.21\times10^{-3}$ m$^2$. We employ 6-node prism elements to discretize the 3D domain while the maximum element size is set as $0.18$ m for reducing computational cost. In addition, the time increment is set as 0.05 s in each simulation. In this 3D example, we only test the influence of the stress in the $z$ direction $\sigma_{z0}$. Therefore,  $\sigma_{z0}$ is set as 0.5, 1, and 5 MPa, respectively, while the stresses in the other two directions are fixed to 1 MPa.
 
\begin{figure}[htbp]
	\centering
	$\sigma_{z0}=0.5$ MPa\subfigure[$t=15$ s]{\includegraphics[width = 4.5cm]{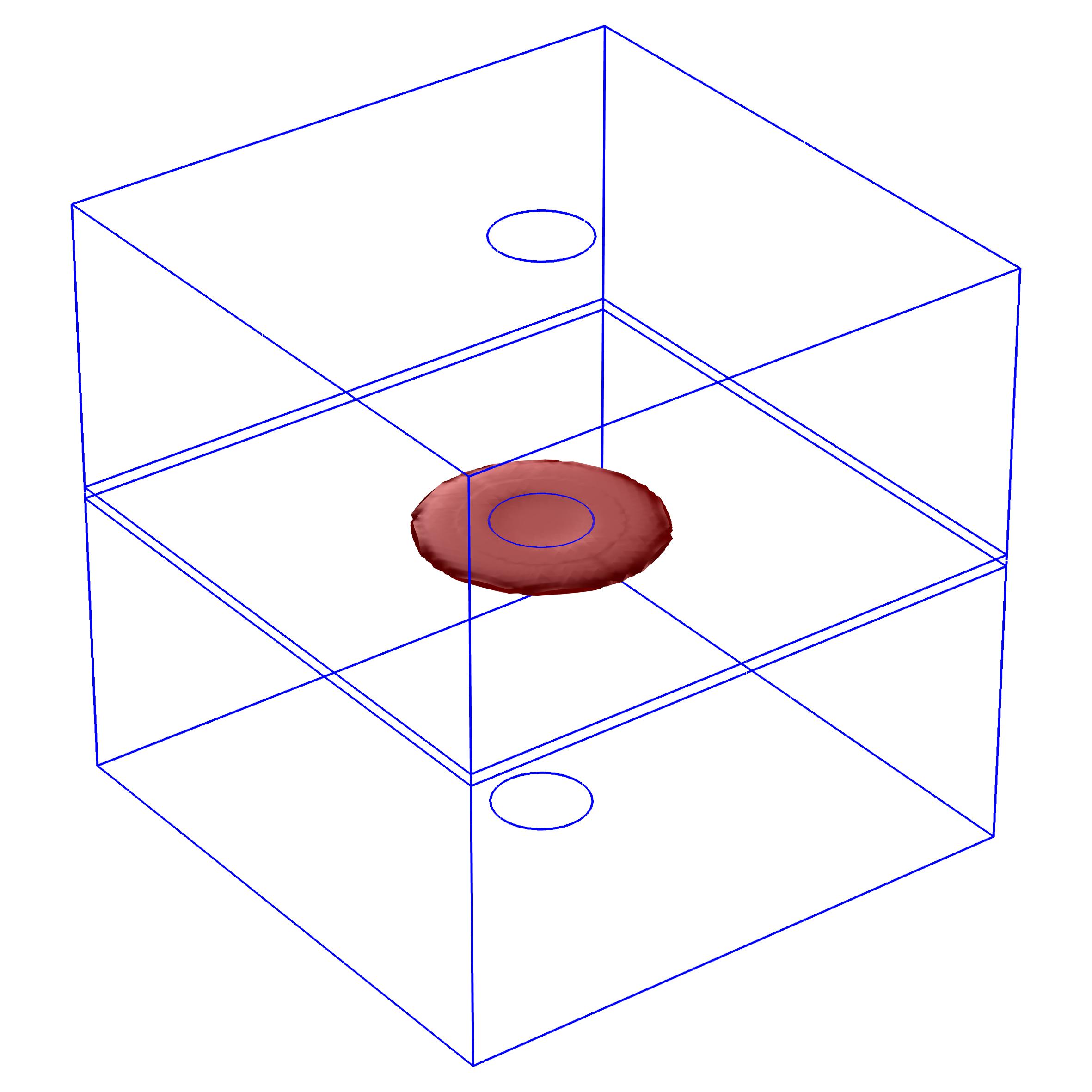}}
	\subfigure[$t=21$ s]{\includegraphics[width = 4.5cm]{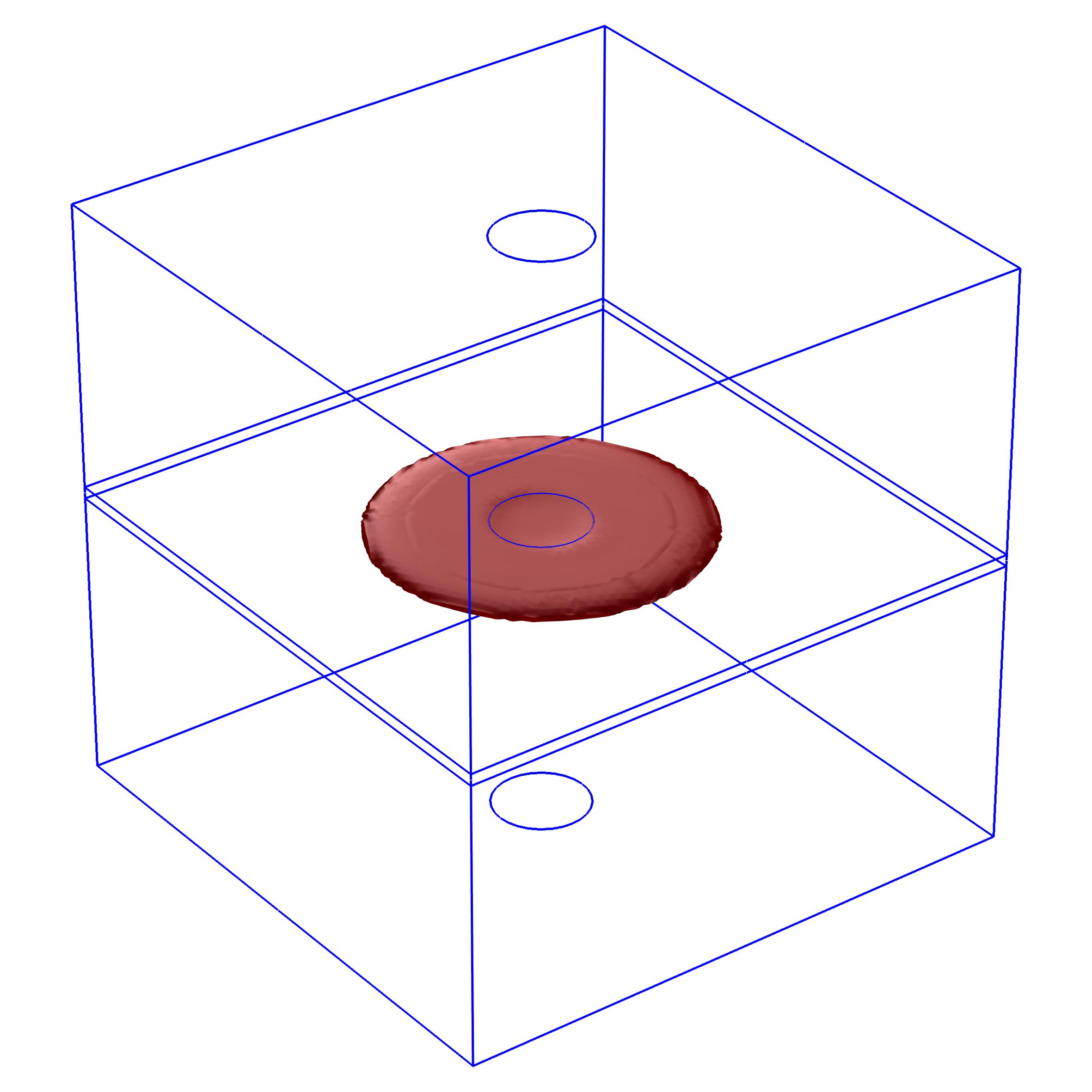}}
	\subfigure[$t=27$ s]{\includegraphics[width = 4.5cm]{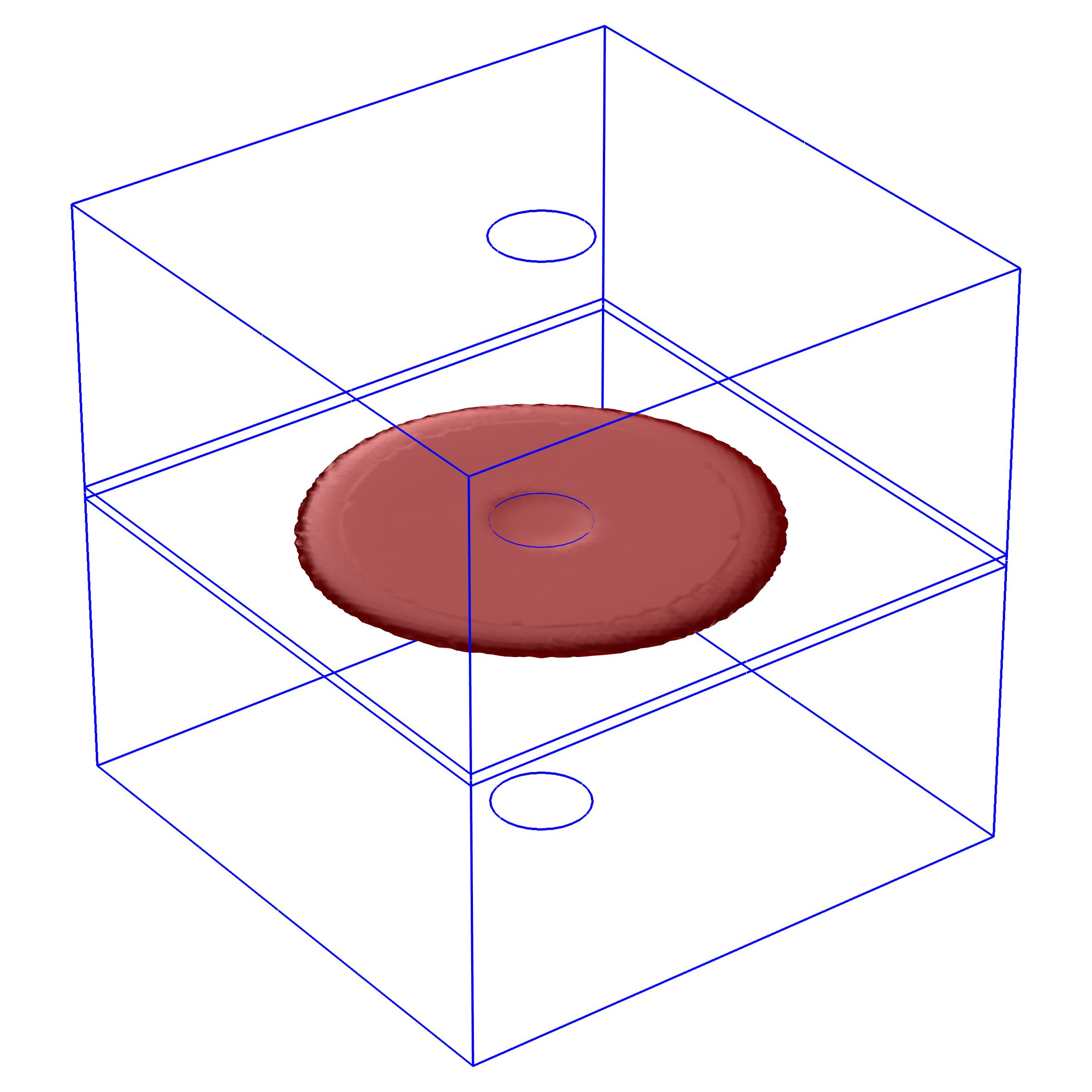}}\\
	$\sigma_{z0}=1$ MPa\subfigure[$t=15$ s]{\includegraphics[width = 4.5cm]{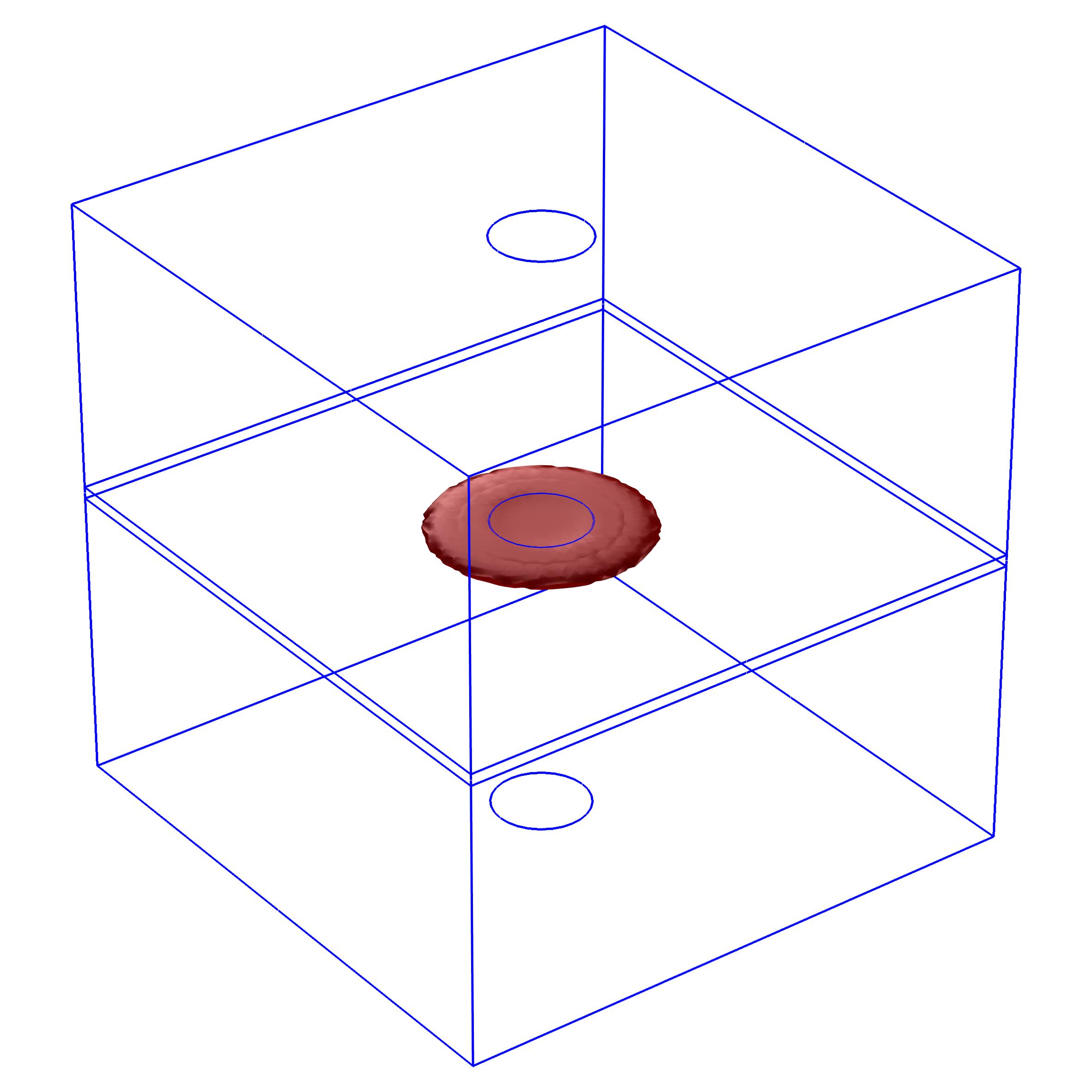}}
	\subfigure[$t=21$ s]{\includegraphics[width = 4.5cm]{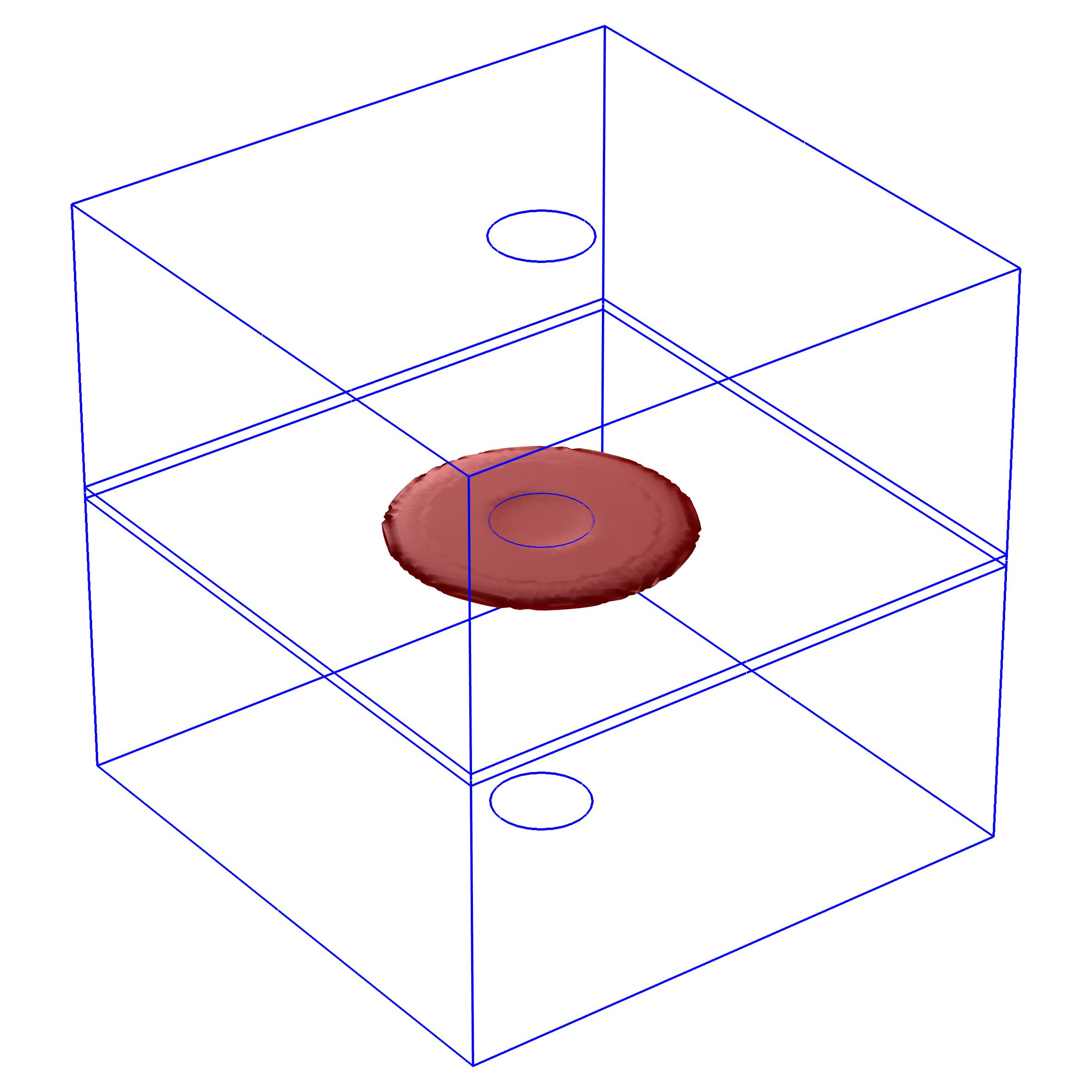}}
	\subfigure[$t=27$ s]{\includegraphics[width = 4.5cm]{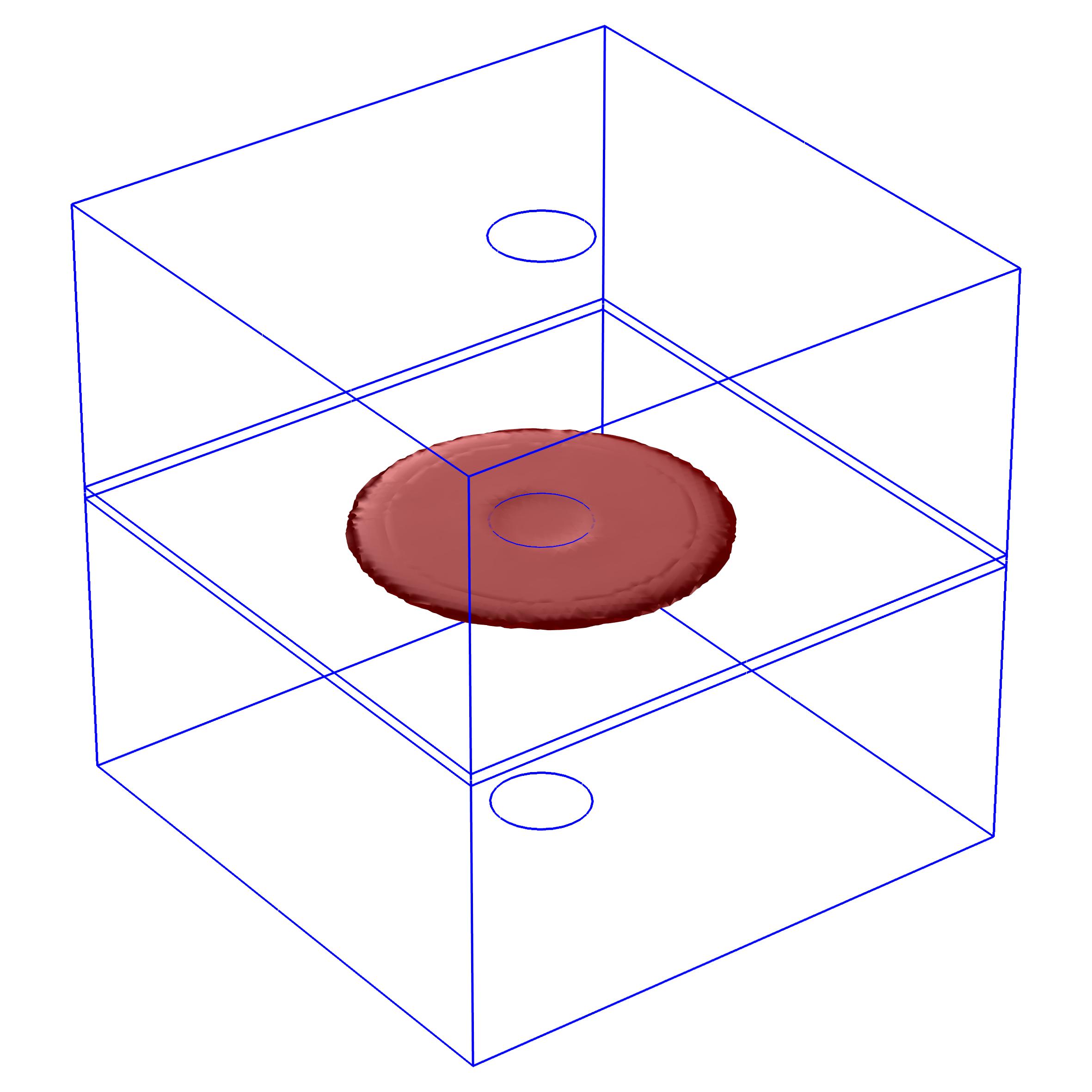}}\\
	$\sigma_{z0}=5$ MPa\subfigure[$t=15$ s]{\includegraphics[width = 4.5cm]{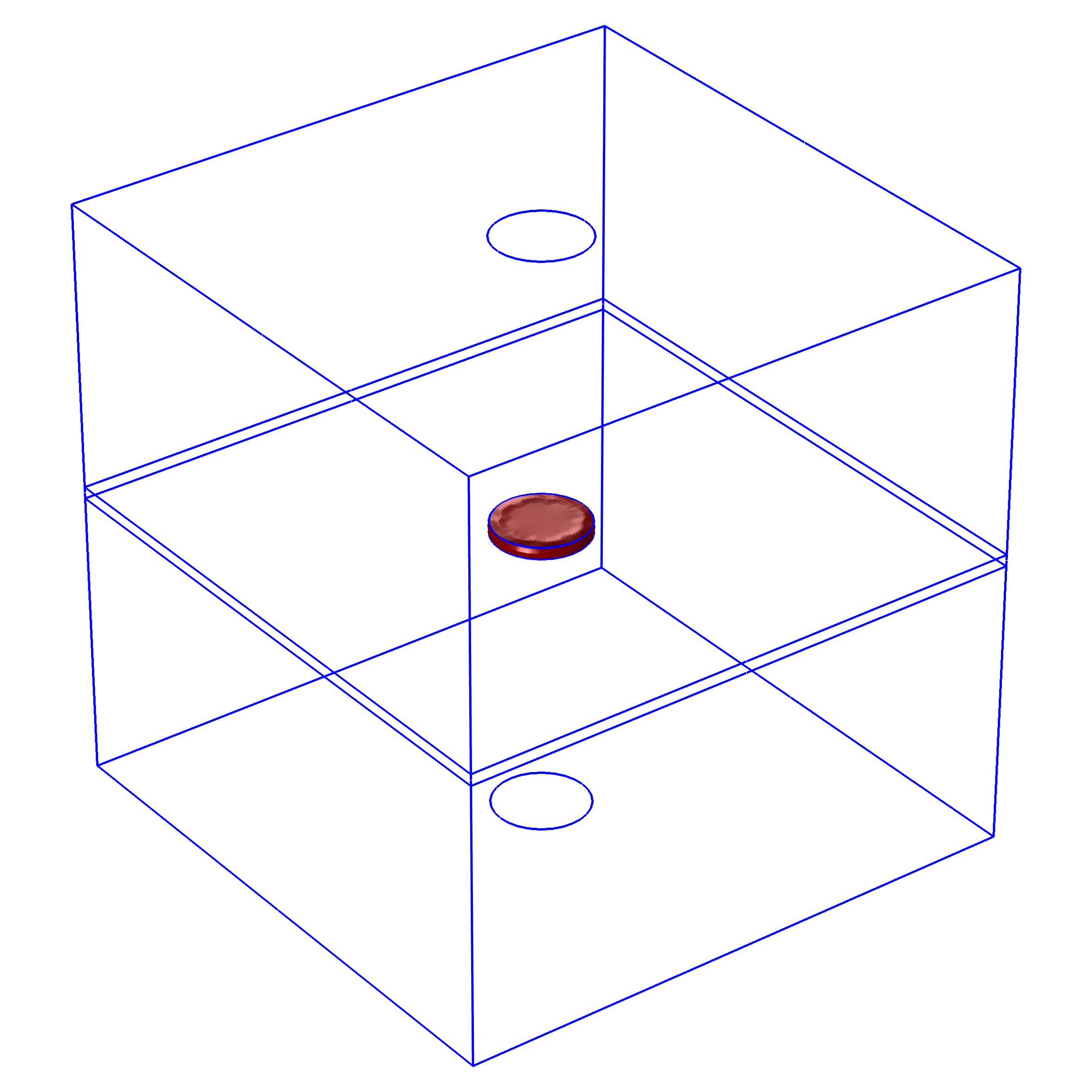}}
	\subfigure[$t=21$ s]{\includegraphics[width = 4.5cm]{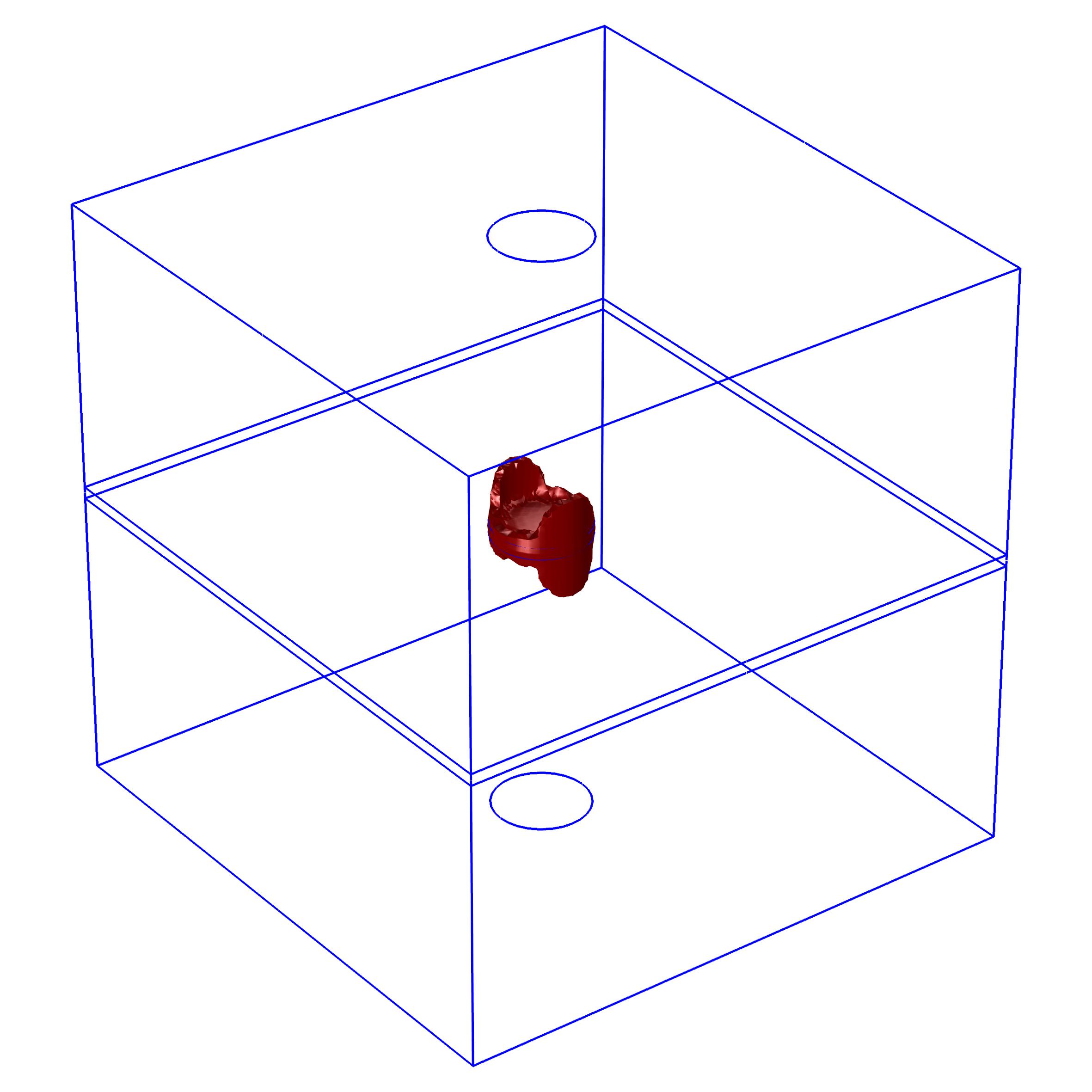}}
	\subfigure[$t=27$ s]{\includegraphics[width = 4.5cm]{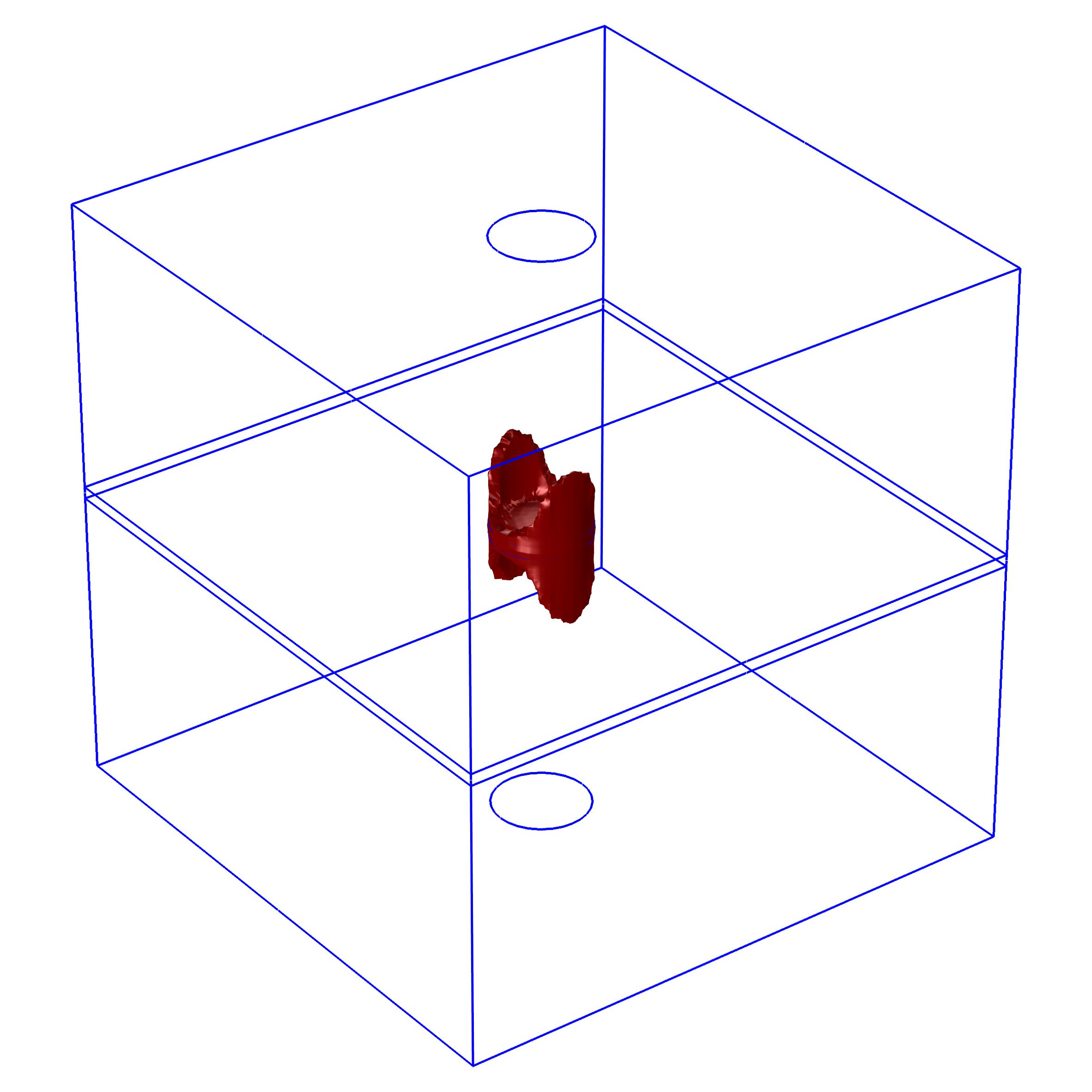}}\\		
	\caption{Fracture evolution of the 3D isotropic porous medium with an interior notch under different $\sigma_{z0}$}
	\label{Fracture evolution of the 3D isotropic porous medium with an interior notch under different}
	\end{figure}

Hydraulic fracture propagation patterns in the 3D isotropic medium are shown in Fig. \ref{Fracture evolution of the 3D isotropic porous medium with an interior notch under different}. It can be seen from Figs. \ref{Fracture evolution of the 3D isotropic porous medium with an interior notch under different}a-f that for $\sigma_{z0}=0.5$ MPa and 1 MPa, the fractures initiate and propagate only in the $x$-$y$ plane while the area of the fractured domain increases as $\sigma_{z0}$ decreases. Figures \ref{Fracture evolution of the 3D isotropic porous medium with an interior notch under different}g-i indicate that when $\sigma_{z0}$ is too large, the fracture propagation in the $x$-$y$ plane is hindered and fractures only propagate in the $x$-$z$ or $y$-$z$ plane.

\begin{figure}[htbp]
	\centering
	\includegraphics[width = 10cm]{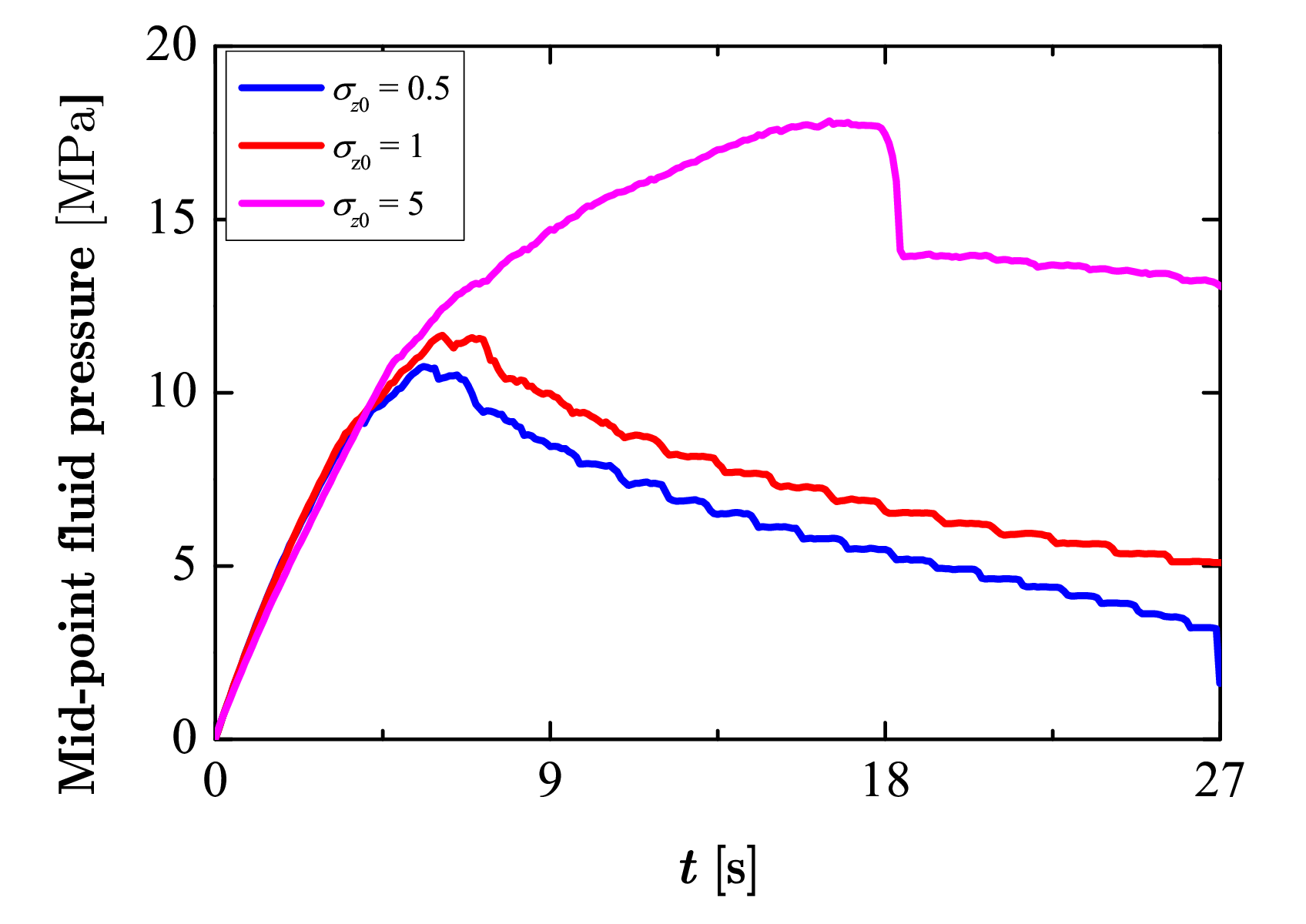}
	\caption{Fluid pressure-time curve for 3D porous medium}
	\label{Fluid pressure-time curve for 3D porous medium}
	\end{figure}

The fluid pressure-time curves under different $\sigma_{z0}$ are shown in Fig. \ref{Fluid pressure-time curve for 3D porous medium}. The fluid pressure is observed to drop after fracture initiation and the maximum fluid pressure increases with the increase in the in-situ stress $\sigma_{z0}$. Comparing Figs. \ref{Fracture evolution of the 3D isotropic porous medium with an interior notch under different} and \ref{Fluid pressure-time curve for 3D porous medium} indicate that the effect of in-situ stress on the 3D porous medium can be well captured by the proposed PFM in a fixed FE mesh without requiring any re-meshing or adaptive techniques. The 3D simulations fully verify the strong capability of our proposed PFM for predicting complex hydraulic fracture propagation in porous media subjected to stress boundary condition.

\section {Conclusions}\label{Conclusions}

A new phase field model for simulating quasi-static hydraulic fracture propagation in porous media subjected to stress boundary conditions is proposed. A new energy functional is established to consider the effect of initial in-situ stress field. This energy functional is then used to achieve the governing equations for the displacement and phase fields through the variational approach. Biot poroelasticity theory is used to couple the fluid pressure field and the displacement field while the phase field is used for determining the fluid properties from the intact domain to the fully broken domain.

The presented numerical examples in this work verify the capability of the proposed PFM in capturing complex hydraulic fracture growth patterns in 2D and 3D. The numerical examples also indicate that under stress boundary conditions the proposed approach can obtain correct displacement distribution and reflect the sensitiveness of hydraulic fracture propagation to stress boundary conditions. In future research, the proposed PFM can be more widely applied in HF practices where the stress boundaries are dominated and can be used to investigate the influence of naturally-layered porous media and multi-zone HF on fracture propagation.

\section*{Acknowledgment}
The authors gratefully acknowledge financial support provided by Deutsche Forschungsgemein-schaft (DFG ZH 459/3-1), and RISE-project BESTOFRAC (734370).

\section*{Appendix Finite element discretization}
We first derive the weak forms of all the governing equations as
	\begin{equation}
	\int_{\Omega}\left[(\bm\sigma^e+g(\phi)\bm\sigma_0-\alpha p\bm I):\delta \bm {\varepsilon}\right] \mathrm{d}\Omega = \int_{\Omega_{h}}\bm f_t \cdot \delta \bm u \mathrm{d}S
	\label{weak form 1}
	\end{equation}
	\begin{equation}
	\int_{\Omega}-2(1-k)H(1-\phi)\delta\phi\mathrm{d}\Omega+\int_{\Omega}G_c\left(l_0\nabla\phi\cdot\nabla\delta\phi+\frac{1}{l_0}\phi\delta\phi\right)\mathrm{d}\Omega=0
	\label{weak form 2}
	\end{equation}
	\begin{equation}
	\int_{\Omega} \rho S \frac{\partial p}{\partial t}\delta p\mathrm{d}\Omega-\int_{\Omega} \rho \bm v \cdot\nabla\delta p \mathrm{d}\Omega=\int_{\partial\Omega_N}M_n\mathrm{d}S+\int_{\Omega}\left(q_m-\rho\alpha\chi_R\frac{\partial \varepsilon_{vol}}{\partial t}\right)\mathrm{d}\Omega
	\label{weak form 3}
	\end{equation}  

In an element with $n$ nodes, the nodal values for the three fields ($\bm u$, $\phi$, and $p$) are defined as $\bm u_i$, $\phi_i$, and $p_i$ ($i=1,2,\cdots,n$). The fields are then discretized as follows,
	\begin{equation}
	\bm u = \sum_i^{n}N_i \bm u_i,\hspace{0.5cm} \phi = \sum_i^{n}N_i \phi_i,\hspace{0.5cm} p = \sum_i^{n}N_i p_i
	\end{equation}
\noindent where $N_i$ is the shape function at node $i$. We then derive the gradients of the three fields as
	\begin{equation}
	\bm \varepsilon = \sum_i^{n}\mathbf B_i^u \bm u_i,\hspace{0.5cm} \nabla\phi = \sum_i^{n}\mathbf B_i^{\phi} \phi_i,\hspace{0.5cm} \nabla p = \sum_i^{n}\mathbf B_i^{p} p_i
	\end{equation}
\noindent where $\mathbf B_i^u$, $\mathbf B_i^\phi$, and $\mathbf B_i^p$ are derivatives of the shape functions:
	\begin{equation}
	\mathbf B_i^u=\left[
	\begin{array}{cccccc}
	N_{i,x}&0&0&N_{i,y}&0&N_{i,z}\\
	0&N_{i,y}&0&N_{i,x}&N_{i,z}&0\\
	0&0&N_{i,z}&0&N_{i,y}&N_{i,x}\\
	\end{array}\right]^{\mathrm{T}},
	\hspace{0.5cm}\mathbf B_i^{\phi}=\mathbf B_i^p=\left[
	\begin{array}{ccc}
	N_{i,x}\\
	N_{i,y}\\
	N_{i,z}
	\end{array}\right]\label{gradient of shape function 3D}
	\end{equation}

For 2D, the components along the $z$ direction is removed from above equations \eqref{gradient of shape function 3D}. Due to the arbitrariness in the test functions, the external force $\mathbf F_i^{u,ext}$ and inner force $\mathbf F_i^{u,int}$ for the displacement field are described by
	\begin{equation}
	\left\{
	\begin{aligned}
	\mathbf F_i^{u,ext} &= \int_{\Omega_{h}}N_i\bm f_t \mathrm{d}S+ \int_{\Omega}[\mathbf B_{i}^{u}]^{\mathrm T}\alpha p \bm I \mathrm{d}\Omega - \int_{\Omega}[\mathbf B_{i}^{u}]^{\mathrm T} g(\phi)\bm \sigma_0 \mathrm{d}\Omega\\
	\mathbf F_i^{u,int} &= \int_{\Omega}[\mathbf B_i^u]^{\mathrm T}\bm\sigma \mathrm{d}\Omega
	\end{aligned}
	\right.
	\end{equation}

The inner force term of the phase field is also obtained by
	\begin{equation} 
	 F_i^{\phi,int} = \int_{\Omega} -2(1-k)(1-\phi)H N_i+G_c\left(l_0[\mathbf B_i^{\phi}]^{\mathrm T}\nabla \phi+\frac 1 {l_0}\phi N_i \right ) \mathrm{d}\Omega
	\end{equation}

Finally, for the pressure field, the inner force $ F_i^{p,int}$, viscous force $F_i^{p,vis}$, and external force $F_i^{p,ext}$ are given by
 
	\begin{equation}
	\left\{
	\begin{aligned}
	F_i^{p,int} &= \int_{\Omega}[\mathbf B_i^p]^{\mathrm T}\frac{\rho K_{eff}}{\mu_{eff}}\nabla p \mathrm{d}\Omega\\
	F_i^{p,vis} &= \int_{\Omega}N_i \rho S \frac{\partial p}{\partial t} \mathrm{d}\Omega\\
	F_i^{p,ext} &=  \int_{\Omega}N_i \left( q_m-\rho\alpha\chi_R\frac{\partial \varepsilon_{vol}}{\partial t} \right) \mathrm{d}\Omega+ \int_{\partial\Omega_{N}}N_i M_N \mathrm{d}S
	\end{aligned}
	\right.
	\end{equation}
	
Thus, contribution of node $i$ to the residual of the discrete equations for the three field is written as
	\begin{equation} 
	\left\{
	\begin{aligned}
	\mathbf R_i^{u}&=\mathbf F_i^{u,ext}-\mathbf F_i^{u,int}\\
	R_i^{\phi}&=- F_i^{\phi,int}\\
	R_i^{p}&=F_i^{p,ext}- F_i^{p,int}- F_i^{p,vis}
	\end{aligned}
	\right.
	\end{equation}

Because the staggered scheme is used to solve the displacement, phase field and fluid pressure sequentially. We also adopt the Newton-Raphson approach sequentially to achieve $\mathbf R_i^{u}=\bm 0$, $ R_i^{\phi}=0$, and $ R_i^{p}=0$ for the three fields. In addition, the tangents on the element level are calculated by
 
	\begin{equation}
	\left\{
	\begin{aligned} 
	\mathbf K_{ij}^{uu}&=\frac{\partial \mathbf F_i^{u,int}}{\partial \bm u_j}=\int_{\Omega}[\mathbf B_i^u]^{\mathrm T}\mathbf D_e [\mathbf B_j^u] \mathrm{d} \Omega\\
	\mathbf K_{ij}^{\phi\phi}&=\frac{\partial F_i^{\phi,int}}{\partial \phi_j}=\int_{\Omega}\left\{[\mathbf B_i^\phi]^{\mathrm T} G_cl_0 [\mathbf B_j^\phi]+N_i\left(2(1-k)H+\frac{G_c}{l_0}\right)N_j\right\} \mathrm{d} \Omega\\
	\mathbf K_{ij}^{pp}&=\frac{\partial F_i^{p,int}}{\partial p_j}=\int_{\Omega}[\mathbf B_i^p]^{\mathrm T} \frac{\rho K_{eff}}{\mu_{eff}} [\mathbf B_j^p] \mathrm{d} \Omega
	\end{aligned}
	\right.
	\end{equation}
\noindent where $\mathbf D_e$ is the elasticity matrices derived from the elasticity tensor $\bm D = {\partial \bm\sigma^e}/{\partial \bm\varepsilon}$.

\bibliography{references}

%\onecolumn
%\tableofcontents
\end{document}